\documentclass{elsart}
\usepackage{graphics}
\usepackage{graphicx}
\usepackage{amssymb}

\def\gsim{\;\raise0.3ex\hbox{$>$\kern-0.75em\raise-1.1ex\hbox{$\sim$}}\;}
\def\lsim{\;\raise0.3ex\hbox{$<$\kern-0.75em\raise-1.1ex\hbox{$\sim$}}\;}

\begin{document}

\begin{frontmatter}
{\hfill \small LAPTH-1131/05, IFIC/05-59}

\title{Massive neutrinos and cosmology}

\author[LAPTH]{Julien Lesgourgues} 
\ead{lesgourg@lapp.in2p3.fr}
and
\author[IFIC]{Sergio Pastor}
\ead{pastor@ific.uv.es}

\address[LAPTH]{Laboratoire de Physique Th\'{e}orique LAPTH
(CNRS-Universit\'{e} de Savoie)\\ B.P.\ 110, F-74941 Annecy-le-Vieux
Cedex, France}

\address[IFIC]{Instituto de F\'{\i}sica Corpuscular (CSIC-Universitat
de Val\`encia)\\ Ed.\ Institutos de Investigaci\'on, Apdo.\ 22085,
E-46071 Valencia, Spain}

\begin{abstract}
The present experimental results on neutrino flavour oscillations
provide evidence for non-zero neutrino masses, but give no hint on
their absolute mass scale, which is the target of beta decay and
neutrinoless double-beta decay experiments. Crucial complementary
information on neutrino masses can be obtained from the analysis of
data on cosmological observables, such as the anisotropies of the
cosmic microwave background or the distribution of large-scale
structure. In this review we describe in detail how free-streaming
massive neutrinos affect the evolution of cosmological perturbations.
We summarize the current bounds on the sum of neutrino 
masses that can be derived from various combinations of cosmological data,
including the most recent analysis by the WMAP team.
We also discuss how future cosmological experiments are expected to be
sensitive to neutrino masses well into the sub-eV range.
\end{abstract}

\end{frontmatter}

\setcounter{tocdepth}{3}
\tableofcontents

\section{Introduction}
\label{sec:intro}

Neutrino cosmology is a fascinating example of the fecund interaction
between particle physics and astrophysics.  At the present time, the
researchers working on neutrino physics know that the advances in this
field need a combined effort in the two areas.

{}From the point of view of cosmologists, the idea that massive
neutrinos could play a significant role in the history of the Universe
and in the formation of structures has been discussed for more than
thirty years, first as a pure speculation. However, nowadays we know
from experimental results on flavour neutrino oscillations that
neutrinos are massive. At least two neutrino states have a large
enough mass for being non-relativistic today, thus making up a small
fraction of the dark matter of the Universe.  At a stage in which
cosmology reaches high precision thanks to the large
amount of observational data, it is unavoidable to take into account
the presence of massive neutrinos. In particular, the observable
matter density power spectrum is damped on small scales by massive neutrinos.
This effect can range from a few per cent for
$0.05-0.1$ eV masses, the minimal values of the total neutrino mass
compatible with oscillation data, up to $10-20\%$ in the limit of
three degenerate masses.

{}From the point of view of particle physicists, fixing the absolute
neutrino mass scale (or, equivalently, the lightest neutrino mass once
the data on flavour oscillations are taken into account) is the target
of terrestrial experiments such as the searches for neutrinoless
double beta decay or tritium beta decay experiments, a difficult task
in spite of huge efforts and very promising scheduled experiments.  At
the moment, the best bounds come from the analysis of cosmological
data, from the requirement that neutrinos did not wash out too much of
the small-scale cosmological structures. Recently the cosmological
limits on neutrino masses progressed in a spectacular way, in
particular thanks to the precise observation of the Cosmic Microwave
Background (CMB) anisotropies by the Wilkinson Microwave Anisotropy
Probe (WMAP) satellite, and to the results of a new generation of very
deep galaxy redshift surveys. Neutrino physicists look with anxiety at
any new development in this field, since in the next years many
cosmological observations will be available with unprecedented
precision.

At this very exciting moment, which might be preceding a breaking
discovery within a few years, our aim is to present here a clear and
comprehensive review of the role of massive neutrinos in cosmology,
including a description of the underlying theory of cosmological
perturbations, a summary of the current bounds and a review of the
sensitivities expected for future cosmological observations.  Our
main goal is to address this manuscript simultaneously to particle
physicists and cosmologists: for each aspect, we try to avoid useless
jargon and to present a self-contained summary. Compared to recent
short reviews on the subject, such as Refs.\
\cite{Hannestad:2004nb,Elgaroy:2004rc,Hannestad:2005ey,Tegmark:2005cy},
we tried to present a more detailed discussion.  At the same time, we
will focus on the case of three flavour neutrinos with masses in
accordance with the current non-cosmological data. A recent review on
primordial neutrinos can be found in \cite{Hannestad:2006zg}, while
for many other aspects of neutrino cosmology we refer the
reader to \cite{Dolgov:2002wy}. Finally, a more general review on
the connection between particle physics and cosmology can be found in
\cite{Kamionkowski:1999qc}.

We begin in Sec.\ \ref{sec:numasses} with an introduction to flavour
neutrino oscillations, and their implications for neutrino masses in
the standard three-neutrino scenario. We also briefly review the
limits on neutrino masses from laboratory experiments. Then, in Sec.\
\ref{sec:CNB}, we describe some basic properties of the Cosmic Neutrino
Background. According to the Big Bang cosmological model, each of the
three flavour neutrinos were in thermal equilibrium in the Early
universe, and then decoupled while still relativistic. We review the
consequences of this simple assumption for the phase-space
distribution and number density of the relic neutrinos, and we
summarize how this standard picture is confirmed by observations from
Big Bang Nucleosynthesis (BBN) and cosmological perturbations.
Finally we explain why neutrinos are expected to constitute a fraction
of the dark matter today.

In Sec.\ \ref{sec:mass_cosmo}, we describe the impact of massive
neutrinos on cosmological perturbations. This section is crucial for
understanding the rest, since current and future bounds rely precisely
on the observation of cosmological perturbations. Although the theory
of cosmological perturbations is a very vast and technical topic,
we try here to be accessible both to particle physicists willing to
learn the field, and to cosmologists willing to understand better the
specific role of neutrinos. We introduce the minimal number
of concepts, technicalities and equations for understanding the main
results related to neutrinos. Still this section is quite long because
we wanted to make it self-contained, and to explain many intermediate
steps that are usually hidden in most works on the subject.

In Sec.\ \ref{subsec:observables}, we define the quantities that can
be observed, i.e.\ those that we want to compute. 
In Sec.\ \ref{subsec:background}, we describe the evolution of
homogeneous quantities and in Sec.\ \ref{subsec:gauge} we define the
general setup for the theory of cosmological perturbations. Then in
Sec.\ \ref{subsec:pert_no_nu} we described in a rather simplified way
what the evolution of perturbations would look like in absence of
neutrinos. For a cosmologist, these four sections are part of common
knowledge and can be skipped.  In Sec.\ \ref{subsec:pert_nu}, we
present a detailed description of the impact of neutrinos on
cosmological perturbations. This long section is the most technical
part of this work, so the reader who wants to avoid technicalities and
to know the results can go directly to Sec.\
\ref{subsec:sum_mass-effect}, for a comprehensive summary of the
effects of neutrino masses on cosmological observables.

Then, in Sec. \ref{sec:present} we review the current existing bounds
on neutrino masses, starting from those involving only CMB
observations, and then adding different sets of data on the
distribution of the Large Scale Structure (LSS) of the Universe, which
include those coming from galaxy redshift surveys and the
Lyman-$\alpha$ forest. We explain why a unique cosmological bound on
neutrino masses does not exist, and why there are significant
variations from paper to paper, depending on the data included and the
assumed cosmological model.

Finally, in Sec.\ \ref{sec:future} we describe the prospects for the
next ten or fifteen years. We summarize the forecasts which can be
found in the literature concerning the sensitivity to neutrino masses
of future CMB experiments, galaxy redshift surveys, weak lensing
observations (either from the CMB or from galaxy ellipticity) and
galaxy cluster surveys. We conclude with some general remarks in Sec.\
\ref{sec:concl}.

\section{Neutrino oscillations and absolute neutrino mass searches}
\label{sec:numasses}

Neutrinos have played a fundamental role in the understanding of weak
interactions since they were postulated by W.\ Pauli in 1930 to
safeguard energy conservation in beta decay processes. These
chargeless leptons are massless in the framework of the succesful
Standard Model (SM) of particle physics. However, this is an
accidental prediction of the SM, and there are many well-motivated
extended models where neutrinos acquire mass and other non-trivial
properties (see e.g.\
\cite{Gonzalez-Garcia:2002dz,Hirsch:2004he,Altarelli:2004za,Mohapatra:2006gs}).
Thus the measurement of neutrino masses could give us some hints on
the new fundamental theory, of which the SM is just the low-energy
limit, and in particular on new energy scales.

It was realized by B.\ Pontecorvo in 1957 that if neutrinos were massive
there could exist processes where the neutrino flavour is not
conserved, that we call neutrino oscillations. For small neutrino
masses, these oscillations actually take place on macroscopic
distances and can be measured if we are able to detect neutrinos from
distant sources, identify their flavour and compare the results with
the theoretical predictions for the initial neutrino fluxes.
Actually, after decades of experimental efforts in underground
facilities, neutrinos have been detected from various natural or
artificial sources. The former include the neutrinos produced in the
nuclear reactions in the Sun, as secondary particles from the
interaction of cosmic rays in the atmosphere of the Earth and even a
few neutrinos were detected from a supernova explosion (SN1987A).
We have also measured the neutrino fluxes originated in artificial
sources such as nuclear reactors or accelerators.

Nowadays there exist compelling evidences for flavour neutrino
oscillations from a variety of experimental data on solar,
atmospheric, reactor and accelerator neutrinos. These are very
important results, because the existence of flavour change implies
that neutrinos mix and have non-zero masses, which in turn requires
particle physics beyond the SM. There are many excellent reviews on
neutrino oscillations and their implications, to which we refer the
reader for more details (see e.g.\ the recent ones
\cite{Maltoni:2004ei,Fogli:2005cq}).

We know that the number of light neutrinos sensitive to weak
interactions (flavour or active neutrinos) equals three from the
analysis of the invisible $Z$-boson width at LEP, $N_\nu=2.994 \pm
0.012$ \cite{Eidelman:2004wy}, and the three flavour neutrinos
($\nu_e,\nu_\mu,\nu_\tau$) are linear combinations of states with
definite mass $\nu_i$, where $i$ is the number of massive neutrinos.

In a three-neutrino scenario flavour and mass eigenstates are related
by the mixing matrix $U$, parametrized as
\cite{Eidelman:2004wy,Fogli:2005cq}
\begin{equation}
\left(
    \begin{array}{ccc}
        c_{12} c_{13}
        & s_{12} c_{13}
        & s_{13}e^{-i\delta} \\
        -s_{12} c_{23} - c_{12} s_{23} s_{13}e^{i\delta}
        & c_{12} c_{23} - s_{12} s_{23} s_{13}e^{i\delta}
        & s_{23} c_{13} \\
        s_{12} s_{23} - c_{12} c_{23} s_{13}e^{i\delta}
        & -c_{12} s_{23} - s_{12} c_{23} s_{13}e^{i\delta}
        & c_{23} c_{13}
\end{array}\right)
\,\,.
\end{equation}
Here $c_{ij}=\cos \theta_{ij}$ and $s_{ij}=\sin \theta_{ij}$ for
$ij=12, 23$ or $13$, and $\delta$ is a CP-violating phase. Together
with the three masses, in total there are seven flavour parameters in
the neutrino sector. This would be all if neutrinos were Dirac
particles, like the charged leptons, and the total lepton number is
conserved. Instead, if neutrinos are Majorana particles (i.e.\ a
neutrino is its own antiparticle), the matrix $U$ is multiplied by a
diagonal matrix of phases that can be taken as ${\rm
diag}(1,e^{i\phi_2/2},e^{i(\phi_3+2\delta)/2})$. These two phases do
not show up in neutrino oscillations, but appear in
lepton-number-violating processes such as neutrinoless double beta
decay, as discussed later.
\begin{figure}[t]
\begin{center}
\includegraphics[width=.75\textwidth]{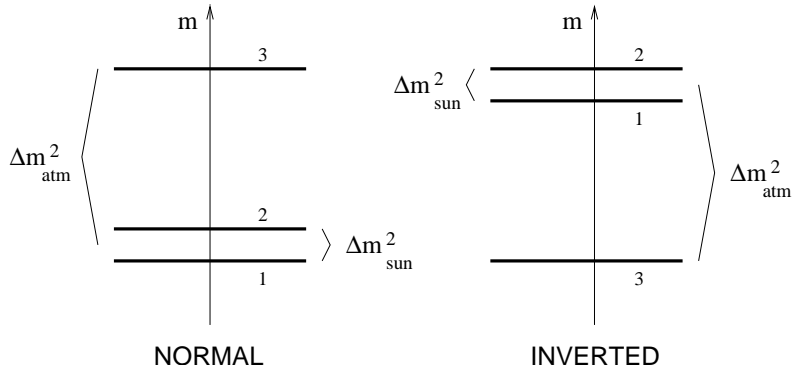}
\caption{\label{fig:nuschemes} The two neutrino schemes allowed if
$\Delta m_{\rm atm}^2\gg \Delta m_{\rm sun}^2$: normal hierarchy (NH)
and inverted hierarchy (IH).}
\end{center}
\end{figure}

\begin{figure}[t]
\begin{center}
\vspace{-2cm}
\hspace{-1.5cm}
\includegraphics[width=.92\textwidth]{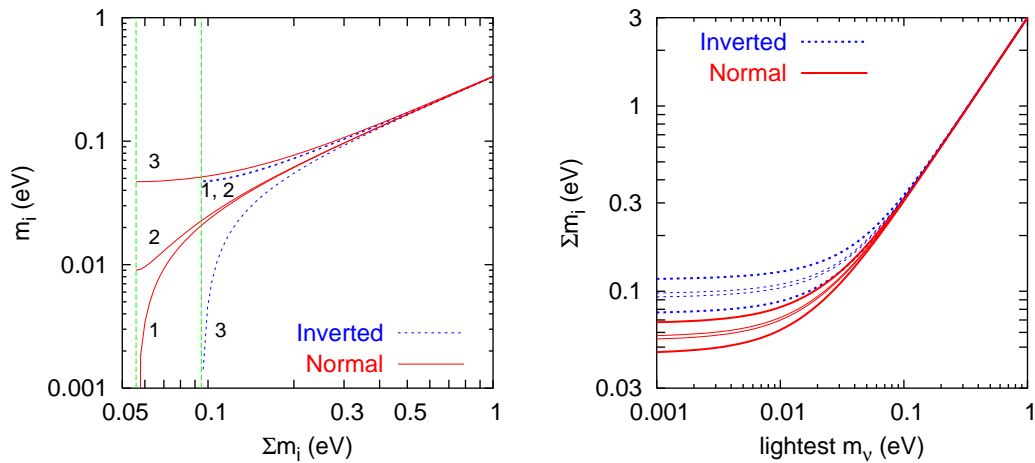}
\caption{\label{fig:numasses} Expected values of neutrino masses
according to the values in Eq.\ (\ref{oscpardef}).  Left: individual
neutrino masses as a function of the total mass for the best-fit
values of the $\Delta m^2$.  Right: ranges of total neutrino mass as a
function of the lightest state within the $3\sigma$ regions (thick lines)
and for a future determination at the $5\%$ level (thin lines).}
\end{center}
\end{figure}

Oscillation experiments can measure the differences of squared
neutrino masses $\Delta m^2_{21} = m^2_2 - m^2_1$ and $\Delta m^2_{31}
= m^2_3 - m^2_1$, the relevant ones for solar and atmospheric
neutrinos, respectively. As a reference, we take the following
$3\sigma$ ranges of mixing parameters from  an update of
ref.\ \cite{Maltoni:2004ei},
\begin{eqnarray}
\Delta m^2_{21} &=& (7.9_{-0.8}^{+1.0})\times 10^{-5}~{\rm eV}^2
\qquad
|\Delta m^2_{31}| = (2.2_{-0.8}^{+1.1})\times 10^{-3}~{\rm eV}^2
\nonumber \\
s_{12}^2  &=&  0.30_{-0.06}^{+0.10}\qquad
s_{23}^2  =  0.50_{-0.16}^{+0.18}\qquad
s_{13}^2  \leq  0.043
\label{oscpardef}
\end{eqnarray}
Unfortunately oscillation experiments are insensitive to the absolute
scale of neutrino masses, since the knowledge of $\Delta m^2_{21}>0$
and $|\Delta m^2_{31}|$ leads to the two possible schemes shown in
Fig.\ \ref{fig:nuschemes}, but leaves one neutrino mass unconstrained
(see e.g.\ the discussion in the reviews
\cite{Fogli:2005cq,Elliott:2002xe,Bilenky:2002aw,Barger:2003qi,McKeown:2004yq}).
These two schemes are known as normal (NH) and inverted (IH)
hierarchies, characterized by the sign of $\Delta m^2_{31}$, positive
and negative, respectively.  For small values of the lightest neutrino
mass $m_0$, i.e.\ $m_1$ ($m_3$) for NH (IH), the mass states follow a
hierarchical scenario, while for masses much larger than the
differences all neutrinos share in practice the same mass and then we
say that they are degenerate. In general, the relation between the
individual masses and the total neutrino mass can be found
numerically, as shown in Fig.\ \ref{fig:numasses}.

It is also possible that the number of massive neutrino states is
larger than the number of flavor neutrinos. In such a case, in order
to not violate the LEP results the extra neutrino states must be
sterile, i.e.\ singlets of the SM gauge group and thus insensitive to
weak interactions. At present, the results of the Liquid Scintillator
Neutrino Detector (LSND) \cite{Aguilar:2001ty}, an experiment that has
measured the appearance of electron antineutrinos in a muon
antineutrino beam, constitute an independent evidence of neutrino
conversions at a larger mass difference than those in Eq.\
\ref{oscpardef}. In such a case, a fourth sterile neutrino is required
with mass of ${\mathcal O}$(eV) \cite{Maltoni:2004ei}. The LSND
results will be checked by the ongoing MiniBoone experiment
\cite{MiniBoone}, whose first data are expected for 2006. In this
review, we will mainly consider the three-neutrino scenario, but we
briefly comment on the cosmological bounds on the four-neutrino
mass schemes that include the LSND results in Sec.\
\ref{subsec:extraparam}.

As we discuss in the next sections, cosmology is at first order
sensitive to the total neutrino mass if all states have the same
number density, providing information on $m_0$ but blind to neutrino
mixing angles or possible CP violating phases. Thus cosmological
results are complementary to terrestrial experiments such as beta
decay and neutrinoless double beta decay, which are
respectively sensitive to the effective masses
\begin{eqnarray}
m_\beta &=& \left (\sum_i |U_{ei}|^2\, m_i^2\right )^{1/2}
= (c_{12}^2c_{13}^2\, m_1^2+s_{12}^2c_{13}^2\, m_2^2
+s_{13}^2\, m_3^2)^{1/2},
\nonumber \\
m_{\beta\beta} &=& \left |\sum_i U_{ei}^2\,  m_i\right |
= |c_{12}^2c_{13}^2\, m_1+s_{12}^2c_{13}^2\, m_2\, e^{i\phi_2}
+s_{13}^2\, m_3\, e^{i\phi_3}|~.
\label{beta_2beta}
\end{eqnarray}

The search for a signal of non-zero neutrino masses in a beta decay
experiment is, in principle, the best strategy for measuring directly the
neutrino mass, since it involves only the kinematics of electrons (see
\cite{Eitel:2005hg} for a recent review). The current limits from
tritium beta decay apply only to the range of degenerate neutrino
masses, so that $m_\beta \simeq m_0$ (if unitarity of $U$ is assumed).
The bound at $95\%$ CL is $m_0<2.05-2.3$ eV from the Troitsk
\cite{Lobashev:2003kt} and Mainz \cite{Kraus:2004zw} experiments,
respectively. This value is expected to be improved by the KATRIN
project \cite{Osipowicz:2001sq}
to reach a discovery potential for
$0.3-0.35$ eV masses (or a sensitivity of $0.2$ eV at $90\%$ CL).

The neutrinoless double beta decay $(Z,A) \to (Z+2,A)+2e^-$ (in short
$0\nu2\beta$) is a rare nuclear processes where lepton number is
violated and whose observation would mean that neutrinos are Majorana
particles. If the $0\nu2\beta$ process is mediated by a light
neutrino, the results from neutrinoless double beta decay experiments
are converted into an upper bound or a measurement of the effective
mass $m_{\beta\beta}$ in Eq.\ (\ref{beta_2beta}). A recent work
\cite{Strumia:2005tc} gives the following upper bounds
at $99\%$ CL
\begin{equation}
|m_{\beta\beta}|<(0.44-0.62)\,h_N~{\rm eV}
\label{mbb_res}
\end{equation}
where $h_N$ is a parameter that characterizes the uncertainties on the
corresponding nuclear matrix elements (see \cite{Rodin:2005dp} for a
recent discussion and \cite{Elliott:2002xe} for a review). The above
range corresponds to the results of recent double beta decay
experiments based on $^{76}$Ge, such as Heidelberg-Moscow
\cite{Klapdor-Kleingrothaus:2000sn} and IGEX \cite{Aalseth:1999ji}, or
$^{130}$Te such as Cuoricino \cite{Arnaboldi:2005cg} (see e.g.\
\cite{Arnold:2005rz} for results from other isotopes). In addition,
there exists a claim of a positive $0\nu2\beta$ signal which would
correspond to the approximate range $0.1<|m_{\beta\beta}|/{\rm
eV}<0.9$ \cite{Klapdor-Kleingrothaus:2004wj}. Future $0\nu2\beta$
projects will improve the current sensitivities down to values of the
order $|m_{\beta\beta}| \sim 0.01-0.05$ eV \cite{Elliott:2002xe}.
An experimental detection of $0\nu2\beta$, in
combination with results from beta decay or cosmology, will also help
us to discriminate between the two neutrino mass spectra and to pin
down the values of the neutrino flavour parameters, as discussed
for instance in \cite{Petcov:2004wz,Pascoli:2005zb}.

There are other ways to obtain information on the absolute scale of
neutrino masses, such as the measurement of the time-of-flight
dispersion of a supernova neutrino signal (at most sensitive to masses
of order some eV, see \cite{Nardi:2004zg} and references therein), but
to conclude this section let us note that the sum of neutrino masses
is restricted to the approximate range
\begin{equation}
0.056\, (0.095)~{\rm eV} \lsim \sum_i m_i \lsim 6~{\rm eV}
\label{sum_range}
\end{equation}
where the upper limit comes exclusively from tritium beta decay
results and the lower limit reflects the minimum values of the
total neutrino mass in the normal (inverted) hierarchy.

\section{The cosmic neutrino background}
\label{sec:CNB}

\subsection{Basics on relic neutrinos, including neutrino decoupling}
\label{sec:basics}

The existence of a relic sea of neutrinos is a generic prediction of
the standard hot big bang model, in number only slightly below that of
relic photons that constitute the CMB. Produced at large temperatures
by frequent weak interactions, cosmic neutrinos were kept in
equilibrium until these processes became ineffective in the course of
the expansion. While coupled to the rest of the primeval plasma,
neutrinos had a momentum spectrum with an equilibrium Fermi-Dirac
form with temperature $T$,
\begin{equation}
f_{\rm eq}(p)=\left
[\exp\left(\frac{p-\mu_\nu}{T}\right)+1\right]^{-1}\,.
\label{FD}
\end{equation}
Here we have included a neutrino chemical potential
$\mu_\nu$ that would exist in the presence of a neutrino-antineutrino
asymmetry, but it was shown in
\cite{Dolgov:2002ab,Wong:2002fa,Abazajian:2002qx} that the stringent
BBN bounds on $\mu_{\nu_e}$ apply to all flavours, since neutrino
oscillations lead to approximate flavour equilibrium before BBN.  The
current bounds on the common value of the neutrino degeneracy
parameter $\xi_\nu\equiv \mu_\nu/T$ are $-0.05<\xi_\nu<0.07$ at
$2\sigma$ \cite{Serpico:2005bc} (relaxed to $-0.13<\xi_\nu<0.3$ in
presence of extra relativistic degrees of freedom
\cite{Barger:2003rt,Cuoco:2003cu}).  Thus the contribution of a relic
neutrino asymmetry can be safely ignored.

As the universe cools, the weak interaction rate $\Gamma_\nu$ falls
below the expansion rate given by the Hubble parameter $H$ and one
says that neutrinos decouple from the rest of the plasma. An estimate
of the decoupling temperature can be found by equating the thermally
averaged value of the weak interaction rate
\begin{equation}
\Gamma_\nu=\langle\sigma_\nu\,n_\nu\rangle \,\, ,
\label{Gamma}
\end{equation}
where $\sigma_\nu \propto G_F^2$ is the cross section of the
electron-neutrino processes with $G_F$ the Fermi constant and
$n_\nu$ is the neutrino number density, with the expansion rate
\begin{equation}
H=\sqrt{\frac{8\pi\rho}{3M_P^2}} \,\, ,
\label{H_MeV}
\end{equation}
where $\rho$ is the total energy density and $M_P$ is the Planck mass.
If we approximate the numerical factors to unity, with $\Gamma_\nu
\approx G_F^2T^5$ and $H \approx T^2/M_P$, we obtain the rough
estimate $T_{\rm dec}\approx 1$ MeV.  More accurate calculations give
slightly higher values of $T_{\rm dec}$ which are flavour dependent
since electron neutrinos and antineutrinos are in closer contact with
$e^\pm$, as shown e.g.\ in \cite{Dolgov:2002wy}.

Although neutrino decoupling is not described by a unique $T_{\rm
dec}$, it can be approximated as an instantaneous process.  The
standard picture of {\it instantaneous neutrino decoupling} is very
simple (see e.g.\ \cite{kt,dodelson}) and reasonably accurate.  In
this approximation, the spectrum in Eq.\ (\ref{FD}) is preserved after
decoupling, since both neutrino momenta and temperature redshift
identically with the universe expansion. In other words, the number
density of non-interacting neutrinos remains constant in a comoving
volume since the decoupling epoch.

We have seen in Sec.\ \ref{sec:numasses} that active
neutrinos cannot possess masses much larger than $1$ eV, so they were
ultra-relativistic at decoupling. This is the reason why the momentum
distribution in Eq.\ (\ref{FD}) does not depend on the neutrino
masses, even after decoupling, i.e.\ there is no neutrino energy in
the exponential of $f_{\rm eq}(p)$. 

When calculating quantities related to relic neutrinos, one must
consider the various possible degrees of freedom per flavour. If
neutrinos are massless or Majorana particles, there are two degrees of
freedom for each flavor, one for neutrinos (one negative helicity
state) and one for antineutrinos (one positive helicity state).
Instead, for Dirac neutrinos there are in principle twice more degrees
of freedom, corresponding to the two helicity states. However, the
extra degrees of freedom should be included in the computation only if
they are populated and brought into equilibrium before the time of
neutrino decoupling. In practice, the Dirac neutrinos with the
``wrong-helicity'' states do not interact with the plasma at
temperatures of the MeV order and have a vanishingly small density
with respect to the usual left-handed neutrinos (unless neutrinos have
masses close to the keV range, as explained in Sec.\ 6.4 of
\cite{Dolgov:2002wy}, but such a large mass is excluded for active
neutrinos). Thus the relic density of active neutrinos does not depend
on their nature, either Dirac or Majorana particles.

Shortly after neutrino decoupling the photon temperature drops below
the electron mass, favouring $e^{\pm}$ annihilations that heat the
photons. If one assumes that this entropy transfer did not affect the
neutrinos because they were already completely decoupled, it is easy
to calculate the ratio between the temperatures of relic photons
and neutrinos $T_\gamma/T_\nu=(11/4)^{1/3}\simeq 1.40102$. Any
quantity related to relic neutrinos can be calculated at later times
with the spectrum in Eq.\ (\ref{FD}) and $T_\nu$. For instance, the
number density per flavour is fixed by the temperature,
\begin{equation}
n_{\nu} = \frac{3}{11}\;n_\gamma =
\frac{6\zeta(3)}{11\pi^2}\;T_\gamma^3~,
\label{nunumber}
\end{equation}
which leads to a present value of $113$ neutrinos and antineutrinos of
each flavour per cm$^{3}$. Instead, the energy density for massive
neutrinos should in principle be calculated numerically, with two
well-defined analytical limits,
\begin{eqnarray}
\rho_\nu (m_\nu \ll T_\nu) & = & 
\frac{7\pi^2}{120}
\left(\frac{4}{11}\right)^{4/3}\;T_\gamma^4~,
\nonumber \\
\rho_\nu (m_\nu \gg T_\nu) & = & m_\nu n_\nu~.
\label{nurho}
\end{eqnarray}
Thus we see that the contribution of massive neutrinos to the energy
density in the non-relativistic limit is a function of the mass (or
the sum of masses if all neutrino states have $m_i \gg T_\nu$).
\begin{figure}[t]
\begin{center}
\vspace{-2cm}
\hspace{-2cm}
\includegraphics[width=.87\textwidth]{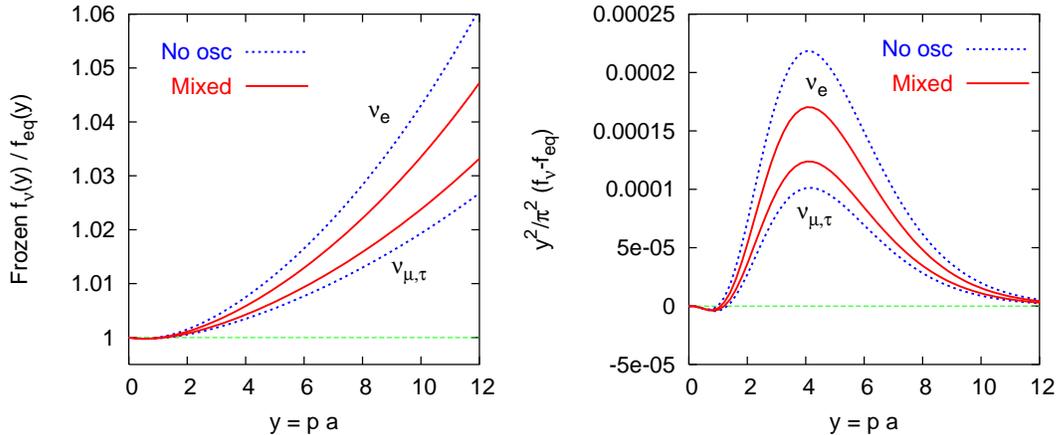}
\caption{\label{fig:finalfnu} The upper (lower) lines are the frozen
distortions of the electron (muon or tau) neutrino spectra as a
function of the comoving momentum $y$, calculated with (solid) and
without (dotted) the effect of flavour oscillations. Left: real
neutrino distribution functions normalized to the equilibrium
one. Right: contribution of the distortions to the comoving number
density. Here the scale factor was normalized so that $a(t)\to
1/T_\gamma$ at large temperatures.}
\end{center}
\end{figure}

In a more accurate analysis of neutrino decoupling, the standard
picture described above is modified: the processes of neutrino
decoupling and $e^{\pm}$ annihilations are sufficiently close in time
so that some relic interactions between $e^{\pm}$ and neutrinos exist.
These relic processes are more efficient for larger neutrino energies,
leading to non-thermal distortions in the neutrino spectra and a
slightly smaller increase of the comoving photon temperature, as noted
in a series of works (see the full list given in the review
\cite{Dolgov:2002wy}).  A proper calculation of the process of
non-instantaneous neutrino decoupling demands solving the Boltzmann
equations for the neutrino spectra, a set of integro-differential
kinetic equations that are difficult to solve numerically.
The momentum-dependent calculations were carried out in refs.\
\cite{Hannestad:1995rs,Dolgov:1997mb,Esposito:2000hi}, while the
inclusion of finite temperature QED corrections to the electromagnetic
plasma was done in \cite{Mangano:2001iu}.

A recent work \cite{Mangano:2005cc} has considered the effect of
flavour neutrino oscillations on the neutrino decoupling process (see
also \cite{Hannestad:2001iy}). The frozen values of the neutrino
distributions are shown in Fig.\ \ref{fig:finalfnu}, for the cases
calculated with and without the effect of flavour oscillations.  One
can see that the distortions grow with the neutrino momentum, and
their contribution to the number density of neutrinos is maximal
around $y=4$.  For the best-fit values of the mixing parameters in
Eq.\ (\ref{oscpardef}), Ref.\ \cite{Mangano:2005cc} found an increase
in the neutrino energy densities of $0.73\%$ and $0.52\%$ for
$\nu_e$'s and $\nu_{\mu,\tau}$'s, respectively. At the same time,
after $e^\pm$ annihilations the comoving photon temperature is a
factor $1.3978$ larger, instead of $1.40102$ in the approximation of
instantaneous decoupling. Note, however, that for any cosmological
epoch when neutrino masses can be relevant one should consider the
spectra of the neutrino mass eigenstates $\nu_{1,2,3}$ instead of
flavour eigenstates $\nu_{e,\mu,\tau}$.  In principle, for numerical
calculations of the cosmological perturbation evolution such as those
done by the Boltzmann codes {\sc cmbfast} \cite{Seljak:1996is} or {\sc
camb} \cite{Lewis:1999bs}, one should include the full distribution
function of each mass eigenstate, including the distortions shown in
Fig.\ \ref{fig:finalfnu}. It is only when the neutrinos are
relativistic or when one is interested in an observable that does not
depend on neutrino masses (such as the number density) that the effect
of the distortion can be simply integrated over momentum.  For
instance, the contribution of relativistic relic neutrinos to the
total energy density is taken into account just by using $N_{\rm eff}
= 3.046$ as defined later in Eq.\ (\ref{neff}).  In practice, the
distortions calculated in Ref.\ \cite{Mangano:2005cc} only have small
consequences on the evolution of cosmological perturbations, and for
many purposes they can be safely neglected.

\subsection{Extra radiation and the effective number of neutrinos}
\label{subsec:neff}

Neutrinos fix the expansion rate during the cosmological era when the
Universe is dominated by radiation. Their contribution to the total
radiation content can be parametrized in terms of the effective number
of neutrinos $N_{\rm eff}$ \cite{Shvartsman:1969mm,Steigman:1977kc},
through the relation
\begin{equation}
\rho_{\rm R} = \left[ 1 + \frac{7}{8} \left( \frac{4}{11}
\right)^{4/3} \, N_{\rm eff} \right] \, \rho_\gamma \,\,,
\label{neff}
\end{equation}
where $\rho_\gamma$ is the energy density of photons, whose value
today is known from the measurement of the CMB temperature. This
equation is valid when neutrino decoupling is complete and holds as
long as all neutrinos are relativistic. 

In Sec.\ \ref{sec:numasses} we saw that from accelerator data the
number of active neutrinos is three, while in the previous subsection
we learned from the analysis of neutrino decoupling these three active
neutrinos contribute as $N_{\rm eff}=3.046$. Any departure of $N_{\rm
eff}$ from this last value would be due to non-standard neutrino
features or to the contribution of other relativistic relics.

A detailed discussion of cosmological scenarios where 
$N_{\rm eff}$ is not fixed to three
can be found in the reviews
\cite{Dolgov:2002wy,Sarkar:1995dd}, while the particular case of
active-sterile neutrino mixing was recently analyzed in
\cite{Cirelli:2004cz}.
Since in the present work we focus on the standard case of three
active neutrinos, here we only give a brief review of the most recent
bounds on $N_{\rm eff}$ from cosmological data. 

The value of $N_{\rm eff}$ is constrained at the BBN epoch from the
comparison of theoretical predictions and experimental data on the
primordial abundances of light elements, which also depend on the
baryon-to-photon ratio $\eta_{\rm b}=n_{\rm b}/n_\gamma$ (or baryon
density). The main effect of $N_{\rm eff}$ is to fix the Hubble
expansion rate through its contribution to the total energy
density. This in turn changes the freezing temperature of the
neutron-to-proton ratio, therefore producing a different abundance of
$^4$He.

\begin{figure}[t]
\begin{center}
\includegraphics[width=.6\textwidth]{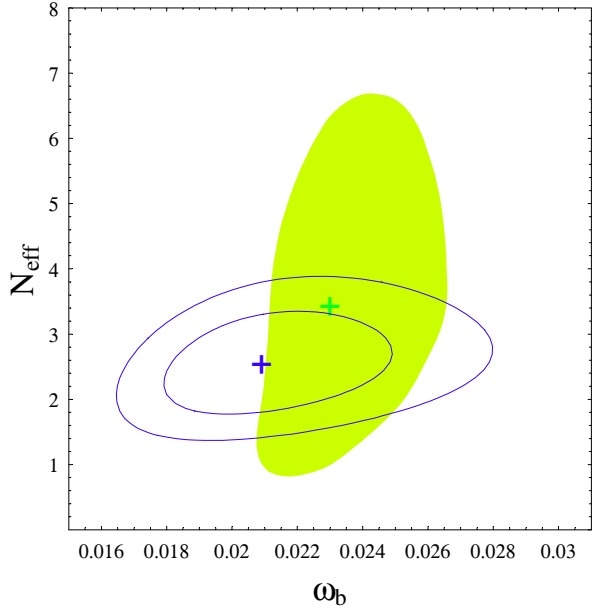}
\caption{\label{fig:likelihoodbbn} Bounds on $N_{\rm eff}$ from BBN
including D and $^4$He (contours at $68\%$ and $95\%$ CL) and from a
combined analysis of BBN (D only) and CMB data. Here $\omega_{\rm
b}=\Omega_{\rm b}h^2\simeq 10^{10}(\eta_{\rm b}/274)$. This Figure is
taken from Ref.\ \cite{Cuoco:2003cu}.}
\end{center}
\end{figure}
The BBN bounds on $N_{\rm eff}$ have been recently reanalyzed taking
in input the value of the baryon density derived from the WMAP first
year data \cite{Spergel:2003cb} $\eta_{\rm CMB}=6.14\pm 0.25$. In
Fig.\ \ref{fig:likelihoodbbn} we show the results from
\cite{Cuoco:2003cu}, where the allowed range $N_{\rm
eff}=2.5^{+1.1}_{-0.9}$ ($95\%$ CL) was inferred from data on light
element abundances (see also
\cite{Hannestad:2003xv,Barger:2003zg,Cyburt:2004yc} and
\cite{Steigman:2005uz} for a recent review on BBN).  This range is
perfectly compatible with the standard prediction of $3.046$. However,
the reader should be cautious in the interpretation of the BBN allowed
range for $N_{\rm eff}$ such as that in Fig.\
\ref{fig:likelihoodbbn}. It is well-known that the main problem when
deriving the primordial abundances from observations in astrophysical
sources is the existence of systematics not accounted for, in
particular for $^4$He (a discussion of the observational situation can
be found in Ref.\ \cite{Olive:2004kq}). For instance, the detailed BBN
analysis in \cite{Serpico:2004gx} gives the allowed range $1.90<
N_{\rm eff}<3.77$ for the $^4$He ``box'' range in \cite{Olive:2004kq}.

Independent bounds on the radiation content of the universe at a later
epoch can be extracted from the analysis of the power spectrum of CMB
anisotropies. We will describe later the neutrino effects on the CMB
spectrum. Assuming a minimal set of cosmological parameters and a flat
universe, Ref.\ \cite{Crotty:2003th} found the value
$N_{\rm eff}=3.5^{+3.3}_{-2.1}$ (95\% CL) (see also
\cite{Hannestad:2003xv,Barger:2003zg,Pierpaoli:2003kw}) from the
combination of WMAP first year data with other CMB and LSS data.  More
recently, it has been shown that the addition of Supernovae and new
CMB data leads to a reduction of the allowed range: $N_{\rm
eff}=4.2^{+1.2}_{-1.7}$ \cite{Hannestad:2005jj} (with the latest
BOOMERANG data) and $N_{\rm eff}=3.3^{+0.9}_{-4.4}$
\cite{WMAP3:Spergel} (with WMAP three year data).

\subsection{Massive neutrinos as dark matter}
\label{subsec:neutrinos_dm}

A priori, massive neutrinos are excellent candidates for
contributing to the dark matter density, in particular because we are
certain that they exist, in contrast with other candidate
particles. Together with CMB photons, relic neutrinos can
be found anywhere in the Universe with a number density of $339$
neutrinos and antineutrinos per cm$^{3}$. Depending on the value of the mass,
this density is enhanced when neutrinos cluster into
gravitational potential wells, although recent analyses show that the
overdensity is limited to small factors (see e.g.\
\cite{Ringwald:2004np}).

In addition, it is easy to have a neutrino contribution of order unity
to the present value of the energy density of the Universe, just by
considering eV neutrino masses. In such a case, we saw before that all
neutrinos should approximately share the same mass $m_0$ and their
energy density in units of the critical value of the energy density
(see Eq.\ (\ref{critical density})) is\footnote{ For high precision,
the relation between $(m_1,m_2,m_3)$ and $\Omega_{\nu} h^2$ must be
evaluated numerically using the distorted distributions described
in \cite{Mangano:2005cc}.}
\begin{equation}
\Omega_{\nu} = \frac{\rho_\nu}{\rho_{\rm c}} =
\frac{\sum_i m_i}{93.14\,h^2~{\rm eV}}~,
\label{omeganu}
\end{equation}
where $h$ is the present value of the Hubble parameter in units of
$100$ km s$^{-1}$ Mpc$^{-1}$ and $\sum_i m_i = 3 m_0$.  Even if the
three neutrinos are non-degenerate in mass, Eq.\ (\ref{omeganu}) can
be safely applied.  Indeed, we know from neutrino oscillation data
that at least two of the neutrino states are non-relativistic today,
since both $(\Delta m^2_{31})^{1/2}\simeq 0.047 $ eV and $(\Delta
m^2_{21})^{1/2}\simeq 0.009$ eV are larger than the temperature $T_\nu
\simeq 1.96$ K $\simeq 1.7\times 10^ {-4}$ eV. If the third neutrino
state is very light and still relativistic, its relative contribution
to $\Omega_{\nu}$ is negligible and Eq.\ (\ref{omeganu}) remains an
excellent approximation of the total density.

If we demand that neutrinos should not be heavy enough to overclose
the Universe ($\Omega_{\nu}<1$), we obtain an upper bound $m_0 \lsim
15$~eV for the absolute neutrino mass scale $m_0= \sum_i m_i/3$
(fixing $h=0.7$). This argument was used many years
ago by Gershtein and Zeldovich \cite{Gershtein:1966gg} (see also
\cite{Cowsik:1972gh}).  Since from present analysis of cosmological
data we know that the approximate contribution of matter is
$\Omega_{\rm m} \simeq 0.3$, the neutrino masses should obey the 
stronger bound
$m_0 \lsim 5$ eV.

Dark matter particles with a large velocity dispersion such as that of
neutrinos are called hot dark matter (HDM). The role of neutrinos as
HDM particles has been widely discussed since the 1970s, and the
reader can find a historical review in Ref.\ \cite{Primack:2001ib}. 
It was realized in the mid-1980s (see e.g.\
\cite{Bond:1980ha,Bond:1983hb,White:1984yj}) that HDM affects the
evolution of cosmological perturbations in a particular way: it erases
the density contrasts on wavelengths smaller than a mass-dependent
free-streaming scale (we will discuss in more detail the effects of
massive neutrinos on the evolution of cosmological perturbations in
Secs.\ \ref{subsec:pert_nu} and
\ref{subsec:sum_mass-effect}). In a universe dominated by HDM, this
suppression is in contradiction with various observations. For
instance, large objects such as superclusters of galaxies form first,
while smaller structures like clusters and galaxies form via a
fragmentation process. This top-down scenario is at odds with the fact
that galaxies seem older than clusters.

Given the failure of HDM-dominated scenarios, the attention then
turned to cold dark matter (CDM) candidates, i.e.\ particles which
were non-relativistic at the epoch when the universe became
matter-dominated, which provided a better agreement with
observations. Still in the mid-1990s it appeared that a small mixture
of HDM in a universe dominated by CDM fitted better the observational
data on density fluctuations at small scales than a pure CDM
model (see e.g. \cite{Valdarnini:1998zy}). 
However, within the presently favoured $\Lambda$CDM model
dominated at late times by a cosmological constant (or some form of
dark energy) there is no need for a significant contribution of
HDM. Instead, one can use the available cosmological data to find how
large the neutrino contribution can be. 
In Sec.\ \ref{sec:mass_cosmo},
we will explain the effect of neutrino masses on cosmological observables.
In Sec.\ \ref{sec:present}, we will review the 
upper bounds on the sum of all neutrino
masses which have been derived from current data.

\section{Massive neutrinos and cosmological perturbations}
\label{sec:mass_cosmo}

Let us start with some generalities on the theory of cosmological
perturbations. This field has been thoroughly investigated over the past
thirty years, and many excellent reviews have been written on the
subject (see for instance Refs.\
\cite{dodelson,Ma:1995ey,Hu:1995em,Liddle:2000cg,Bertschinger:2001is}).
Many equations in this section are reminiscent of those in Ma \&
Bertschinger\ \cite{Ma:1995ey} and Bertschinger\
\cite{Bertschinger:2001is} -- however, with several sign differences,
since we choose the metric signature to be $(+,-,-,-)$. This
signature tends to be the most popular nowadays in cosmology.

In our Universe, the metric and the energy-momentum
tensor are inhomogeneous. Their perturbations, given by
\begin{eqnarray}
\label{def_perturb}
\delta g_{\mu \nu}({\bf x},t) &=
& g_{\mu \nu}({\bf x},t) - \bar{g}_{\mu \nu}(t)~,
\\
\delta T_{\mu \nu}({\bf x},t) &=
& T_{\mu \nu}({\bf x},t) - \bar{T}_{\mu \nu}(t)~,
\end{eqnarray}
are known to be small in the early Universe, typically $10^{5}$ times
smaller than the background quantities, as shown by CMB anisotropies.
As we shall see in this section, after photon decoupling, the matter
perturbations grow by gravitational collapse and reach the non-linear
regime, starting with the smallest scales. However, the linear
perturbation theory is a good tool both for describing the early
Universe at any scales, and the recent universe on the largest scales.
The most reliable observations in cosmology are those involving mainly
linear (or quasi-linear) perturbations. In particular, current
cosmological neutrino mass bounds are based on such
observations. Therefore, our goal in this section is to describe
the evolution of linear cosmological perturbations. The great
advantage of linear theory is, as usual, to obtain independent
equations of evolution for each Fourier mode. Note that the Fourier
decomposition must be performed with respect to the comoving
coordinate system: so, the quantity $(2 \pi / k)$ is the {\it comoving
wavelength} of a perturbation of wavevector ${\bf k}$, while the {\it
physical wavelength} is given by
\begin{equation}
\lambda = a(t) \frac{2 \pi}{k}~,
\end{equation}
where $a(t)$ is the scale factor of the Universe. For each mode ${\bf
k}$, the perturbation amplitudes evolve under some equations of motion
(which depend only on the modulus $k$, since the background is
isotropic), and on top of this evolution, the physical wavelength is
stretched according to the Universe expansion.

For concision, we will restrict ourselves to the main-stream standard
cosmological model: the inflationary $\Lambda$CDM scenario, whose
background evolution is reviewed in Sec.\ \ref{subsec:background}. In
Sec.\ \ref{sec:present} we will briefly comment on some results based
on more exotic scenarios. However, the theory of cosmological
perturbations is so rich that in order to present a short and
pedagogical summary, it is necessary to start from strong assumptions
concerning the underlying cosmological background. Therefore, we will
assume that at early times (e.g.\ at BBN) the Universe contains a
stochastic background of Gaussian, adiabatic\footnote{If we relax the
assumption of adiabaticity, i.e. if we allow for isocurvature modes in
the early Universe, the neutrino background can play a very specific
role and provides some non-trivial initial conditions for the
evolution of cosmological perturbations.  In this review, we will not
cover this possibility and refer the interested reader to Refs.\
\cite{Bucher:1999re,Trotta:2001yw,Lyth:2002my,Crotty:2003rz,Bucher:2004an,Moodley:2004nz,Beltran:2004uv}.}
and nearly scale-invariant primordial perturbations, as predicted by
inflation, and as strongly suggested by CMB observations.  Since the
perturbations are stochastic and Gaussian, what we will call
``amplitude'' should often be understood as ``variance'': the linear
equations of evolution give the evolution of each individual mode
$\delta \! f_{\bf k}$ of a perturbation $\delta \! f$, and also, more
interestingly, that of the root mean square $\langle |\delta \! f|^2
\rangle^{1/2}_k$, obtained by averaging over all modes of fixed
wavenumber $k$.  It is this last quantity which is the target of most
observations, and which carries information on the cosmological
parameters.

The perturbations defined in Eqs.\ (\ref{def_perturb}) contain many
degrees of freedom: the homogeneity and isotropy of the background
implies that $\bar{g}_{\mu \nu}$ and $\bar{T}_{\mu \nu}$ are diagonal,
but in general this is not true at the level of perturbations.
However, we will see in Sec.\ \ref{subsec:gauge} that some of
these degrees of freedom are just artifacts of the relativistic
perturbation theory set-up; moreover, only a fraction of the physical
degrees of freedom contribute significantly to the CMB and LSS
observables. So, we will see at the end of Sec.\
\ref{subsec:gauge} that the problem can be reduced to the integration
of a small number of linear equations of evolution.

In order to perform this integration, one must specify the properties
of each fluid contributing to the energy density.  Apart from the
cosmological constant, the standard $\Lambda$CDM scenario includes
contributions from photons, neutrinos, baryons and cold dark matter.
In Sec.\ \ref{subsec:pert_no_nu}, we will review the behavior of
the most relevant perturbations under the assumption that neutrinos
are absent, in order to better identify the specific role of neutrinos
in Sec.\ \ref{subsec:pert_nu}. In Sec.\ \ref{subsec:sum_mass-effect},
we will summarize the effects of neutrino masses on cosmological
observables, which are important for understanding the present
bounds. Thus, the reader already familiar with cosmological
perturbation theory in $\Lambda$CDM and $\Lambda$MDM ($\Lambda$ Mixed
Dark Matter) models can go directly to the last subsection.

\subsection{Observables targets: definition of the power spectra}
\label{subsec:observables}

Our goal is to compute observable quantities like
\begin{itemize}
\item
the CMB temperature anisotropy power spectrum, defined as the angular
two-point correlation function of CMB maps $\delta T/\bar{T}(\hat{n})$
($\hat{n}$ being a direction in the sky). This function is usually
expanded in Legendre multipoles
\begin{equation}
\left\langle {\delta T\over\bar{T}}(\hat{n}) 
{\delta T \over \bar{T}} (\hat{n}') \right\rangle
= \sum_{l=0}^{\infty} {(2l+1)\over4\pi} C_l P_l(\hat{n}\cdot\hat{n}')~,
\end{equation}
where $P_l(x)$ are the Legendre polynomials. So, for Gaussian
fluctuations, all the information is encoded in the multipoles $C_l$
which probe correlations on angular scales $\theta=\pi/l$.  Since for
each neutrino family the mass is already known to be at most of the
order of $1$ eV (see Sec.\ \ref{sec:present}), the transition to the
non-relativistic regime is expected to take place after the time of
recombination between electrons and nucleons, i.e. after photon
decoupling. The shape of the CMB spectrum is related mainly to the
physical evolution {\it before} recombination. Therefore, the CMB will
only be marginally affected by the neutrino mass (we will see later
that there is only an indirect effect through the modified background
evolution). For this reason, we will only mention in this review some
essential aspects of the CMB temperature spectrum, without entering
into details. For the effects on CMB of heavier neutrinos (with masses
of a few eV), we refer the reader to Ref.\ \cite{Dodelson:1995es}. \\

\item
the CMB polarization anisotropy power spectra bring interesting
complementary information to the temperature one, but for the same reasons
as above, plus the fact that current polarization measurements are still
rather imprecise, we will not discuss polarization in this section.\\

\item
the matter power spectrum, observed with
various techniques described in the next section (directly or indirectly,
today or in the near past), probes the current
Large Scale Structure of the Universe. It is defined as the 
two-point correlation function of non-relativistic
matter fluctuations in Fourier space
\begin{equation}
P(k,z)=\langle | \delta_{\rm m}(k,z) |^2 \rangle~,
\end{equation}
where $\delta_{\rm m}=\delta \rho_{\rm m}/\bar{\rho}_{\rm m}$.  In the
case of several fluids (e.g. CDM, baryons and non-relativistic
neutrinos), the total matter perturbation can be expanded as
\begin{equation}
\delta_{\rm m}= 
\frac{\sum_i \, \bar{\rho}_i \, \delta_i}{\sum_i \, \bar{\rho}_i}~.
\end{equation}
Since the energy density
is related to the mass density of non-relativistic matter through
$E=mc^2$, $\delta_{\rm m}$ represents indifferently the energy or mass
power spectrum.  When the redshift $z$ is not explicitly
written, we assume that $P(k)$ refers to the matter power spectrum
evaluated today (at $z=0$).  The shape of the matter power spectrum is
the key observable for constraining small neutrino masses with
cosmological methods. Therefore the main purpose of this section is to
review the physics responsible for its shape, in a pedagogical but
sufficiently detailed way, starting with linear structure formation in
absence of neutrinos (Sec.\ \ref{subsec:pert_no_nu}), 
and then including the impact of free-streaming
massive neutrinos (Secs.\ \ref{subsec:pert_nu},
\ref{subsec:sum_mass-effect}).\\

\item
finally, one can build some cross-correlation power spectra (which
measure the angular or spatial correlation between two different types
of observables). These correlations are in principle very interesting
to extract from observations, since they provide some additional
information beyond that contained in the self-correlation power
spectra $C_l$ or $P(k)$. For instance, the cross-correlation between
CMB temperature and polarization maps is non-zero, since the two
signals are affected by the same physical inhomogeneities close to the
last-scattering surface. The CMB temperature maps are also correlated
to some extent to the matter distribution at low redshift, due to the
late integrated Sachs-Wolfe effect (that we will briefly introduce in
Sec.\ \ \ref{subsec:params}).  Finally, different measurements of the
matter distribution in different overlapping redshift ranges or with
different techniques are likely to be correlated with each other. The
measurement of various cross-correlation power spectra is expected to be
crucial in the future, but since nowadays it only plays a
marginal role in neutrino mass determinations\ \cite{Ichikawa:2005hi},
we will not introduce the corresponding formalism in this review.
\end{itemize}

\subsection{Background evolution}
\label{subsec:background}

In most of this section, we will use the conformal time $\tau$ instead
of the proper time $t$. The two are related through $d\tau = dt /
a(t)$.  A dot will denote a derivative with respect to $\tau$. For
instance, the Hubble parameter reads $H=d(\ln a)/ dt = \dot{a} / a^2$.
In absence of spatial curvature, the background Friedmann metric
reduces to
\begin{equation}
ds^2 = a(\tau)^2 [ d\tau^2 - \delta_{ij} dx^i dx^j ]~,
\label{Friedmann_metric}
\end{equation}
and the Friedmann equation reads
\begin{equation}
H^2 = \frac{8 \pi G}{3} 
(\rho_{\gamma} + \rho_{\rm cdm} + \rho_{\rm b} 
+  \rho_{\nu} + \rho_{\Lambda})~,
\end{equation}
where the homogeneous density of photons $\rho_{\gamma}$ scales like
$a^{-4}$, that of non-relativistic matter ($\rho_{\rm cdm}$ for CDM
and $\rho_{\rm b}$ for baryons) like $a^{-3}$, that of neutrinos
$\rho_{\nu}$ interpolates between the two behaviors at the time of the
non-relativistic transition, and finally the cosmological constant
density $\rho_{\Lambda}$ is of course time-independent. At any time,
the critical density $\rho_{c}$ is defined as
$\rho_{\rm c}=3H^2/8\pi G$, and the current value $H_0$ (or
reduced value $h$) of the Hubble parameter gives the critical density
today
\begin{equation}
\label{critical density}
\rho_{\rm c}^0 = 1.8788 \times 10^{-29}h^2 ~\mathrm{g~cm}^{-3}~.
\end{equation}

\begin{figure}[t]
\begin{center}
\vspace{-2cm}
\hspace{-1.5cm}
\includegraphics[width=.95\textwidth]{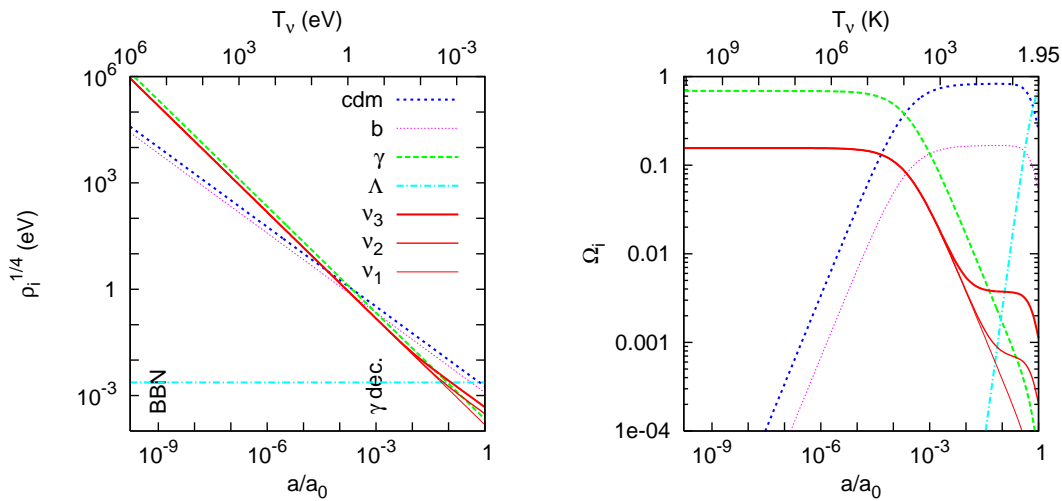}
\caption{\label{fig:rho_i} Evolution of the background densities from
the time when $T_{\nu}=1$ MeV (soon after neutrino decoupling) until
now, for each component of a flat $\Lambda$MDM model with $h=0.7$ and
current density fractions $\Omega_{\Lambda}=0.70$, $\Omega_{\rm
b}=0.05$, $\Omega_{\nu}=0.0013$ and $\Omega_{\rm
cdm}=1-\Omega_{\Lambda}-\Omega_{\rm b} -\Omega_{\nu}$. The three
neutrino masses are distributed according to the Normal Hierarchy
scheme (see Sec.\ \ref{sec:numasses}) with $m_1=0$, $m_2 =
0.009$ eV and $m_3 = 0.05$ eV. On the left plot we show the densities
to the power $1/4$ (in eV units) as a function of the scale factor. On
the right plot, we display the evolution of the density fractions (i.e.,
the densities in units of the critical density).  We also show on the
top axis the neutrino temperature (on the left in eV, and on the right
in Kelvin units). The density of the neutrino mass states $\nu_2$ and
$\nu_3$ is clearly enhanced once they become
non-relativistic. On the left plot, we also display the characteristic
times for the end of BBN and for photon decoupling or recombination.}
\end{center}
\end{figure}

The normalization of the scale factor $a$ is arbitrary. If we first
neglect neutrino masses, the photon and neutrino densities can be
combined into a total radiation density $\rho_{\rm r} = \rho_{\gamma}
+ \rho_{\nu}$ scaling like $a^{-4}$, while the CDM and baryons combine
into a matter density $\rho_{\rm m} = \rho_{\rm cdm} + \rho_{\rm b}$
scaling like $a^{-3}$.  Then, we can choose the normalization of $a$
such that $\rho_{\rm m} = 3 / (8 \pi G a^{3})$ in order to obtain a
particularly simple Friedmann equation valid during radiation
domination (RD) and matter domination (MD)
\begin{equation}
\left(\dot{a}\right)^2 = a + a_{\rm eq}~,
\end{equation}
where $a_{\rm eq}$ is the scale factor at equality, when $\rho_{\rm m}
= \rho_{\rm r}$.  There is a simple solution,
\begin{equation}
a = \frac{\tau^2}{4} + \sqrt{a_{\rm eq}} \, \tau =
\frac{\tau}{4} \left( \tau + 2 (\sqrt{2}+1) \tau_{\rm eq} \right)~,
\label{back_RDMD}
\end{equation}
where $\tau_{\rm eq}=2(\sqrt{2}-1)\sqrt{a_{\rm eq}}$ is the value of
the conformal time at equality.  At low redshift (typically $z <
0.5$), the cosmological constant density takes over, causing a
departure from the above solution, with an acceleration of the scale
factor.  Finally, if we include the effect of small neutrino masses,
the solution is also slightly modified, since the non-relativistic
transition of each neutrino species amounts in converting a fraction
of radiation into matter. This can be seen in Fig.\ \ref{fig:rho_i},
where we plot the evolution of background densities for a $\Lambda$MDM
model in which the three neutrino masses follow the Normal Hierarchy
scheme (see Sec.\ \ref{sec:numasses}) with $m_1=0$, $m_2 =
0.009$~eV and $m_3 = 0.05$~eV.

\subsection{Gauge transformations and Einstein equations}
\label{subsec:gauge}

In the real Universe all physical quantities (densities, curvature...)
are functions of time and space. Thanks to the covariance of general
relativity, they can be described in principle in any coordinate
system, without changing the physical predictions. The problem is that
in order to obtain simple equations of evolution, we wish to use a
linear perturbation theory, in which the true physical quantities are
artificially decomposed into a homogeneous background and some small
perturbations. This is artificial because the homogeneous quantities
are defined as spatial averages over hypersurfaces of simultaneity:
$\bar{f}(t) = \langle f(t,{\bf x}) \rangle_{\bf x}$. Any change of
coordinate system which
\begin{enumerate}
\item mixes time and space (therefore, redefining
hypersurfaces of simultaneity, and changing the way to perform spatial
averages), and\\

\item remains small everywhere, so that the differences between
true quantities and spatial averages are still small perturbations,
\end{enumerate}
gives a new set of perturbations (new equations of evolution, new
initial conditions), although the physical quantities (i.e., the total
ones) are the same.  This ambiguity is called the {\it gauge freedom}
in the context of relativistic perturbation theory.

Of course, using a linear perturbation theory is only possible when
there exists at least one system of coordinates in which the Universe
looks approximately homogeneous. We know that this is the case at
least until the time of photon decoupling: in some reference frames,
the CMB anisotropies do appear as small perturbations. It is a
necessary condition for using linear theory to be in such a frame;
however, this condition is vague and leaves a lot of gauge freedom,
i.e.\ many possible ways to slice the spacetime into hypersurfaces of
simultaneity.

We can also notice that the definition of hypersurfaces of
simultaneity is not ambiguous at small distances, as long as different
observers can exchange light signals in order to synchronize their
clocks.  Intuitively, we see that the gauge freedom is an infrared
problem, since on very large distances (larger than the Hubble
distance) the word ``simultaneous'' does not have a clear meaning.
The fact that the gauge ambiguity is only present on large scales
emerges naturally from the mathematical framework describing gauge
transformations.

Formally, a gauge transformation is described by a quadrivector field
$\epsilon^{\mu}({\bf x},t)$ (see e.g.\ Ref.\ \cite{Ma:1995ey}). When
the latter is infinitesimal, the Lorentz scalars, vectors and tensors
describing the perturbations are shifted by the Lie derivative along
$\epsilon$,
\begin{equation}
\delta A_{\mu \nu ...}({\bf x},t) \rightarrow 
\delta A_{\mu \nu ...}({\bf x},t) + 
{\mathcal L}_{\epsilon}
[\delta A_{\mu \nu ...}({\bf x},t)]~.
\end{equation}
Since there are four degrees of freedom (d.o.f.) in this
transformation - the four components of $\epsilon^{\mu}$ - we see that
among the ten d.o.f. of the perturbed Einstein equation $\delta G_{\mu
  \nu}=8\pi{G} \, \delta T_{\mu \nu}$, four represent gauge
modes, and six represent physical degrees of freedom.

In addition, it can be shown that this equation contains three
decoupled sectors. In other words, when the metric and the
energy-momentum tensor are parametrized in an adequate way, the ten
equations can be decomposed into three systems independent of each
other:
\begin{enumerate}
\item four equations relate four scalars in the perturbed metric
$\delta g_{\mu \nu}$ to four scalars in $\delta T_{\mu \nu}$,\\

\item four equations relate two transverse 3-vectors
in the perturbed metric (4 d.o.f. in total)
to two transverse 3-vectors in $\delta T_{\mu \nu}$,\\

\item two equations relate one transverse traceless 
$3\! \times\! 3$-tensor in the perturbed metric
(2 d.o.f.\ in total)
to a similar tensor in $\delta T_{\mu \nu}$.
\end{enumerate}
Moreover, this decomposition is left invariant by gauge
transformations.  These three types of variables are called scalar,
vector and tensor modes; physically, they describe respectively the
generalization of Newtonian gravity, gravitomagnetism, and
gravitational waves.  It can be shown that two scalar d.o.f. and two
vector d.o.f. are gauge modes, so each of the three sectors contains
only two physical d.o.f.  In principle, all modes can contribute to
the CMB anisotropy maps, but CMB observations show that the vector and
tensor contributions are negligible (at least for temperature
anisotropies). As for the LSS of the Universe, it is related to the
mass/energy density distribution of non-relativistic matter, i.e.,
again to the scalar sector.  Therefore, in this report, we will focus
only on scalar perturbations, not because the cosmological backgrounds
of vector/tensor modes are insensitive to neutrino properties, but
because they are not likely to be observed with great precision in a
near future.

It is possible to build some gauge-invariant combinations, and to
reduce the Einstein equation into a set of gauge-invariant equations.
This is not the most economic way to proceed: one can simply choose an
arbitrary gauge-fixing condition, i.e., a prescription that will limit
the number of effective degrees of freedom to that of physical modes
only, and make all calculations into that gauge. When the same problem
is studied in two different gauges, the solutions can look very
different on large wavelength; for instance, the total energy density
perturbation $\delta \rho (t,k)$ for a given time and wavenumber can
appear as growing in one gauge, and constant in another gauge,
although the two solutions describe the same Universe. However,
physical observables - like the matter density perturbations probed by
galaxy surveys or temperature/polarization anisotropies probed by CMB
experiments - are always limited to small scales, at most of the order
of the Hubble length. On those scales, the predictions arising from
different gauge choices always coincide with each other.

Throughout our discussion, we choose to work in the longitudinal
gauge, which is probably the most popular for studying cosmological
perturbations. In this gauge, one requires that the non-diagonal
metric perturbations vanish. This eliminates two scalar degrees of
freedom, and the remaining ones are defined in such a way that Eq.\
(\ref{Friedmann_metric}) reads
\begin{equation}
ds^2 = g_{\mu \nu} dx^{\mu} dx^{\nu} =
a^2( \tau) [(1+2\phi) d\tau^2 - (1-2 \psi)\delta_{ij} dx^i dx^j]~.
\end{equation}
For the energy-momentum tensor, the four scalar degrees of freedom
can be identified as
\begin{eqnarray}
\delta T^0_0 &=& \delta \rho ~, \\
\delta T^0_i &=& (\bar{\rho} + \bar{p}) v^{||}_i ~,\\
\delta T^i_j &=& - \delta p \, \delta^i_j + \Sigma^{i||}_j~,
\end{eqnarray}
where $\delta \rho$ is the energy density perturbation, $\delta p$ the
pressure perturbation, $v^{||}_i= \partial_i \tilde{v}$ the
longitudinal component of the velocity field, and $\Sigma^{i||}_j =
(\partial_i \partial_j - \frac{1}{3}\delta_{ij}
\nabla^2)\tilde{\sigma}$ the traceless and longitudinal-divergence
component of the $3 \! \times \!3$ tensor $\delta T^i_j$.  For these
last two degrees of freedom, we could write the equations in terms of
the two potentials $\tilde{v}$ and $\tilde{\sigma}$.  However, it is
more conventional to deal with the two quantities $(\theta, \sigma)$
which represent respectively the velocity divergence and the shear
stress (or anisotropic stress)
\begin{eqnarray}
\label{deftheta}
\theta &\equiv& \sum_{i} \partial_i v_i = \nabla^2 \tilde{v}~, \\
(\bar{\rho}+\bar{p}) \nabla^2 \sigma &\equiv& - \sum_{i,j} (\partial_i
\partial_j - \frac{1}{3} \nabla^2 \delta_{ij}) \Sigma^{i||}_{j} = -
\frac{2}{3}\nabla^4 \tilde{\sigma}~.
\end{eqnarray}
Let us recall that the four scalar components of the energy-momentum
tensor are not gauge-invariant: we define the above quantities {\it in
the longitudinal gauge}, but in another gauge, for instance, $\delta
T_0^0$ could behave differently than the above $\delta \rho$,
especially on super-Hubble scales.

We can now write the perturbed Einstein equation for the scalar sector
in the longitudinal gauge,
\begin{eqnarray}
\delta G^0_0
&=&
2 a^{-2} \left\{ - 3  \left( \frac{\dot{a}}{a} \right)^2 \phi
- 3 \frac{\dot{a}}{a} \dot{\psi}
+ \nabla^2 \psi \right\}
= 8 \pi {G}~\delta \rho~,
\\
\delta G^0_i
&=&
2 a^{-2} \partial_i \left\{ \frac{\dot{a}}{a} \phi + \dot{\psi} \right\}
= 8 \pi {G}~(\bar{\rho} + \bar{p}) ~v_i~,
\\
\delta G^i_j
&=&
- 2 a^{-2} \left\{ \left[
\left( 2 \frac{\ddot{a}}{a} - 
\left(\frac{\dot{a}}{a} \right)^2 \right) \phi + 
\frac{\dot{a}}{a} (\dot{\phi} + 2 \dot{\psi}) + \ddot{\psi} 
+ \frac{1}{3} \nabla^2 (\phi - \psi)
\right] \delta^i_j \right.
\nonumber \\
&&
\left.
- \frac{1}{2} \left(
\partial^i \partial_j - \frac{1}{3} \nabla^2 \delta^i_j 
\right) (\phi - \psi) \right\} \\
&=& 8 \pi {G}~(- \delta p ~ \delta^i_j 
+ \Sigma^i_j)~.
\end{eqnarray}
Using the variables $\delta \equiv \delta \rho / \bar{\rho}$,
$\delta p$,  $\theta$, $\sigma$, and switching to comoving
Fourier space, we obtain
\begin{eqnarray}
- 3  \left( \frac{\dot{a}}{a} \right)^2 \phi
- 3 \frac{\dot{a}}{a} \dot{\psi}
- k^2 \psi
&=& 4 \pi {G}~a^2~\bar{\rho}~\delta~,
\label{Einstein1}
\\
- k^2 \left( \frac{\dot{a}}{a} \phi + \dot{\psi} \right)
&=& 4 \pi {G}~a^2~(\bar{\rho} + \bar{p}) ~\theta~,
\label{Einstein2}
\\
\left( 2 \frac{\ddot{a}}{a} - 
\left(\frac{\dot{a}}{a} \right)^2 \right) \phi + 
\frac{\dot{a}}{a} (\dot{\phi} + 2 \dot{\psi}) + \ddot{\psi} 
- \frac{k^2}{3} (\phi - \psi)
&=& 4 \pi {G}~a^2~\delta p~,
\label{Einstein3}
\\
k^2 (\phi - \psi)
&=& 12 \pi {G}~a^2~ (\bar{\rho} + \bar{p}) ~\sigma~.
\label{Einstein4}
\end{eqnarray}
Through the Bianchi identities, the Einstein equation implies the
conservation of the total energy-momentum tensor. Actually, the
energy-momentum tensor of each uncoupled fluid is conserved, and obeys
the continuity equation
\begin{equation}
\dot{\delta} = (1+w) (\theta + 3 \dot{\psi})~,
\label{continuity}
\end{equation}
and the Euler equation
\begin{equation}
\label{Euler}
\dot{\theta} =
\frac{\dot{a}}{a} (3w -1) \theta
- \frac{\dot{w}}{1 + w} \theta
- k^2 \phi
- k^2 \sigma
- \frac{w}{1+w} k^2 \delta~,
\end{equation}
where we assumed an equation of state $p(x,t) = w \rho(x,t)$ for the
fluid, so that $w = \bar{p} / \bar{\rho} = \delta p / \delta \rho$.

\subsection{Linear perturbation theory in a neutrinoless Universe 
(pure $\Lambda$CDM)}
\label{subsec:pert_no_nu}

\subsubsection{Perfect fluids}

Intuitively, in perfect fluids, microscopic interactions enforce a ``collective
behavior'' of the particles. More precisely, they
guarantee that the stress
tensor $T_{ij}$ is isotropic and diagonal, and that the energy-momentum
tensor can be simply described in terms of functions of time and space:
the density $\rho$, the pressure $p$ and the bulk velocity $U^{\mu}$,
\begin{equation}
T^{\mu \nu}=-p~g^{\mu \nu} + (\rho + p) U^{\mu} U^{\nu}~.
\label{tmunuperfect}
\end{equation}
Here, $U^{\mu}= dx^{\mu}/\sqrt{ds^2}$ is the 4-velocity. 
As long as the bulk 3-velocity $v^i=d x^i / d \tau$
remains first-order in perturbations, we can 
approximate the 4-velocity by $U^{\mu}= dx^{\mu}/ [a (1+\phi){d\tau}]$
and obtain:
\begin{equation}
U^{\mu}=\left(a^{-1}[1 - \phi], \, a^{-1} v^i \right)~, \qquad
T^0_0 = \rho~, \quad
T_0^i = v^i~, \quad
T_i^i = -p~,
\label{umu}
\end{equation}
where $T^i_i$ is {\it not} summed over the repeated index.  In a pure
$\Lambda$CDM model, the Universe contains photons, baryons and cold
dark matter. Strictly speaking, the perfect fluid approximation does
not apply to this situation, since on the one hand photons become
collisionless after decoupling, and on the other hand cold dark matter
is collisionless at any time (so it cannot even be called a
fluid). However, we will see in the next paragraphs that we can use
Eqs.\ (\ref{tmunuperfect}, \ref{umu}) in an effective way.

First, after its own decoupling time, 
the cold dark matter component is always non-relativistic and
collisionless. As long as the bulk motion of CDM particles can be described
by a single-valued flow, the CDM energy-momentum tensor is 
formally identical to
that of a perfect fluid with density $\rho_{\rm m}$ and zero pressure.

Second, let us discuss the case of photons and baryons.
If our purpose was to review the CMB physics, it would be crucial to
introduce separately the exact form of the photon and baryon
energy-momentum tensors. However, in this report, we only wish to
provide a simplified description of large scale structure
formation. For this purpose, we can view the baryon and photon
components before recombination as forming a single tightly-coupled
fluid, obeying to Eq.\ (\ref{tmunuperfect}) with the equation of state
of ordinary radiation: $p_{\rm r} = \frac{1}{3} \rho_{\rm r}$ (this
assumes that the baryon density is negligible with respect to the
photon density, which is only a crude approximation around the time of
equality and recombination). The local value of $\rho_{\rm r}$ is
related to the local value of the photon blackbody temperature.
The bulk velocity of this fluid remains small enough to use Eqs.\
(\ref{umu}).  After recombination, the photons are irrelevant for
structure formation, and we will simply discard this component; the
baryons behave like a non-relativistic collisionless medium, but for
simplicity we will neglect their density with respect to that of cold
dark matter.

Finally, since we are only interested in scalar perturbations, we can
reduce the bulk velocities of radiation and CDM into irrotational vector
fields, and define two scalars $\theta_{\rm r}$ and $\theta_{\rm m}$ as in 
Eq.~(\ref{deftheta}).

In summary, we can solve the Einstein equations with the following
perturbed energy-momentum tensor
\begin{eqnarray}
\delta T^0_0 &=& \delta \rho_{\rm r} + \delta \rho_{\rm m}~, \\
\partial^i (\delta T^0_i) &=& 
(\bar{\rho}_{\rm r} + \bar{p}_{\rm r}) \theta_{\rm r} 
+ \bar{\rho}_{\rm m} \theta_{\rm m} = 
\frac{4}{3} \bar{\rho}_{\rm r} \theta_{\rm r} +
\bar{\rho}_{\rm m} \theta_{\rm m}~, \\
\delta T^i_i &=& - \delta p_{\rm r} = - \frac{1}{3} \delta \rho_{\rm r}~.
\end{eqnarray}

\subsubsection{Jeans length}
\label{sec:JeansLength}

In a spatially flat Friedmann Universe, a physical process switched on
at a time $t_i$ and propagating at a velocity $v$ along radial
geodesics ($v \, dt = a(t) \, dx$) can only affect wavelengths
smaller than a causal horizon, defined as
\begin{equation}
d(t_i, t) = a(t) \int_{t_i}^t dx = 
a(t) \int_{t_i}^t \frac{v \, dt'}{a(t')}~.
\end{equation}
This horizon is simply the maximal physical distance on which the
signal can propagate between $t_i$ and $t$ (sometimes, is it defined
with a factor two bigger than above).  The so-called {\it particle
horizon} $d_H$ obeys this definition in the particular case of a
signal traveling at the speed of light, $v=1$ (we adopted units in which
$c=1$).  If both $t_i$ and $t$ are chosen during the matter or
radiation dominated stage, and if $t \gg t_i$, it is straightforward
to show that the particle horizon can be approximated by the Hubble
length $R_H(t) = 1/H(t)$, up to a numerical factor of order
one. Indeed, for a power-law expansion $a(t) \propto t^n$ with $n<1$,
one has
\begin{equation}
R_H(t) = \frac{t}{n} ~, \qquad d_H(t \gg t_i) \simeq \frac{t}{1-n}~.
\end{equation}
Similarly, the characteristic velocity $c_s$ under which acoustic
perturbations propagate before photon decoupling defines a
characteristic length $d_s(t_i,t)$, called the {\it sound
horizon}. For a constant sound speed, and under the conditions
described above for $t_i$ and $t$, the sound horizon is obviously
given (up to a numerical factor of order one) by the ratio $c_s/H(t)$,
which is called the {\it Jeans length}.  The exact definition of the
Jeans wavenumber and Jeans length is usually
\begin{equation}
\label{defJeans}
k_{J}(t) = \left(\frac{4 \pi G \bar{\rho}(t) a^2(t)}{c_s^2(t)}\right)^{1/2}~,
\qquad
\lambda_J(t) 
= 2 \pi \frac{a(t)}{k_{J}(t)}
= 2 \pi \sqrt{2 \over 3} \frac{c_s(t)}{H(t)}~,
\end{equation}
where the numerical factors --which are unimportant for the purpose of
understanding the physics-- are simply dictated by the particular form
of the Newtonian equation of evolution of the density contrast of a
single uncoupled perfect fluid with constant sound speed,
\begin{equation}
\ddot{\delta} + \frac{\dot{a}}{a} \, \dot{\delta}
+ (k^2 - k_J^2) \, c_s^2 \, \delta = 0~,
\end{equation}
which derives from the combination of the continuity, Euler and
Poisson equations.

Our physical expectation is that for a fluid of sound speed $c_s$,
modes with $k>k_J$ will oscillate with a pulsation $\omega = k c_s$,
due to the competition between the gas pressure and the gravitational
compression. These modes are said to be Jeans stable. On the other
hand, for modes $k<k_J$, pressure cannot causally resist to
gravitational compression, and density perturbations should in
principle grow monotonically. This {\it Jeans instability} explains
some basic phenomena in the inhomogeneous Universe: before
recombination, the photon-baryon fluid has a sound speed of order $c_s
\sim c / \sqrt{3}$ and oscillates on scales smaller than $\lambda_J$;
after recombination, $c_s$ and $\lambda_J$ become vanishingly small,
$k_J$ grows to infinity and structures form.

\subsubsection{Radiation domination}

Deep inside the radiation era, we have seen that the radiation
component can be approximated by a perfect fluid of sound speed
$c_s$, due to the tight coupling between baryon and photons. This
self-gravitating fluid oscillates inside the Jeans length (or sound
horizon). The cold dark matter evolves as a test fluid in this
background, until the time (close to equality) at which its
gravitational back-reaction becomes important. However, for
simplicity, we will work at first order in the expansion parameter
$\bar{\rho}_{\rm m} / \bar{\rho}_{\rm r}$.

In this limit, it is possible to derive an analytic expression for the
evolution of $\delta_{\rm r}$ and $\delta_{\rm m}$. The absence of
anisotropic stress implies that the two metric fluctuations are equal,
$\phi=\psi$.  The full continuity, Euler and Einstein equations for
the radiation and matter fluid read
\begin{eqnarray}
\dot{\delta}_{\rm m} = \theta_{\rm m} + 3 \dot{\phi}~, &\qquad&
\dot{\delta}_{\rm r} = \frac{4}{3} \theta_{\rm r} + 4 \dot{\phi}~,
\label{A1} \\
\dot{\theta}_{\rm m} = - \frac{\dot{a}}{a} \theta_{\rm m} - k^2 \phi~,
&\qquad& \dot{\theta}_{\rm r} = - \frac{k^2}{4} \delta_{\rm r} - k^2
\phi~, \label{A2} \\
-3 \frac{\dot{a}}{a} \dot{\phi}
- \left( 3 \left(\frac{\dot{a}}{a}\right)^2 + k^2 \right) \phi &=& 4 \pi G
a^2
\left(\bar{\rho}_{\rm m} \delta_{\rm m}
+ \bar{\rho}_{\rm r} \delta_{\rm r}\right)~, \label{A5} \\
- k^2 \left( \dot{\phi} + \frac{\dot{a}}{a} \phi \right) &=& 4 \pi G
a^2 \left(\bar{\rho}_{\rm m} \theta_{\rm m} + \frac{4}{3}
\bar{\rho}_{\rm r} \theta_{\rm r}\right)~,
\label{A6} \\
\ddot{\phi} + 3 \frac{\dot{a}}{a} \dot{\phi} + \left( 2
\frac{\ddot{a}}{a} - \left( \frac{\dot{a}}{a} \right)^2 \right) \phi
&=& 4 \pi G a^2 \left(\frac{1}{3} \bar{\rho}_{\rm r} \delta_{\rm
r}\right)~, \label{A7}
\end{eqnarray}
and only five out of these seven equations are independent. In
addition, one out of the five independent equations is just a
constraint equation, so the system admits four independent
solutions. In order to find them, let us first notice that since we
have two fluids, we can have differences in the number density
contrasts, usually parametrized by the entropy perturbation
\begin{equation}
\label{def_eta}
\eta = \frac{3}{4} \delta_{\rm r} - \delta_{\rm m}~.
\end{equation}
The pressure perturbation $\delta p = \delta p_{\rm r}$ can be
uniquely decomposed in the basis of the total density perturbation
$\delta \rho = \delta \rho_{\rm r} + \delta \rho_{\rm m}$ and entropy
perturbation $\eta$,
\begin{equation}
\delta p = \left[ 3 \left( 1 + \frac{3 \bar{\rho}_{\rm m}}{4
\bar{\rho}_{\rm r}} \right) \right]^{-1} \left( \delta \rho +
\bar{\rho}_{\rm m} \, \eta \right)~.
\end{equation}
This defines the effective sound speed in the two-fluid as
\begin{equation}
c_{\rm eff}^2 = \left[ 3 \left( 1 + \frac{3 \bar{\rho}_{\rm m}}{4
\bar{\rho}_{\rm r}} \right) \right]^{-1} =
\frac{\dot{\bar{p}}}{\dot{\bar{\rho}}}~.
%= \left[ 3 \left( 1 + \frac{3 a}{4 a_{\rm eq}} \right) \right]^{-1}
\end{equation}
%
%where $a_{\rm eq}$ is the scale factor at equality, when
%$\rho_{\rm m} = \rho_{\rm r}$.
We can reduce the system (\ref{A1})-(\ref{A7}) to a pair of
second-order coupled equations, obtained from the combinations
$[(\ref{A7})-c_{\rm eff}^2(\ref{A5})]$ and $[k^2(\ref{A5})-3\dot{a}/a
(\ref{A6})]$, as follows
\begin{eqnarray}
\ddot{\phi} + 3 \frac{\dot{a}}{a} (1 + c_{\rm eff}^2) \dot{\phi} +
\frac{1}{a} \left[ 3 \left( 1+ \frac{\bar{\rho}_{\rm
r}}{\bar{\rho}_{\rm m}} \right) c_{\rm eff}^2 - \frac{\bar{\rho}_{\rm
r}}{\bar{\rho}_{\rm m}}\right] \phi + k^2 c_{\rm eff}^2 \phi &=&
\frac{3}{2a} c_{\rm eff}^2 \eta~,
\label{sys_phi} \\ 
\frac{1}{3 c_{\rm eff}^2} \ddot{\eta} + \frac{\dot{a}}{a} \dot{\eta} +
\frac{k^2}{4} \frac{\bar{\rho}_{\rm m}}{\bar{\rho}_{\rm r}} \eta &=&
\frac{k^4}{6} a \frac{\bar{\rho}_{\rm m}}{\bar{\rho}_{\rm r}} \phi~.
\label{sys_eta}
\end{eqnarray}
Once the four independent solutions have been found, the density
contrasts can be obtained e.g.\ from combining Eqs.\ (\ref{A7}) and
(\ref{sys_phi})
\begin{eqnarray}
\delta_{\rm r} &=& \frac{3 c_{\rm eff}^2}{4 \pi G a^2 \bar{\rho}_{\rm
r}} \left(\frac{3}{2 a} \eta - 3 \frac{\dot{a}}{a} \dot{\phi} - \left(
3 \left( \frac{\dot{a}}{a} \right)^2 + k^2 \right) \phi \right)~,
\label{deltar_evol} \\
\delta_{\rm m} &=& \frac{9 c_{\rm eff}^2}{16 \pi G a^2 \bar{\rho}_{\rm
r}} \left(\frac{3}{2 a} \eta - 3 \frac{\dot{a}}{a} \dot{\phi} - \left(
3 \left( \frac{\dot{a}}{a} \right)^2 + k^2 \right) \phi \right) -
\eta~,
\label{deltam_evol}
\end{eqnarray}
and the velocity gradients follow from Eqs.\ (\ref{A2}).  The
evolution of the scale factor can be taken from equation
(\ref{back_RDMD}).  Deep inside the radiation era, we can write $a$,
$c_{\rm eff}^2$ and the two equations $[\tau^2 \times
(\ref{sys_phi})]$ and (\ref{sys_eta}) at first order in
$\bar{\rho}_{\rm m}/\bar{\rho}_{\rm r} = \tau/\tau_{\rm eq}$,
\begin{equation}
\tau^2 \ddot{\phi} + 4 \tau \dot{\phi} + 
(\tau^2 \omega^2) \phi = 0~, \qquad 
\ddot{\eta} + \frac{1}{\tau} \dot{\eta}
= \frac{3 \omega^4 \tau^2}{2} \phi~,
\label{eta_limit}
\end{equation}
where $\omega=k c_{\rm eff} = k/\sqrt{3}$.  In this limit, the first
equation decouples: this just confirms that the radiation fluid is
self-gravitating, and the CDM behaves as a test fluid. Two independent
solutions of the system, both satisfying $\eta(\tau \rightarrow 0)=0$,
read
\begin{eqnarray}
\phi = (\omega \tau)^{-2} && \left\{  
C(k) \left( \frac{\sin \omega
\tau}{\omega \tau} - \cos \omega \tau \right) + D(k) \left( - \sin
\omega \tau - \frac{\cos \omega \tau}{\omega \tau}\right) \right\}~,
\nonumber \\ 
\eta = - 3  && \left\{  
C(k) \left( \int_0^{\omega \tau} \frac{\cos x-1}{x} dx - \frac{1}{2}
(\cos \omega \tau - 1) \right) \right. \nonumber \\
&& \left. + D(k) \left( \int_0^{\omega \tau} \frac{\sin x}{x} dx + \frac{1}{2}
\sin \omega \tau \right) \right\}~.
\end{eqnarray}
The solution proportional to $C(k)$ represents the growing adiabatic
(or isentropic) mode, the other one is the decaying adiabatic mode.
There are two other obvious solutions $(\phi,\eta)=(0,A(k))$ and
$(\phi,\eta)=(0,B(k)\ln\tau)$ which stand respectively for the
isocurvature growing and decaying mode. For the growing adiabatic
mode, we get from Eqs.\ (\ref{deltar_evol}) and (\ref{deltam_evol})
that the radiation and matter density contrasts evolve like
\begin{eqnarray}
\delta_{\rm r} &=& -2 C(k) \left\{  
\frac{2 \sin \omega \tau - \omega \tau \cos \omega \tau}{\omega \tau}
-2 
\frac{\sin \omega \tau - \omega \tau \cos \omega \tau}{(\omega \tau)^3}
\right\}~, \label{pert_r_rd}
\nonumber \\
\delta_{\rm m} &=& 3 C(k) \left\{  
- \frac{\sin \omega \tau}{\omega \tau}
+ \frac{\sin \omega \tau - \omega \tau \cos \omega \tau}{(\omega \tau)^3}
+ \frac{1}{2} + \int_0^{\omega \tau} \frac{\cos x-1}{x} dx
\right\}~.
\label{pert_m_rd}
\end{eqnarray}
The integral is given by
\begin{equation}
\int_0^{\omega \tau} \frac{\cos x-1}{x} dx
= \mathrm{Ci}(\omega \tau) - \mathcal{C} - \ln(\omega \tau)~,
\end{equation}
where $\mathrm{Ci}(x)$ is the cosine integral and $\mathcal{C}\simeq
0.5772$ is the Euler constant. For $x>1$, the function
$\mathrm{Ci}(x)$ describes damped oscillations around zero.  Let us
summarize the physics described by these solutions. During the
radiation era and far outside the sound horizon, all quantities appear
as frozen in the longitudinal gauge, with
\begin{equation}
\delta_{\rm m} = \frac{3}{4} \delta_{\rm r} = - \frac{3}{2} \phi = -
\frac{1}{2} C(k)~.  \quad \quad \quad [\mathrm{RD}, \quad \omega \tau
= k \tau / \sqrt{3} \ll 1]
\label{deltam_RD_lw}
\end{equation}
Then, inside the sound horizon, the evolution of the radiation and
matter perturbations are radically different from each other.  The
self-gravitating radiation component oscillates on scales smaller than
its Jeans length, and Eq.\ (\ref{pert_r_rd}) represents a first-order
approximation to the acoustic oscillations observed in the CMB
anisotropy spectrum.  Instead, the cold dark matter has a considerably
smaller Jeans length and does not fluctuate on its own.  The
gravitational driving force only induces damped oscillations on
$\delta_{\rm m}$, on top of a net gravitational clustering
effect. This clustering is not very efficient, because the
gravitational potential fluctuations are damped by the acoustic
oscillations and do not feel significant back-reaction from the CDM
component. In average, the growth of the CDM perturbations is
proportional to the logarithm of conformal time
\begin{equation}
\delta_{\rm m} \simeq 3 C(k) [1/2 + \mathcal{C} + \ln(\omega \tau)]~.
\quad \quad \quad [\mathrm{RD}, \quad \omega \tau = k \tau / \sqrt{3}
\gg 1]
\label{deltam_RD_sw}
\end{equation}  

\subsubsection{Matter domination}

Deep into the matter domination era, when $a \propto \tau^2$, the
effective sound speed is vanishingly small and Eq.\ (\ref{sys_phi})
simply reduces to
\begin{equation}
\ddot{\phi} + 3 \frac{\dot{a}}{a} \dot{\phi} = 0~,
\end{equation}
with solutions
\begin{equation}
\phi = \tilde{C}(k) + \tau^{-5} \tilde{D}(k)~.
\end{equation}
The matter density contrast then follows from Eq.\ (\ref{A5}), which
simplifies into
\begin{equation}
- \left( \frac{12}{\tau^2}+k^2 \right) \phi - \frac{6}{\tau} \dot{\phi} 
= \frac{3}{\tau^2} \delta_{\rm m}~,
\end{equation}
and we finally obtain
\begin{equation}
\delta_{\rm m} = - \left( 4 + \frac{k^2 \tau^2}{3} \right) \tilde{C}(k) 
+ \left( \frac{k^2 \tau^2}{3} - 6 \right) \tau^{-5} \tilde{D}(k)~.
\label{pert_md}
\end{equation}
We can identify the solution proportional to $\tilde{C}(k)$
(resp.\ $\tilde{D}(k)$) with the adiabatic growing (resp.\ decaying)
mode.  A detailed and precise matching between the solutions
(\ref{deltam_RD_lw}), (\ref{deltam_RD_sw}) --valid deep inside the
radiation era-- and the solution (\ref{pert_md}) --valid deep inside
the matter era-- would require some significant work in order to take
into account various effects close to the time of equality.  However,
a crude matching is sufficient for catching the essential behavior of
the matter power spectrum. {}From (\ref{deltam_RD_lw}),
(\ref{deltam_RD_sw}) and (\ref{pert_md}) we obtain that at any time
$\tau$ belonging to the MD stage,
\begin{equation}
\delta_{\rm m} (k,\tau) = \left(4 + \frac{k^2 \tau^2}{3}\right)
\frac{\delta_{\rm m} (k,\tau_{\rm eq})}{4 + k^2 \tau^2_{\rm eq} / 3}~,
\label{deltam_rdmd}
\end{equation}
where $\delta_{\rm m} (k,\tau_{\rm eq})$ is equal to
Eq.\ (\ref{pert_m_rd}) evaluated at $\tau=\tau_{\rm eq}$.  Since we are
interested only in modes which are inside the Hubble radius in the
recent Universe, we can take the limit $k^2 \tau^2 \gg 1$. During
matter domination, it is also possible to replace $\tau$ by
$2/(aH)$. Then, we obtain
\begin{equation}
\delta_{\rm m} (k,\tau_M) = \frac{k^2}{3 (a H)^2}
\, \times \, 
\frac{\delta_{\rm m} (k,\tau_{\rm eq})}{1+k^2 \tau^2_{\rm eq} /12}~.
\end{equation}

\subsubsection{Dark energy domination}

After the matter dominated era, the dark energy component becomes
the dominant contribution to the energy density of the Universe. 
In this era, the evolution of the gravitational potential is still given
by Eq.~(\ref{A7}), but now the effective mass term does not cancel.
When the dark energy corresponds to the contribution of a
cosmological constant with equation of state parameter $w=-1$
($P_\Lambda=-\rho_\Lambda)$,
\begin{equation}
\ddot{\phi} + 3 \frac{\dot{a}}{a} \dot{\phi} 
+ (8 \pi G a^2 \bar{\rho}_{\Lambda})
\, \phi= 0~,
\label{dif_phi_lambda}
\end{equation}
and the potential decays proportionally to a scale-independent damping factor
$g(\tau)=\phi(\tau)/\phi(\tau_M)$, where $\tau_M$ can be any time during
matter domination. This factor can be found by numerical integration.
In the case of a
cosmological constant, a fairly good analytic approximation of
$g(\tau_0)$ -- where $ \tau_0$ is the present time -- is provided in
Ref.\ \cite{Kofman:1993ag}
\begin{equation}
g(\tau_0) \simeq \Omega_{\rm
m}^{0.2}/[1+0.003(\Omega_{\Lambda}/\Omega_{\rm m})^{4/3}]~.
\end{equation}

The modes probed by LSS experiments are all deep inside the Hubble
radius during the end of matter domination and during $\Lambda$
domination.  On those scales, the Poisson equation of Newtonian
gravity (which corresponds to Eq.~(\ref{A5}) with $k \gg aH$) gives a
simple relation between the gravitational fluctuations and the matter
perturbations,
\begin{equation}
\Delta \phi = 4 \pi G \bar{\rho}_{\rm m} \delta_{\rm m}
\qquad
\Longleftrightarrow
\qquad
\delta_{\rm m} = - { k^2 \phi \over 4 \pi G a^2 \bar{\rho}_{\rm m}}
= - { 2 k^2 \phi \over 3 \Omega_{\rm m} (aH)^2}~.
\label{poisson_LD}
\end{equation}
So, $\delta_{\rm m}$ is proportional to $a \phi$, which proves that
the decay of the gravitational potential slows down the growth of
matter fluctuations as a function of the scale factor.  Also, we
immediately obtain a relation between $\delta_{\rm m}$ during
$\Lambda$ domination and matter domination
\begin{equation}
\delta_{\rm m} (\tau) = \left[ 
\frac{a(\tau_M) H(\tau_M)}{a(\tau) H(\tau)} \right]^2
\frac{g(\tau)}{\Omega_{\rm m}(\tau)} \delta_{\rm m} (\tau_M)~.
\label{m_evol_lambda}
\end{equation}
Finally, Eqs.\ (\ref{pert_m_rd}), (\ref{deltam_rdmd}) and
(\ref{m_evol_lambda}) give us the expression of matter perturbations
on sub-Hubble scales during $\Lambda$ domination,
\begin{eqnarray}
&&\!\!\!\!\!\!\!\! \delta_{\rm m} (k,\tau) = 
\frac{ g(\tau) k^2 C(k)}{\Omega_{\rm m}(\tau)[a(\tau)H(\tau)]^2} \times
\\
&&\!\!\!\! \!\!\!\! \frac{\left[ - (\sin kx)/(kx)
+ (\sin kx - kx \cos kx)/(kx)^3
+ \frac{1}{2} + \mathrm{Ci}(kx) - \mathcal{C} - \ln(kx)
\right]
}{[1+(kx) ^2 /12]}~, \nonumber
\end{eqnarray}
where $x \equiv \tau_{\rm eq} / \sqrt{3}$.  The matter power spectrum
is defined as $P(k,\tau)=\langle|\delta_{\rm
m}(k,\tau)|^2\rangle$. The previous equation allows us to relate it to
the primordial power spectrum, usually parametrized by a power-law
\begin{equation}
k^3 \langle|\phi(k, \tau \rightarrow 0)|^2\rangle = \frac{k^3}{9} 
\langle|C(k)|^2\rangle
\equiv A k^{n_s-1}~,
\label{eq:primordial_ps}
\end{equation}
where the index $n_s$ is the scalar tilt.
We can compute the two asymptotes for the modes which were far
outside/inside the sound horizon at the time of equality,
\begin{eqnarray}
\!\!\!\!\!\!\!\!\!\!P(k,\tau) &\simeq& \frac{ g(\tau)^2}{\Omega_{\rm
m}(\tau)^2 [a(\tau)H(\tau)]^4} \,\times \, \frac{A k^{n_s}}{4}~,
\qquad \qquad \qquad ~~~~k \tau_{\rm eq} / \sqrt{3} \ll 1~,
\label{eq:pk_smallk} \\
\!\!\!\!\!\!\!\!\!\!P(k,\tau) &\simeq& \frac{g(\tau)^2}{\Omega_{\rm
m}(\tau)^2 [a(\tau)H(\tau)]^4} \,\times \, \frac{9 \,[12]^2 A k^{n_s}
[\ln kx]^2}{[kx]^4}~, ~~~~~ k \tau_{\rm eq} / \sqrt{3} \gg 1.
\label{eq:pk_largek}
\end{eqnarray}
Because of the crude approximations performed around the time of
equality, this solution is not accurate, in particular for the global
amplitude and for the behavior near $k \sim \sqrt{3} / \tau_{\rm
eq}$. However, the shape of the two asymptotes reflects the behavior of
the true power spectrum obtained by a full numerical simulation, which
is close to the BBKS \cite{Bardeen:1985tr} fitting formula
\begin{eqnarray}
\!\!\!\!\!\!\!\!P(k,\tau) &\simeq& \frac{ g(\tau)^2}{\Omega_{\rm
m}(\tau)^2 [a(\tau)H(\tau)]^4} \, \frac{9 A k^{n_s}}{25}~
T(q)^2~,\\
\!\!\!\!\!\!\!\!T(q) &\equiv& \frac{\ln (1 + 2.34~q)}{2.34 q} [1 +
3.89~q + (16.1~q)^2 + (5.46~q)^3 + (6.71~q)^4 ]^{-1/4}~, \nonumber
\\
\!\!\!\!\!\!\!\!q &\equiv&\frac{k}{\Omega_{\rm m} h^2
\exp(-2\Omega_b)~{\rm Mpc}}~.  \nonumber
\end{eqnarray}
Note also that the cosmological constant not only damps the
perturbations $\delta \rho_{\rm m}$ but also implies a smaller
fraction $\Omega_{\rm m}$.  The net effect is that today {\it and for
fixed} $a_0 H_0$, the power spectrum normalization is enhanced by the
cosmological constant (i.e.\ $g(\tau_0)/\Omega_{\rm m}(\tau_0)>1$ if
$\Omega_{\Lambda}>0$).

\subsubsection{Numerical results}
\label{sec:num_res}
\begin{figure}[t]
\begin{center}
\vspace{-1cm}
\hspace{-1.cm}
\includegraphics[width=.95\textwidth]{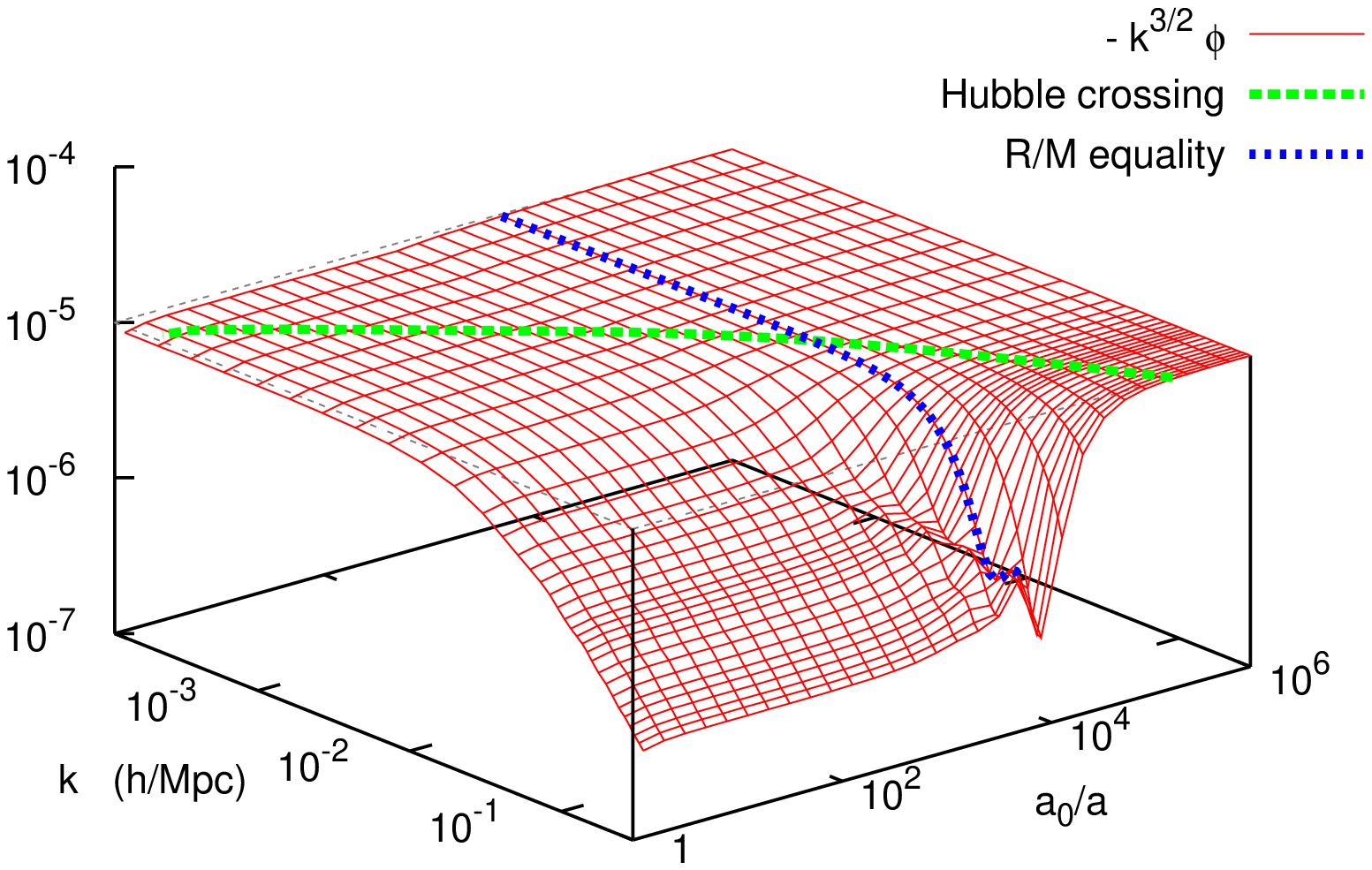}\\
\vspace{-2cm}
\includegraphics[width=.95\textwidth]{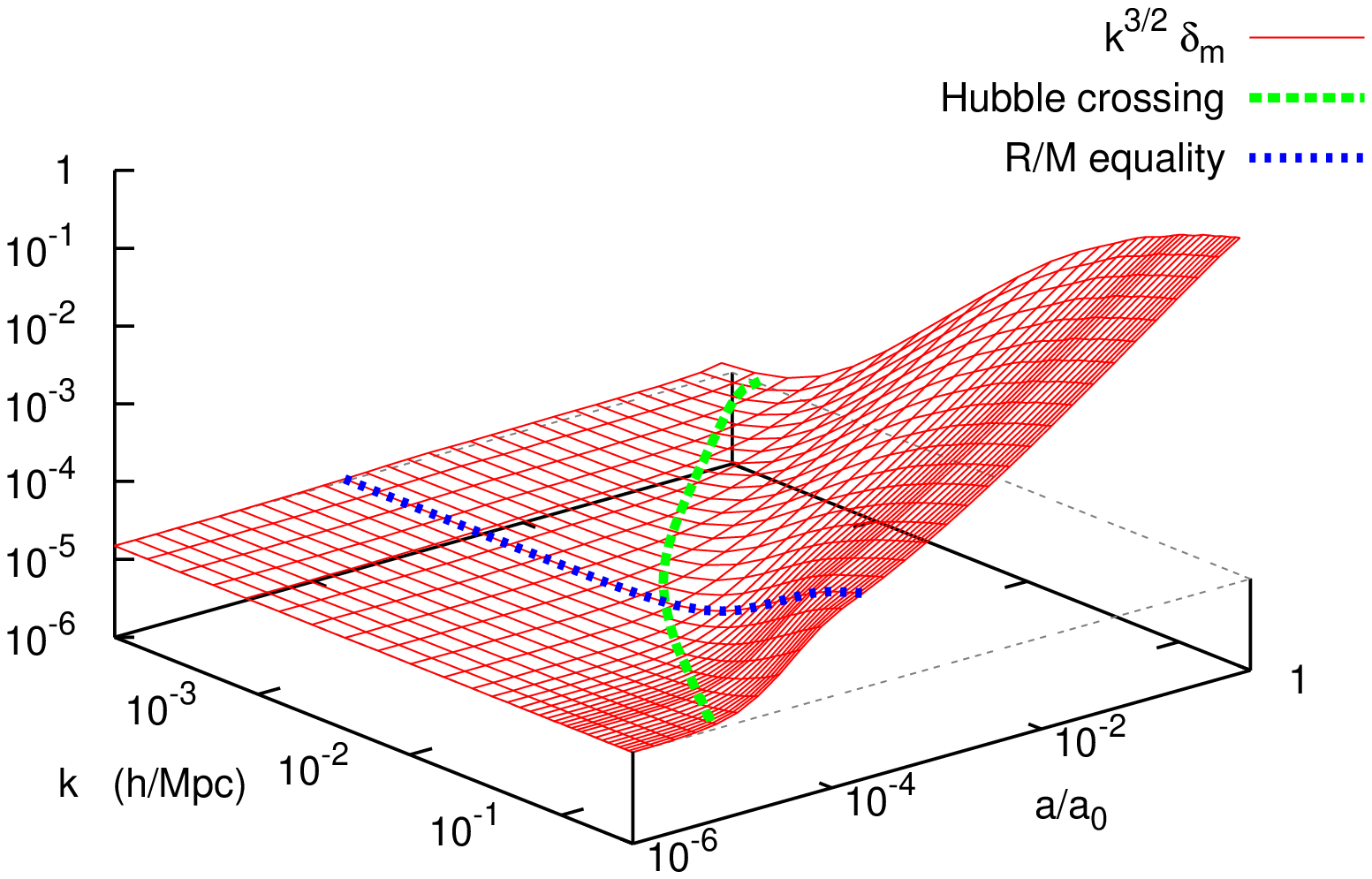}\\
\vspace{-1cm}
\caption{\label{surf_LCDM} Evolution of the longitudinal metric
perturbations $\phi=\psi$ (top) and the matter density
perturbations $\delta_{\rm m}=\delta \rho_{\rm m}/\bar{\rho}_{\rm m}$
(bottom) in a neutrinoless $\Lambda$CDM model, obtained
numerically with a Boltzmann code, as a function of the scale factor
$a$ and Fourier wavenumber $k$. In order to get a better view, we
chose opposite time axes in the two plots: time is evolving from back
to front in the upper plot, and from front to back in the lower plot.
The initial condition was set arbitrarily to $k^{3/2} \phi = - 10^{-5}$.
}
\end{center}
\end{figure}

\begin{figure}[t]
\begin{center}
\vspace{-1cm}
\hspace{-1cm}
\includegraphics[width=0.52\textwidth]{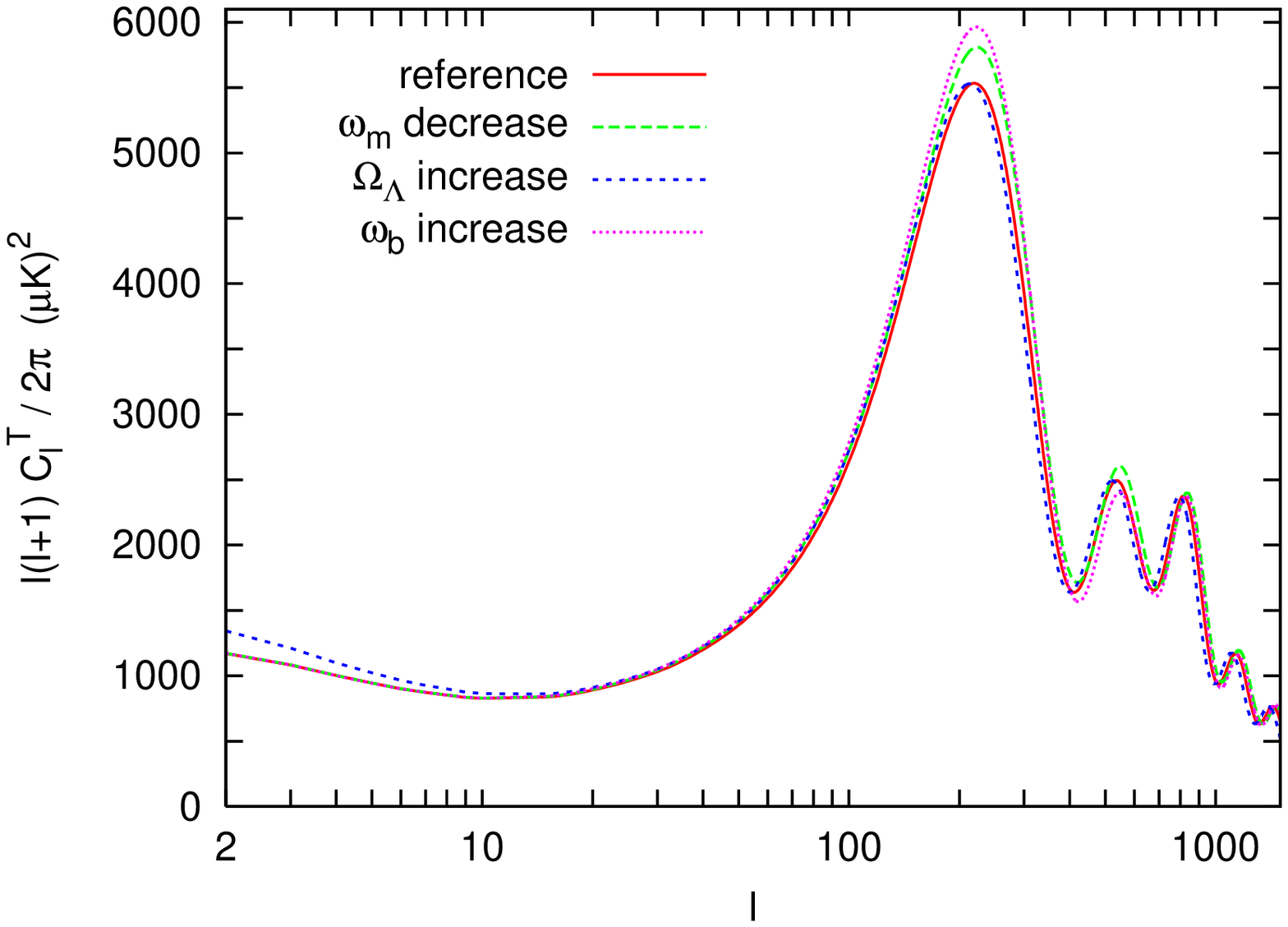}
\includegraphics[width=0.52\textwidth]{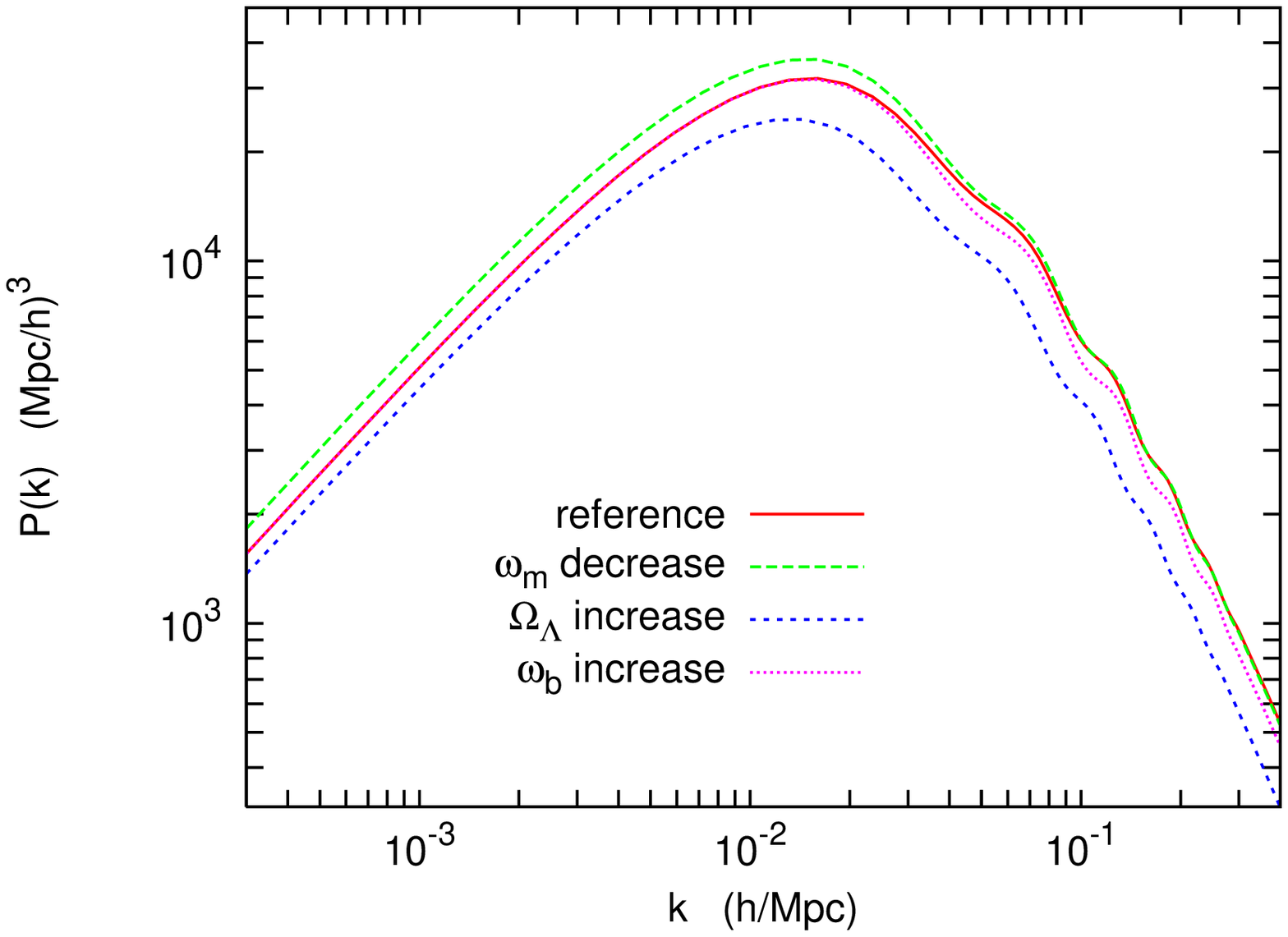}
\caption{\label{fig_powspec}
{\it(Red solid lines)} The power spectrum of CMB temperature
anisotropies in multipole space (left), and the LSS matter power
spectrum evaluated today in Fourier space (right), for a reference
neutrinoless $\Lambda$CDM model with $\omega_{\rm m}=0.13$,
$\Omega_{\Lambda}=0.74$, $\omega_{\rm b}=0.022$, $n_s=0.95$,
$\tau=0.08$.  {\it(Dashed and dotted lines)} Same power spectra for
three models in which we keep all parameters fixed but one: either
$\omega_{\rm m}=0.12$, or $\Omega_{\Lambda}=0.80$, or $\omega_{\rm
b}=0.026$. The changes in the power spectra illustrate the effects 
labeled from (1)
to (4) in Sec.\ \ref{subsec:params}, as explained in details at the
end of this subsection.}
\end{center}
\end{figure}

We show in Fig.\ \ref{surf_LCDM} the evolution of $\psi=\phi$ and
$\delta_{\rm m}$ for a neutrinoless $\Lambda$CDM model, obtained
numerically by solving the exact Einstein and conservation
equations. We run the public code {\sc cmbfast} \cite{Seljak:1996is}
based on the synchronous gauge, and convert the output into the
longitudinal gauge variables used throughout this section.
We start from a flat
primordial spectrum with amplitude $k^{3/2} \phi= 10^{-5}$
(let us recall that the ``amplitudes'' calculated by the code should
be interpreted in the real Universe as the variance of random Gaussian
perturbations). We show in Fig.\ \ref{fig_powspec} the CMB temperature
anisotropy and LSS power spectra for the same cosmological model (but
now with the primordial spectrum amplitude preferred by current data)

The global evolution is consistent with the analytical solutions found
in the previous subsections.  Indeed, for metric perturbations
$\psi=\phi$ (upper plot, time evolving from back to front), we see
that during radiation domination, super-Hubble modes are constant, and
sub-Hubble modes are damped (with small oscillations). During matter
domination, all modes are constant, so the shape imprinted near the
time of equality survives until today (at least, as long as
perturbations remain in the linear regime).  During $\Lambda$
domination, there is a small scale-independent damping of all
perturbations, hardly seen on the figure.  For matter density
perturbations $\delta_{\rm m}$ (lower plot, time evolving from front
to back), we see that during radiation domination, super-Hubble modes
are constant, and sub-Hubble modes grow slowly (asymptotically like
the logarithm of $a$).  During matter domination, super-Hubble modes
are still constant, while sub-Hubble modes grow linearly with $a$. The
smallest scales are the first ones to reach the condition of
non-linearity $\delta_{\rm m}\sim1$, and large scales are still linear
today.  During $\Lambda$ domination, $\delta_{\rm m}$ grows a bit more
slowly as a function of $a$.

\subsubsection{Parameter dependence}
\label{subsec:params}

A minimal flat $\Lambda$CDM models without neutrinos can be described
with six parameters: the cosmological constant fraction
$\Omega_{\Lambda}$, the total non-relativistic matter density
$\omega_{\rm m}=\Omega_{\rm m} h^2$, the baryon density
$\omega_b=\Omega_b h^2$, the primordial spectrum
amplitude $A$ and tilt $n_s$, the optical depth to reionization
$\tau$. These parameters control various physical effect which are
responsible for the shape of the observable power spectra:
\begin{enumerate}
\item
{\bf The time of Radiation/Matter equality}. In the above parameter
basis, the time of equality between $\rho_{\rm m}$ and $\rho_{\rm r}$
is fixed by $\omega_{\rm m}$ only, since $(a_{\rm eq}/a_0)=\rho_{\rm
r}^0/\rho_{\rm m}^0$ (the script $^0$ means ``evaluated today''), and
$\rho_{\rm r}^0$ is fixed by the CMB temperature. A late equality
implies more metric perturbation damping for modes entering the Hubble
radius during radiation domination (see Fig.\ \ref{surf_LCDM}). 
For
the LSS matter power spectrum, the consequence is the following: since on
sub-Hubble scales $\delta_{\rm m}$ grows more efficiently during MD
than during RD (because $\psi$ does not decay), it is clear that when
$\omega_{\rm m}$ decreases (late equality) the matter
power spectrum is suppressed on small scales relatively to large scales.
Also, the global normalization increases, because of the factor 
$\omega_{\rm m}^{-2}$ in Eqs.\ (\ref{eq:pk_smallk},\ref{eq:pk_largek}). 
So, the net effect of 
postponing equality is to amplifies $P(k)$ only for small $k$.
For the CMB, the effect is opposite and
counter-intuitive: small-scale perturbations are boosted.  This is
related to the dilation effect before recombination, and to the early
Integrated Sachs-Wolfe effect just after recombination. In simple
words, the rapid decay of $\psi(k,\tau)$ just after Hubble crossing
during RD tends to boost the beginning of the first acoustic
oscillation through gravitational redshift effects.  So, a later
equality induces higher CMB peaks, especially for the first one.\\

\item
{\bf The time of Matter/$\Lambda$ equality}. If the cosmological
constant is larger, equality between matter and $\Lambda$ takes place
earlier.  We have seen that after that time, $\psi$ decays and
$\delta_{\rm m}$ grows more slowly. Thus the matter power spectrum
normalization is slightly suppressed by a long $\Lambda$ domination.
The CMB spectrum is affected in a more subtle way: the time-variation
of the metric perturbations leads to a net redshifting of the CMB
photons traveling across a gravitational potential well at small $z$
(late Integrated Sachs-Wolfe effect). Therefore, the nearby
distribution of galaxy cluster leaves an imprint in the large-scale
CMB spectrum: the small-$l$ multipoles are enhanced (the larger is
$\Lambda$, the more significant is the enhancement). This effect is
difficult to measure experimentally due to cosmic variance (on large
scales, we see only a few realization of the stochastic mode
amplitudes, and we expect a large scattering of the data points around
the theoretically predicted variance).\\

\item
{\bf The physical scale of the sound horizon at equality, and the
angular scale of the sound horizon at recombination}.  {}From the
previous sections it is clear that the characteristic scale in the
matter power spectrum is fixed by the sound horizon at the time of
equality (i.e., essentially, the Hubble scale $R_H(\tau_{\rm eq})$
divided by $\sqrt{3}$).  Since this scale can be computed by
extrapolating the present value of the Hubble parameter back in time,
it depends on $h$, $\Omega_{\Lambda}$, and on the time of equality. In
our parameter basis, it means that changing any of the two parameters
$\omega_{\rm m}$, $\Omega_{\Lambda}$ shifts the matter power spectrum
horizontally.  The characteristic scale of the oscillations in the CMB
power spectrum is set by the sound horizon $d_s(\tau_{\rm rec})$ at
recombination.  The time $\tau_{\rm rec}$ is more or less fixed by
thermodynamics, but the sound horizon is an integral over $c_s dt /
a(t)$ between $0$ and $\tau_{\rm rec}$. This integral depends on the
time of equality, and on the baryon density at late times (through
$c_s$).  In addition, the actual observable quantity is the {\it
angular scale} of this sound horizon on the last scattering surface,
which is given, for fixed $d_s(\tau_{\rm rec})$, by $h$ and
$\Omega_{\Lambda}$.  We conclude that the observed angular scale of
the peaks constrains a combination of the three parameters
$\Omega_{\Lambda}$, $\omega_{\rm m}$ and $\omega_b$.\\

\item
{\bf The balance between gravity and pressure in the tightly-coupled
photon-baryon fluid}. In this section, for simplicity, we assumed that
before recombination the energy density of the tightly-coupled
photon-baryon fluid was dominated by that of photons.  Actually, when
the baryon density becomes significant (near the time of equality),
the fluid is more affected by the gravitational compression --which
is more efficient for non-relativistic baryons than for relativistic
photons-- relatively to the competing photon pressure. Inside a
gravitational potential well of given amplitude, the zero-point of
oscillations (i.e. the instantaneous value of $\delta T/T$ for which
gravity and pressure exactly cancel each other) is displaced. The main
effect in the CMB spectrum is an increase of the first and third
peaks, but not of the second one. As far as the matter power spectrum
is concerned, we should remember that $P(k)$ probes the total
contribution from baryonic and cold dark matter perturbations. In the
limit of a small CDM contribution ($\omega_b/\omega_{\rm m}\rightarrow
1$), the small-scale matter power spectrum has a low amplitude,
because it does not benefit from the slow growth of CDM perturbations
before equality on sub-Hubble scales (visible in Fig.\ \ref{surf_LCDM}).
Also, in this limit, the oscillatory structure of $\delta_b$ at the
time of recombination survives until now, and the matter power
spectrum exhibits visible wiggles.  In the opposite limit
$\omega_b/\omega_{\rm m}\rightarrow 0$, which we adopted in our
analytical approximate solution, the small-scale power spectrum is
higher, and the baryonic oscillations in $P(k)$ are small (still, they
are visible in Fig.\ \ref{fig_powspec} between $k=0.05~h^{-1}$ Mpc and
$k=0.2~h^{-1}$ Mpc, and they have actually been observed recently, see
Ref.\ \cite{Eisenstein:2005su}).\\

\item
{\bf The amplitude of primordial perturbations} ($A$), defined in Eq.\
(\ref{eq:primordial_ps}), obviously fixes the global normalization of
all spectra.\\

\item
{\bf The tilt of primordial perturbations} ($n_s$), defined in Eq.\
(\ref{eq:primordial_ps}), fixes the balance between large and small
scale amplitudes. Together with the previous parameter it provides
interesting constraints on inflationary models.\\

\item
{\bf The optical depth to reionization} ($\tau$) is a rather
simplified way of parameterizing the reionization of the Universe at
redshift of order 10 to 20 (depending on the model). This parameter
has no effect on the matter power spectrum, but it damps the CMB
temperature fluctuations for scales which are inside the Hubble
horizon at the time of reionization, leaving small-$l$ temperature
multipoles unchanged.
\end{enumerate}

In Fig.\ \ref{fig_powspec}, we illustrate graphically the consequence
of varying either $\omega_{\rm m}$, $\Omega_{\Lambda}$ or $\omega_{\rm
b}$, while keeping all the other cosmological parameters fixed. When
we decrease $\omega_{\rm m}$, the relevant effects are (1) and (3): 
equality is postponed, boosting the first CMB peak as well as $P(k)$ for
small $k$; simultaneously, there is a tiny horizontal shift of both
spectra.  When we increase $\Omega_{\rm \Lambda}$, we can see the
effects (2) and (3): the smallest CMB temperature multipoles increase,
the $P(k)$ normalization decreases and both spectra are shifted to the
left. When $\omega_{\rm b}$ is increased, the effects (3) and (4) are
taking place: the first CMB peak increases, while the second one
decreases slightly; $P(k)$ is slightly suppressed on small scales;
finally, the scale of CMB peaks is slightly modified.

We have seen that ($A$, $n_s$, $\tau$) have very specific effects,
while ($\Omega_{\Lambda}$, $\omega_{\rm m}$, $\omega_b$) have
intricate effects. Nevertheless, since there are seven effects for six
parameters,
it is in principle easy to measure all of them from CMB
and LSS data, assuming a flat $\Lambda$CDM model.  However, the
neutrinoless model described in this section for pedagogical purposes
is unable to provide a good fit to the data: it is time now to include
neutrinos.

\subsection{Linear perturbation theory in presence of neutrinos ($\Lambda$MDM)}
\label{subsec:pert_nu}

We will now restore the neutrino component, which is expected to
behave at late times (after the non-relativistic transition) as an
extra sub-dominant dark matter component: the standard cosmological
scenario should actually be called $\Lambda$ Mixed Dark Matter
($\Lambda$MDM).

The impact of neutrinos on cosmological perturbations was first
clearly explained in 1980 by Bond, Efstathiou and Silk
\cite{Bond:1980ha}. Later, during the 1980's, a very large number of
papers investigated this issue in more details, and it would be
difficult to provide an exhaustive list of references
(see e.g.\ those given in \cite{Dolgov:2002wy,Khlopov:1999rs}). 
Despite their interest, many of these papers are not of practical
use nowadays because they generally assumed a neutrino-dominated
universe with no CDM or $\Lambda$ components. For instance, the
interested reader can take a look at Refs.\
\cite{Bond:1983hb,White:1984yj} and go through the historical summary
of Ref.\ \cite{Primack:2001ib}.

Let us first introduce some notations and formalism --which are very
close to those of Ma \& Bertschinger \cite{Ma:1995ey}, but have been
adapted to our metric signature $(+,-,-,-)$. The momentum of individual
neutrinos can be labeled by $P_i$, which is the canonical conjugate of
the comoving coordinate $x^i$. It differs from the proper momentum
$p_i$ measured by a comoving observer (with fixed $x^i$ coordinates)
by
\begin{equation}
P_i = a (1-\psi) p_i~.
\end{equation}
Note that in absence of metric perturbations, $P_i$ would remain
constant, while $p_i$ would decrease with the expansion like
$a^{-1}$. Usually, the formulas are conveniently expressed in terms of
the third variable $q_i \equiv a p_i$, which can be viewed as the
proper momentum ``corrected'' from the effect of homogeneous
expansion.  The proper energy of a single neutrino measured by a
comoving observer is equal to $(p^2+m^2)^{1/2}$. We define the related
quantity $\epsilon$
\begin{equation}
\epsilon = a (p^2+m^2)^{1/2} = (q^2 + a^2 m^2)^{1/2}~.
\end{equation}
{}From $P_{\mu} P^{\mu}=m^2$ and $P_i=(1-\psi)q_i$, we get $P_0 = (1 +
\phi) \epsilon$.  Finally, we recall that by definition the
phase-space distribution $f$ gives the number of particles in an
infinitesimal phase-space volume: $dN = f(x^i,P_j,\tau) \, d^3 \! x^i
\, d^3 \! P_j$.

\subsubsection{Collisionless fluids}

In the approximation of a homogeneous Universe, the phase-space distribution
of neutrinos would be perfectly isotropic,
\begin{equation}
f(x^i,P_j,\tau)=f_0(P,\tau)~,
\end{equation}
where $f_0$ is the Fermi-Dirac distribution already introduced in 
Eq.\ (\ref{FD}). Then, the energy-momentum tensor would be diagonal
with an isotropic pressure term
\begin{eqnarray}
T_0^0 = \bar{\rho}_{\nu} &=&  \frac{4 \pi}{a^4} \int q^2dq\,
		\epsilon \, f_0(q)~, \\
T_i^i = - \bar{p}_{\nu} &=& - \frac{4 \pi}{3 a^4} \int q^2dq\,
		{q^2\over\epsilon} \, f_0(q)~.
\end{eqnarray}
Note that $f_0$ as a function of $q$ is time-independent, 
\begin{equation}
f_0(q) = \frac{1}
{e^{q/aT_{\nu}}+1}~,
\end{equation}
because after neutrino decoupling the product of the neutrino
temperature $T_{\nu}$ by the scale factor $a$ is a constant
number. 

Spatial perturbations in the metric will induce variations in the
neutrino phase-space distribution depending on time, space and
momentum.  The energy-momentum tensor will be perturbed accordingly.
With respect to a perfect fluid, the main difference is is that, in
absence of microscopic interactions, there is no reason for the stress
tensor $\delta T_{ij}$ to be isotropic at first order in
perturbations.  Therefore, the scalar sector of $\delta T_{\mu \nu}$
contains an extra degree of freedom, the anisotropic stress $\sigma$.

This can be understood in physical terms.  First, strongly-interacting
fluids have bulk motions: the coarse-grained fluid has a unique
velocity in a given point. In the local rest-frame, an observer can
measure the pressure e.g.\ by enclosing some particles in a small
rubber balloon. At first order in perturbation, the balloon will
expand or contract isotropically, because anisotropies arise from
velocity gradients which can be shown to be second order in
perturbations. In contrast, in collisionless fluids, there are no bulk
motions: at a given point, there can be flows of particles in all
directions.  Inhomogeneities and anisotropies are described
perturbatively by a function $\Psi \ll 1$ such that
\begin{equation}
f(x^i,P_j,\tau) = f_0(q)
[1+\Psi(x^i,q_j,\tau)] \qquad {\rm with} \qquad P_j = (1-\Psi)q_j~,
\label{f_exp}
\end{equation}
and anisotropic pressure can now appear at first order in
perturbations. Indeed, the energy-momentum tensor can be computed from
the phase-space distribution $f$ and the 4-momentum $P_{\mu}$
\begin{equation}
\label{tmunu}
T_{\mu\nu}= \int dP_1 dP_2 dP_3\,(-g)^{-1/2}\,
{P_\mu P_\nu\over P^0} f(x^i,P_j,\tau) \,,
\end{equation}
where $(-g)^{-1/2} = a^{-4} (1 - \phi + 3 \psi)$. 
At first order in perturbations, this gives
\begin{eqnarray}
\label{tmunu2}
T^0_0(x^i) &=& a^{-4} \int q^2dq\,d\Omega\,
\epsilon \,f_0(q)\,[1+\Psi(x^i,q\hat{n}_j,\tau)] \,,\nonumber\\
T^0_i(x^i) &=& a^{-4} \int q^2dq\,d\Omega\,
q\,\hat{n}_i\,f_0(q)\,\Psi(x^i,q\hat{n}_j,\tau) \,,\\
T^i_j(x^i) &=& - a^{-4} \int q^2dqd\Omega
\,{q^2\over \epsilon}\,\hat{n}_i \hat{n}_j\,f_0(q)\,
[1+\Psi(x^i,q\hat{n}_j,\tau)]~,
\nonumber
\end{eqnarray}
where $d\Omega$ is the differential of the momentum direction
$\hat{n}_j = q_j / q$. It is then straightforward to obtain the
perturbed components of the energy-momentum tensor
\begin{eqnarray}
\label{deltanu}
\!\!\!\!\!\!\!\!   \delta\rho_{\nu} &=& a^{-4} \int q^2dq\,d\Omega\,\epsilon f_0 \Psi
	\,,~~~~~~~  \delta P_{\nu} = {1\over 3} a^{-4}
        \int q^2dq\,d\Omega\,{q^2\over \epsilon} f_0 \Psi \,,\\
\!\!\!\!\!\!\!\!      \delta T^0_{i\,\nu} &=& a^{-4}
	\int q^2dq\,d\Omega\,q \hat{n}_i\,f_0 \Psi \,,~~~
     \Sigma^i_{j\,\nu} = - a^{-4}
	\int q^2dq\,d\Omega\,{q^2\over\epsilon}
	(\hat{n}_i \hat{n}_j-{1\over 3}\delta_{ij})\,f_0 \Psi \,.
     \nonumber 
\label{thetanu}
\end{eqnarray}
%
%When the neutrinos are relativistic, $\epsilon=q$ and we immediately
%get that $\bar{\rho}_h=\bar{P}_h$ (i.e. $w=c_s^2=1/3$). Otherwise
%$\epsilon=am$ and $w=c_s^2=1/3$ decrease towards zero.

\subsubsection{Free-streaming}
\label{subsubsec:free-streaming}

We have seen that in perfect fluids, sound waves can propagate at the
sound speed on scales smaller than the sound horizon. Sound waves
cannot propagate in a collisionless fluid, but the individual
particles free-stream with a characteristic velocity --for neutrinos,
in average, the thermal velocity $v_{\rm th}$. So, it is possible to
define an horizon as the typical distance on which particles travel
between time $t_i$ and $t$. During MD and RD and for $t \gg t_i$, this
horizon is, as usual, asymptotically equal to $v_{\rm th}/H$, up to a
numerical factor of order one (see section \ref{sec:JeansLength}).
Exactly as we defined the Jeans length, we can define the {\it
free-streaming length} by taking Eq.\ (\ref{defJeans}) and replacing
$c_s$ by $v_{\rm th}$
\begin{equation}
\!\!\!\!\!\! k_{FS}(t) = \left(\frac{4 \pi G \bar{\rho}(t) 
a^2(t)}{v_{\rm th}^2(t)}\right)^{1/2},
\qquad
\lambda_{FS}(t) 
= 2 \pi \frac{a(t)}{k_{FS}(t)}
= 2 \pi \sqrt{2 \over 3} \frac{v_{\rm th}(t)}{H(t)}~.
\end{equation}
As long as neutrinos are relativistic, they travel at the speed of
light and their free-streaming length is simply equal to the Hubble
radius. When they become non-relativistic, their 
thermal velocity decays like
\begin{equation}
v_{\rm th}\equiv\frac{\langle p \rangle}{m}
\simeq\frac{3 T_{\nu}}{m}=\frac{3 T_{\nu}^0}{m}
\left( \frac{a_0}{a} \right) 
\simeq 150 (1+z) \left( \frac{1 \, \mathrm{eV}}{m} \right)
{\rm km}\,{\rm s}^{-1}~,
\end{equation}
where we used for the present neutrino temperature $T_{\nu}^0 \simeq
(4/11)^{1/3} T_{\gamma}^0$ and $T_{\gamma}^0 \simeq 2.726$ K. This
gives for the free-streaming wavelength and wavenumber during matter
or ${\Lambda}$ domination
\begin{eqnarray}
\lambda_{FS}(t) 
&=& 
7.7 
\frac{1+z}{\sqrt{\Omega_{\Lambda}+ \Omega_{m} (1+z)^3}}
\left( \frac{1 \, \mathrm{eV}}{m} \right) h^{-1}\mathrm{Mpc}~, \\
k_{FS}(t) &=& 0.82 
\frac{\sqrt{\Omega_{\Lambda}+ \Omega_{m} (1+z)^3}}{(1+z)^2}
\left( \frac{m}{1 \, \mathrm{eV}} \right) h\,\mathrm{Mpc}^{-1},
\end{eqnarray}
where $\Omega_{\Lambda}$ (resp.\ $\Omega_{m}$) is the cosmological
constant (resp.\ matter) density fraction evaluated today.  So, after
the non-relativistic transition and during matter domination, the
free-streaming length continues to increase, but only like
$(aH)^{-1}\propto t^{1/3}$, i.e.\ more slowly than the scale factor $a
\propto t^{2/3}$.  Therefore, the comoving free-streaming length
$\lambda_{FS} / a$ actually decreases like $(a^2 H)^{-1} \propto
t^{-1/3}$. As a consequence, for neutrinos becoming non-relativistic
during matter domination, the comoving free-streaming wavenumber
passes through a minimum $k_{\rm nr}$ at the time of the transition,
i.e.\ when $m = \langle p \rangle = 3 T_{\nu}$ and $a_0/a=(1+z)=
2.0\times 10^3 (m/ 1 \, \mathrm{eV})$. This minimum value is found to be
\begin{equation}
k_{\rm nr}
\simeq 0.018 \,\, \Omega_{\rm m}^{1/2} 
\left( \frac{m}{1 \, \mathrm{eV}} \right)^{1/2}
h\,\mathrm{Mpc}^{-1}~.
\end{equation}
The physical effect of free-streaming is to damp small-scale neutrino
density fluctuations: neutrinos cannot be confined into (or kept
outside of) regions smaller than the free-streaming length, for
obvious kinematic reasons.  We will see that the metric perturbations
are also damped on those scales by gravitational back-reaction. On
scales much larger than the free-streaming scale, the neutrino
velocity can be effectively considered as vanishing, and after the
non-relativistic transition the neutrino perturbations behave like
cold dark matter perturbations. In particular, modes with $k<k_{\rm
nr}$ are never affected by free-streaming and evolve like in a pure
$\Lambda$CDM model.

\subsubsection{Vlasov equation}

The neutrino equation of evolution is the collisionless Boltzmann
equation, or Vlasov equation, which just expresses the conservation of
the number of particles along the phase-space trajectories
\begin{equation}
{Df \over d\tau} = {\partial f \over \partial \tau}
+ {dx^i \over d\tau}{\partial f\over \partial x^i}
+ {dq \over d\tau}{\partial f\over \partial q}
+ {dn_i \over d\tau}{\partial f\over \partial n_i}
= 0\,.
\end{equation}
We can expand $f$ like in Eq.~(\ref{f_exp}) and keep only first-order
perturbations. Using the fact that $dx^i /
d\tau=P^i/P^0=-P_i/P_0=-q_i/\epsilon$ at order zero, that $dq / d\tau=
q \, \dot{\psi} + \epsilon \, \hat{n_i} \, \partial_i \phi$ at order
one (from the geodesic equation $P^0 \dot{P}^0 + \Gamma^0_{\mu \nu}
P^{\mu} P^{\nu}=0$), and that $dn_i / d\tau=0$ at order zero (in
absence of metric perturbations, the momenta keep a fixed
orientation), we obtain
\begin{equation}
f_0 \dot{\Psi}
- {q_i \over \epsilon} \, f_0 \, \partial_i \Psi
+ (q \, \dot{\psi} + \epsilon \, \hat{n_i} \, \partial_i \phi) 
  {\partial f_0 \over \partial q}
= 0\,,
\end{equation}
or in Fourier space and after dividing by $f_0$
\begin{equation}
\dot{\Psi}
- i \, {q \over \epsilon} \, (\vec{k}\cdot\hat{n}) \, \Psi
= - \left( 
\dot{\psi} + i \, {\epsilon \over q} \, (\vec{k}\cdot\hat{n}) 
\, \phi \right) 
{\partial \ln f_0 \over \partial \ln q} \,.
\end{equation}
This equation gives the response of the neutrino phase-space
distribution to metric perturbations.

\subsubsection{Neutrino perturbations during the relativistic regime}
\label{subsec:r}

When neutrinos are ultra-relativistic ($\epsilon = q$), the Vlasov
equation reduces to
\begin{equation}
  \dot{\Psi}
  - i \, (\vec{k}\cdot\hat{n}) \, \Psi
  = - \left( 
  \dot{\psi} + i \, (\vec{k}\cdot\hat{n}) 
  \, \phi \right) 
     {\partial \ln f_0 \over \partial \ln q} \,.
\label{vlasov_rel}
\end{equation}
The homogeneous equation does not involve $q$, while the source term has
a definite $q$-dependence, fixed once and for all by the shape of the
homogeneous distribution. This simply reflects the fact that neutrinos
at a given point and in a given direction are all redshifted or blueshifted
by the same amount, so that their spectrum always remain that of a blackbody,
with a temperature shift $\delta T_{\nu}(x^i,\hat{n}_j,\tau)$
at a given point and in a given direction
\begin{eqnarray}
f(x^i,q,\hat{n}_j,\tau)
&=&\left(\exp{\left[\frac{q}{a (T_{\nu} + \delta T_{\nu})}\right]}
+1\right)^{-1}
\nonumber \\
&=&
f_0(q) \, + \, \frac{\partial f_0}{\partial T_{\nu}}(q)  \, \, \, 
\delta T_{\nu}(x^i,\hat{n}_j,\tau)~.
\label{perturbed_but_thermal_1}
\end{eqnarray}
Since $f_0$ is a function of $q/aT_{\nu}$, one has $\partial
f_0/\partial T_{\nu} = - (q/T_{\nu}) \, \partial f_0/\partial q$, and
\begin{equation}
f(x^i,q,\hat{n}_j,\tau) 
= f_0(q) \, - \, \frac{\partial f_0}{\partial \ln q}(q)  \,\,\, 
\frac{\delta T_{\nu}}{T_{\nu}}(x^i,\hat{n}_j,\tau)~.
\label{perturbed_but_thermal_2}
\end{equation}
Identifying this result with the definition of the phase-space
perturbation $\Psi$, we get
\begin{equation}
\Psi(x^i,q,\hat{n}_j,\tau) = - \frac{\partial \ln f_0}{\partial \ln q}(q)  
\,\,\, 
\frac{\delta T_{\nu}}{T_{\nu}}(x^i,\hat{n}_j,\tau)~.
\end{equation}
We now see explicitly that the presence of the function
$\frac{\partial \ln f_0}{\partial \ln q}$ in the Vlasov equation means
precisely that the spectrum remains Planckian.  Since for relativistic
neutrinos the local $q$-dependence is trivial, we only need to compute
the evolution of the {\it mean momentum}, which is a function of
position, direction and time. One can define the quantity
$F_{\nu}(x^i,\hat{n}_j,\tau)$ as
\begin{equation}
\label{deffnu}
F_{\nu}(x^i,\hat{n}_j,\tau) 
\equiv {\int q^2 dq\,q f_0(q)\,[1+ \Psi(x^i,\hat{n}_j,\tau)]
\over \int q^2 dq\,q f_0(q)} - 1 ~,
\end{equation}
for which the Vlasov equation reads in Fourier space (after integrating by
part)
\begin{equation}
  \dot{F}_{\nu}
  - i \, (\vec{k}\cdot\hat{n}) \, F_{\nu}
  = 4 \left( 
  \dot{\psi} + i \, (\vec{k}\cdot\hat{n}) 
  \, \phi \right)~. 
\end{equation}
We can further reduce the dimensionality of the problem by noticing
that the evolution depends only on the direction $\hat{n}$ through
the angle $\vec{k}\cdot\hat{n}$, which is a natural consequence of the
isotropy of the homogeneous background. We perform a Legendre expansion
with respect to this angle,
\begin{equation}
\label{deffnul}
F_\nu(\vec{k},\hat{n},\tau) =
\sum_{l=0}^\infty(-i)^l
(2l+1)F_{\nu\,l}(\vec{k},\tau)P_l(\hat{k}\cdot\hat{n})\,.
\end{equation}
By comparing Eqs.\ (\ref{deltanu}) and (\ref{thetanu}) (with
$\epsilon=q$) with the above definitions, it is straightforward to
show that the density contrast (as well as $\delta p_{\nu}=
\frac{1}{3} \delta \rho_{\nu}$) is given by the monopole:
$\delta_{\nu}=F_{\nu\,0}$, the velocity gradient by the dipole:
$\theta_{\nu} = \frac{3}{4} k F_{\nu\,1}$, and the shear (or
anisotropic stress) by the quadrupole: $\sigma_{\nu} = - \frac{1}{2}
F_{\nu\,2}$.  The Vlasov equation now reduces to its most
``concentrated'' form: an infinite hierarchy of coupled equation for
the multipoles $F_{\nu\,l}(\vec{k},\tau)$
\begin{eqnarray}
\label{vlasov_legendre}
\dot{\delta}_{\nu} &=& {4\over 3} \theta_\nu
+ 4\dot{\psi} \,, \nonumber\\
\dot{\theta}_\nu &=& - \frac{k^2}{4}\delta_\nu
- k^2 \sigma_\nu
- k^2\phi \,, \nonumber\\
\dot{F}_{\nu\,l} &=& {k\over 2\, l+1}\left[
  (l+1) F_{\nu\,(l+1)} 
  - l \, F_{\nu\,(l-1)} \right]\,, \quad l \geq 2 \,.
\end{eqnarray}
Note that the first two equations consistently reproduce the
continuity and Euler equations (\ref{continuity})-(\ref{Euler}).

After introducing all this formalism, it is possible to understand the
physical evolution of relativistic neutrino perturbations. We assume
isentropic (adiabatic) initial conditions during radiation domination
and on super-Hubble scales: so, we start from
$\delta_{\nu}=\delta_{\gamma}$ and from time-independent metric
fluctuations. On super-Hubble scales, the neutrino distribution is
static, with a local density (and temperature) perturbation
proportional to the metric fluctuation. An observer would see locally
a quasi-isotropic temperature distribution: the dipole is suppressed
by a factor $k/aH=k\tau$, and the $l$-th multipole by $(k\tau)^l$. At
leading order in $(k\tau)$, the Einstein equations (\ref{Einstein1})
and (\ref{Einstein2}) immediately give
\begin{equation}
\delta_{\nu} = \delta_{\gamma} = - 2 \phi = 4 {\delta T_{\nu} \over T_{\nu}}~,
\qquad \theta_{\nu} = \theta_{\gamma} = - {1\over2} k^2 \tau \phi~,
\end{equation}
while the expressions of $\psi$ and $\sigma_{\nu}$ require more work
(actually, the usual way to obtain them is to solve the Einstein
equations in the long wavelength limit in the synchronous gauge, and
then to convert the result into longitudinal gauge variables). The
result is found to be (see Ref.\ \cite{Ma:1995ey})
\begin{equation}
\psi = \left(1 + {2\over5} R_{\nu} \right)  \phi~, \quad
\sigma_{\nu} = - {1\over15} (k \tau)^2 \phi~, \quad
\mathrm{with}~~R_{\nu} \equiv 
{\bar{\rho}_{\nu} \over \bar{\rho}_{\nu}+ \bar{\rho}_{\gamma}}~.
\end{equation}
For three standard non-relativistic neutrinos, on has $R_{\nu}\simeq
0.405$ and $\psi \simeq 1.162 \, \phi$.  After Hubble crossing,
i.e.\ on scales smaller than the free-streaming length, the various
multipoles are populated one after each other, at the expense of the
low multipoles which get reduced. Physically, this corresponds to the
fact that an observer would see locally the superposition of a growing
number of neutrino flows, coming from all directions. Since the flows
come from a random distribution of over- and under-densities, the
local density contrast tends to average to zero.
\begin{figure}[t]
\begin{center}
\vspace{-1cm}
%\hspace{-1.cm}
\includegraphics[width=\textwidth]{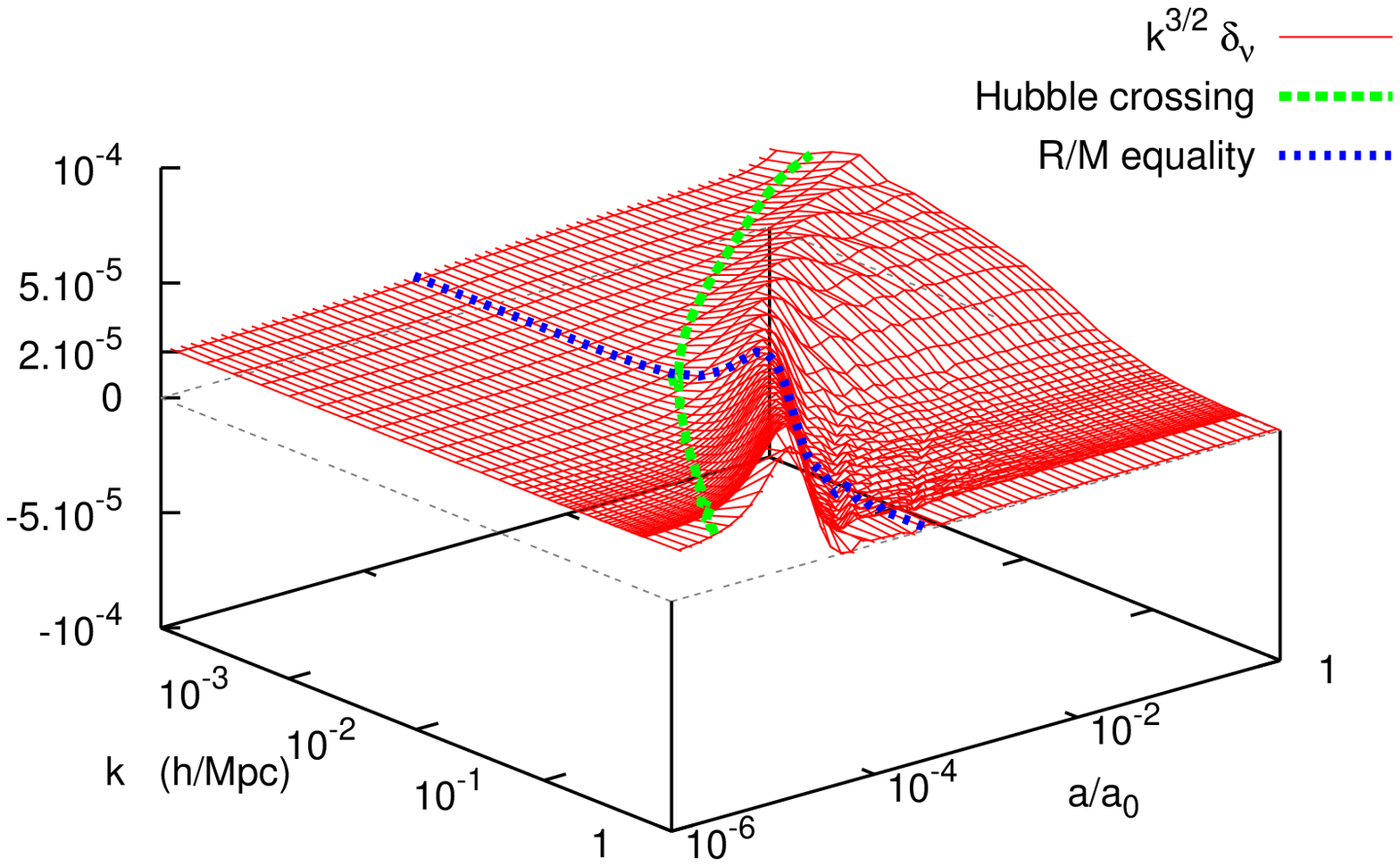}\\
\vspace{-2cm}
\includegraphics[width=\textwidth]{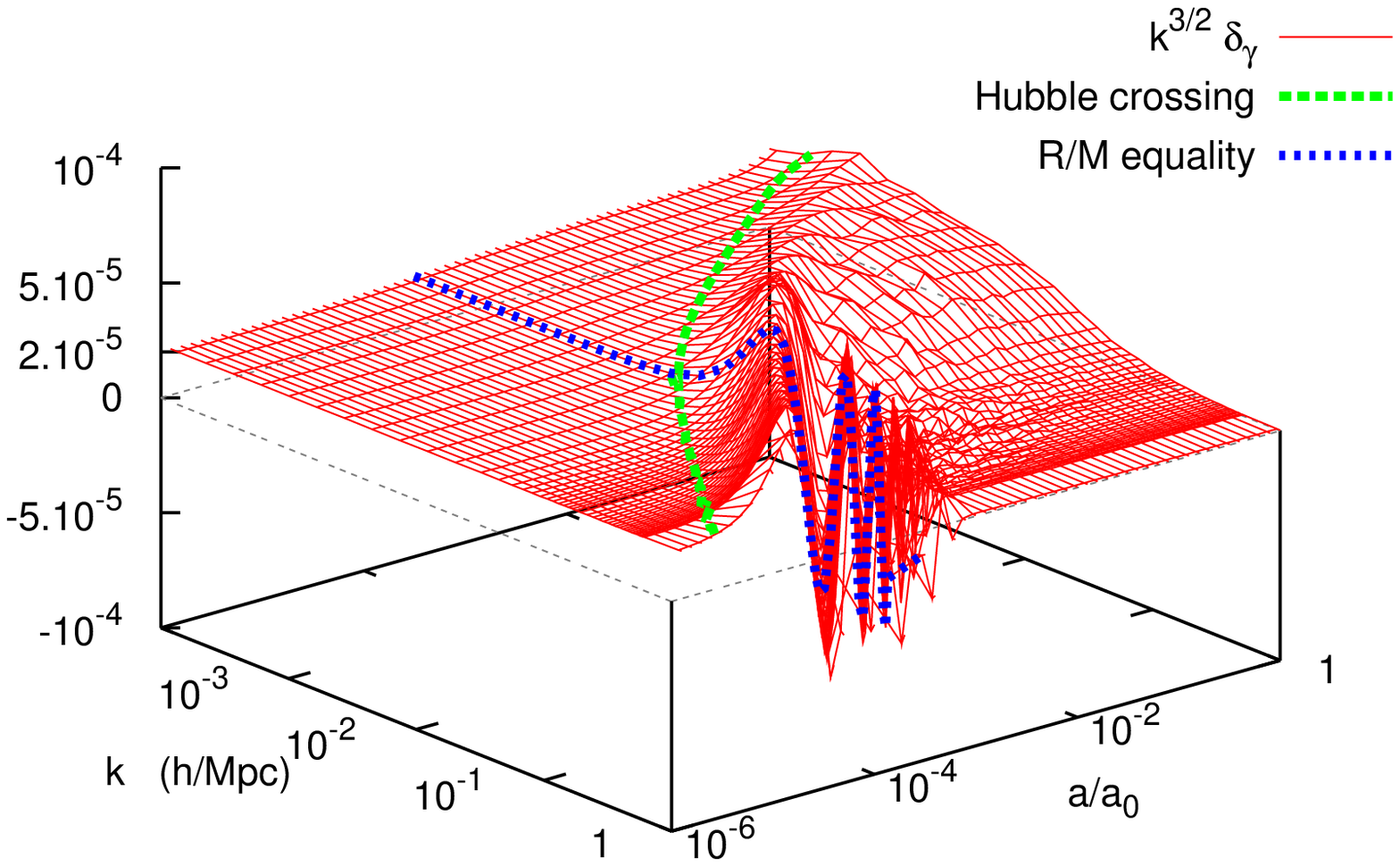}\\
\vspace{-1cm}
\caption{\label{surf_massless} Evolution of (longitudinal gauge)
massless neutrino density perturbations $\delta_{\nu}=\delta
\rho_{\nu}/\bar{\rho}_{\nu}$ (top), compared with those of the photon
density $\delta_{\gamma}=\delta \rho_{\gamma}/\bar{\rho}_{\gamma}$
(bottom), in a $\Lambda$CDM model with three massless neutrinos, as a
function of the scale factor $a$ and Fourier wavenumber $k$, obtained
numerically with a Boltzmann code. Time is evolving from left ($a/a_0
= 10^{-6}$) to right ($a/a_0=1$), and the initial condition is set
arbitrarily to $k^{3/2} \phi = -10^{-5}$.  Unlike the photon density
contrast, which oscillates between Hubble crossing and equality, the
neutrino density contrast is immediately damped after the first
half-oscillation. After equality, $\delta_{\nu}$ and $\delta_{\gamma}$
remain constant: they are preserved from gravitational collapse by
free-streaming and relativistic pressure.  }
\end{center}
\end{figure}
\begin{figure}[t]
\begin{center}
\includegraphics[width=\textwidth]{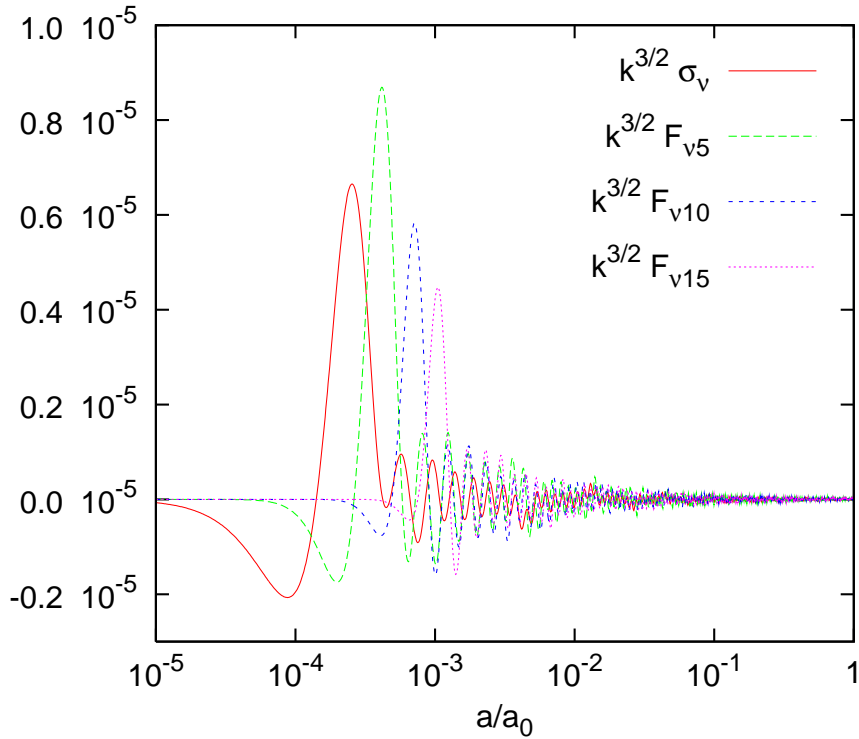}\\
\caption{\label{massless_multipoles} Evolution of some of the first
neutrino multipoles $F_{{\nu} l}$ as a function of the scale factor,
for the mode $k=0.1 \,h\,$Mpc$^{-1}$ which enters into the Hubble
radius close to the end of radiation domination, when
$a/a_0=2.5\times10^{-5}$. We imposed an initial condition
$k^{3/2}\phi=-10^{-5}$. We see that after Hubble crossing, the
multipoles are populated one after each other, reflecting the fact
that an observer would see locally a superposition of a growing number
of neutrino flows, coming from all directions.  }
\end{center}
\end{figure}

It is useful for pedagogical purpose to compare the evolution equation
of neutrinos with that of photons before recombination.  For this
purpose, we combine the continuity and Euler equations into a
second-order equation
\begin{equation}
\ddot{\delta}_{\nu} = - \frac{1}{3} k^2 \delta_{\nu}
- \frac{4}{3} k^2 \phi + 4 \ddot{\psi} - \frac{4}{3} k^2 \sigma_{\nu}~.
\end{equation}
In absence of shear, this equation would be similar to the photon
equation of evolution before recombination and in the limit of small
baryon-to-photon ratio $R_b=\bar{\rho}_b/\bar{\rho}_{\gamma}$. The
terms are easy to identify: $-\frac{1}{3} k^2 \delta_{\nu}$ is the
force arising from relativistic pressure, $- \frac{4}{3} k^2 \phi$ the
gravitational force, and $4 \ddot{\psi}$ the dilation effect: a local
increase of $\psi$ is equivalent to a local decrease of the scale
factor and of the photon/neutrino wavelength, and hence to an increase
of the blackbody temperature and density. In the case of photons, in
the tight-coupling limit, the microscopic interactions guarantee that
$\sigma_{\gamma}$ vanishes (even if $R_b \ll 1$), and the equivalent
of the above equation is that of an harmonic oscillator driven by
metric fluctuations.  In the case of neutrinos, the shear becomes
important inside the Hubble length, and couples the above equation to
the whole multipole hierarchy; qualitatively, the shear acts like
viscosity in a fluid; instead of experiencing driven oscillations, the
neutrino overdensity $\delta_{\nu}$ is damped, while the power is
transfered to higher multipoles.  This behavior is illustrated in
Figs.\ \ref{surf_massless} and \ref{massless_multipoles} which derive
from exact numerical simulations (using the code {\sc cmbfast}
\cite{Seljak:1996is}). In particular, one can see explicitly on
Fig.\ \ref{surf_massless} that in the region where the photon density
contrast experiences acoustic oscillations, the neutrino density
contrasts is damped.

It is possible to gain even more intuition by going back to the
relativistic Vlasov equation (\ref{vlasov_rel}), for which the full
solution can be decomposed into those of the homogeneous and
inhomogeneous differential equations. The general solution of the
homogeneous equation is actually trivial
\begin{equation}
\Psi (\vec{k},\hat{n}, q, \tau) = - \, \frac{\delta T_{\nu}}{T_{\nu}}
(\vec{k},\hat{n}) \,\,\, e^{i(\vec{k}\cdot\hat{n})\tau} \,\,\,
\frac{\partial \ln f_0}{\partial \ln q}~.
\end{equation}
It corresponds to the free propagation of plane waves of temperature
perturbations, i.e.\ to free-streaming. In the $F_{\nu\,l}$ basis,
this yields $F_{\nu\,l}(\vec{k},\tau) \propto j_l(k \tau)$, where the
$j_l$'s are the spherical Bessel functions of the first kind, and the
contribution to $\delta_{\nu}$ is equal to $[\sin(k\tau)/(k \tau)]$
times a constant of integration. As expected, this contribution tends
to zero after Hubble crossing (because of the averaging between
various flows), while higher multipoles are populated one after each
other, with a peak around $\tau \sim l/k$ (the numerical solution of
Fig.\ \ref{massless_multipoles} looks slightly more complicated than
plain Bessel functions: this is due to the small time-variation of
$\phi$ and $\psi$ caused by neutrino shear). On top of these terms,
the full solution receives a contribution from the solution of the
inhomogeneous equation.  During radiation domination, $\psi$ and
$\phi$ decay, and this solution is complicated. However, it is clear
that it should also decay.  Instead, during matter domination, $\psi$
and $\phi$ are frozen, and the solution is obvious: $\Psi= \phi \,
\frac{\partial \ln f_0}{\partial \ln q}$.  In the $F_{\nu\,l}$ basis,
this term only contributes to $F_{\nu 0}=\delta_{\nu}$, for which the
full solution reads
\begin{equation}
\delta_{\nu} = -4 \phi(k) + \alpha(k) \sin(k\tau)/(k \tau)~, \qquad \qquad
\tau > \tau_{\rm eq}~,
\end{equation}
where the constant of integration $\alpha(k)$ could be found my
matching with the radiation-dominated solution.  The first term
corresponds to the stationary equilibrium configuration for which, in
any potential well, the relativistic pressure compensates exactly the
gravitational force. This behavior can be clearly seen on Fig.\
\ref{surf_massless} (note that after decoupling, the photons
free-stream like the relativistic neutrinos, so $\delta_{\gamma}$ also
tends towards $-4 \, \phi$).

In conclusion, during the relativistic regime, the combined effects of
free-streaming and relativistic pressure result in a suppression of
the neutrino density contrast inside the Hubble radius: during
radiation domination, $\delta_{\nu}$ is driven to zero instead of
oscillating like $\delta_{\gamma}$ and $\delta_{\rm b}$, and during
matter domination, it remains constant instead of growing like
$\delta_{\rm b}$ and $\delta_{\rm cdm}$.

\subsubsection{Neutrino perturbations during the non-relativistic regime}
\label{subsec:nr}

\begin{figure}[t]
\begin{center}
\includegraphics[width=\textwidth]{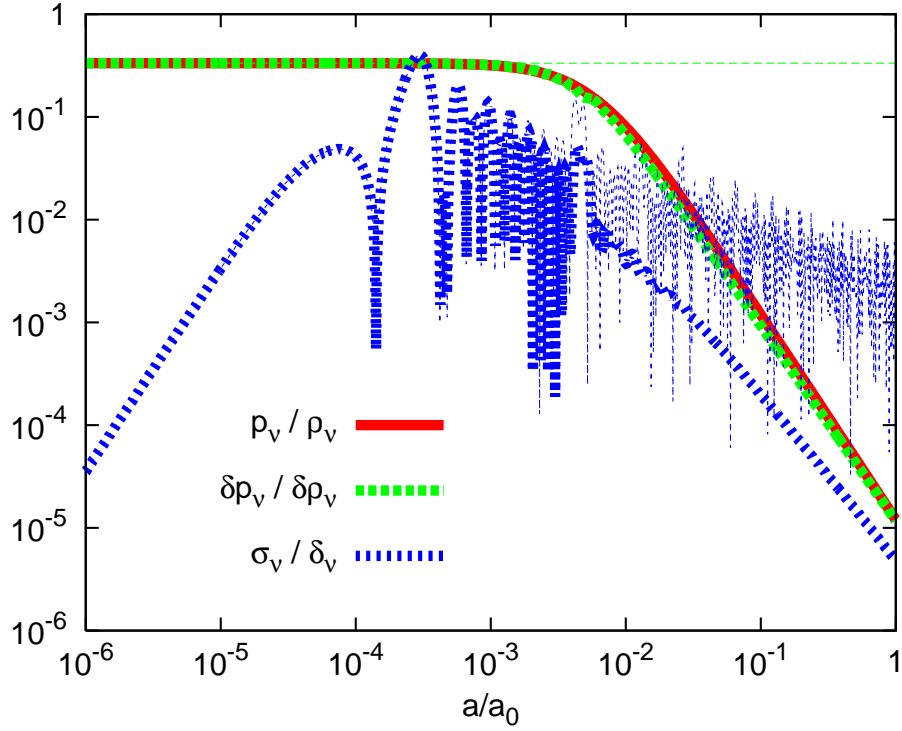}\\
\caption{\label{fig_tmunu} Evolution of the neutrino isotropic
pressure perturbation $\delta p_{\nu}$ and anisotropic stress
$\sigma_{\nu}$ in units of the density perturbation (for the mode
$k=0.1 \,h\,$Mpc$^{-1}$ which enters into the Hubble radius when
$a/a_0=2.5\times10^{-5}$) compared to the evolution of the background
equation of state parameter $\bar{p}_{\nu}/\bar{\rho}_{\nu}$. The thin
curves show the case of massless neutrinos for which
$\bar{p}_{\nu}/\bar{\rho}_{\nu}=\delta p_{\nu}/\delta \rho_{\nu}=1/3$
and $\sigma_{\nu}/\delta_{\nu}$ decays slowly after Hubble crossing.
The thick lines correspond to neutrinos with $m_\nu=0.1$ eV, which
become non-relativistic around $a/a_0=5\times10^{-3}$. After the
transition, pressure and shear perturbations become negligible with
respect to density perturbations, just like for cold dark matter.  }
\end{center}
\vspace{0.5cm}
\end{figure}

After the non-relativistic transition, the $q$-dependence of the phase-space
distribution becomes non-trivial. The geodesic equation
\begin{equation}
\frac{dq}{d\tau} = q \dot{\psi} + (q^2 + a^2 m^2)^{1/2} \hat{n}_i \,
\partial_i \phi
\end{equation}
shows that the amount of blueshift or redshift $dq/q$ experienced by a
neutrino when it travels over metric fluctuations depends on how
non-relativistic it is.  In a given point and direction $(x^i,
\hat{n}_j)$, there is a superposition of neutrinos of different
momenta which are more or less deep inside the non-relativistic
regime. Lowest and highest energy neutrinos experience different
redshifts, and the thermal spectrum of Eqs.\
(\ref{perturbed_but_thermal_1}) and (\ref{perturbed_but_thermal_2}) is
inevitably distorted at first order in perturbation (but not at zero
order, since the distortion is induced by metric perturbations and not
by the background geometry).
Therefore, it is impossible to eliminate the $q$-dependence from the
problem, as we did in the relativistic regime.  However, it is still
possible to get rid partially of the direction dependence by expanding
$\Psi$ in multipole space
\begin{equation}
\Psi(\vec{k},q,\hat{n},\tau)
= \sum_{l=0}^\infty (-i)^l(2l+1) \Psi_l(\vec{k},q,\tau)
P_l(\hat{k}\cdot\hat{n})\,.
\end{equation}
Then, the Vlasov equation becomes
\begin{eqnarray}
\dot{\Psi}_0 &=& {q\,k\over\epsilon} \, \Psi_1
 -\dot{\psi} \, {\partial \ln f_0\over \partial \ln q} \,, \nonumber\\
\dot{\Psi}_1 &=& {q\,k\over 3\,\epsilon} \left(\Psi_2
 - 2 \Psi_0 \right)
  + {\epsilon\,k\over 3\,q} \, \phi \, 
  {\partial \ln f_0\over \partial \ln q} \,,\\
  \dot{\Psi}_l &=& {q\,k \over (2l+1)\,\epsilon} \left[
  (l+1)\Psi_{l+1} - l\Psi_{l-1}\right]\,,
\quad l \geq 2 \,. \nonumber
\end{eqnarray}
The perturbed energy density, pressure, energy flux divergence
and shear stress in $k$-space are given by
\begin{eqnarray}
 \delta\rho_{\nu} &=& 4\pi a^{-4}
   \int q^2 dq\,\epsilon f_0(q) \Psi_0 \,, \nonumber\\
 \delta p_{\nu} &=& {4\pi \over 3} a^{-4}
   \int q^2 dq\,{q^2\over \epsilon} f_0(q) \Psi_0
\,, \nonumber\\
  (\bar{\rho}_{\nu} +\bar{p}_{\nu}) \theta_{\nu} &=& 4\pi k a^{-4}
\int q^2 dq\,q f_0(q)\Psi_1 \,, \nonumber\\
 (\bar{\rho}_{\nu} +\bar{p}_{\nu}) \sigma_{\nu} &=& - {8\pi\over 3} a^{-4}
\int q^2 dq\,{q^2\over\epsilon} f_0(q) \Psi_2 \,.
\end{eqnarray}
When for a given family the neutrinos are well inside the
non-relativistic regime, i.e.\ when most of the momenta for which
$f_0(q)$ is non-negligible satisfy $q \ll \epsilon \sim a\,m$, the
above equations show that both $\delta p_{\nu}$ and $(\bar{\rho}_{\nu}
+\bar{p}_{\nu}) \sigma_{\nu}$ are suppressed with respect to
$\delta\rho_{\nu}$ (as can be checked in Fig.\ (\ref{fig_tmunu})).
As expected, the continuity and Euler equations (\ref{continuity}) and
(\ref{Euler}) become gradually identical to those of ordinary
non-relativistic matter
\begin{eqnarray}
\dot{\delta}_{\nu} &=& \theta_{\nu} + 3 \dot{\psi}~, \\
\dot{\theta}_{\nu} &=& -\frac{\dot{a}}{a} \theta_{\nu} - k^2 \phi~.
\end{eqnarray}
Like for CDM (and baryons after recombination), this system leads to
the second-order equation of evolution
\begin{equation}
\ddot{\delta}_{\nu} + \frac{\dot{a}}{a} \dot{\delta}_{\nu}
= - k^2 \phi + 3 (\ddot{\psi} + \frac{\dot{a}}{a} \dot{\psi})~.
\label{evol_nu_nonrel}
\end{equation}
We are interested in neutrinos becoming non-relativistic during matter
domination, when $\phi$ and $\psi$ are approximately constant and $a
\propto \tau^2$. In this case, the solution of (\ref{evol_nu_nonrel})
reads
\begin{eqnarray}
\delta_{\nu} &=& A \, \ln \tau + B - \frac{(k \tau)^2}{6} \, \phi \\
&=& \tilde{A} \, \ln a + \tilde{B} - \frac{2}{3} 
\left(\frac{k}{aH}\right)^2 \phi~,
\label{sol_nu_nonrel}
\end{eqnarray}
where $A$, $B$ (or $\tilde{A}$, $\tilde{B}$) are some constants 
of integration. The last term (the solution of the inhomogeneous
equation) grows like $\tau^2$, {\it i.e.} like the scale factor.
This solution has a simple interpretation:
\begin{itemize}
\item
first, for $k>k_{\rm nr}$, i.e.\ for modes which are inside the Hubble
radius at the time of the non-relativistic transition, the density
contrast $\delta_{\nu}$ is much smaller than $\delta_{\rm cdm}$ at
that time (as a consequence of free streaming), and then grows faster
than $\delta_{\rm cdm}\propto a$ because of the term $(\tilde{A} \ln
a)$. At some point the first and third terms in
Eq.\ (\ref{sol_nu_nonrel}) will be equal, and then asymptotically
$\delta_{\nu}$ will be dominated by
\begin{equation}
\delta_{\nu} \longrightarrow - \frac{2}{3} 
\left(\frac{k}{aH}\right)^2 \phi \, \, \propto a~,
\label{asymp_delta_nu}
\end{equation}
which corresponds exactly to the solution of the Poisson equation in a
matter-dominated Universe (see Eq.\ (\ref{poisson_LD}) with
$\Omega_{\rm m}$=1). Indeed, this asymptotic value is simply the one
imposed to all non-relativistic species by Newtonian gravitation, and
after an infinite amount of time $\delta_{\nu}$ would be equal to
$\delta_{\rm cdm}$ on all scales.  The same is also true for baryons
after photon decoupling, but since baryons did not experience
free-streaming, the equality between $\delta_{\rm b}$ and $\delta_{\rm
cdm}$ is reached quickly after $\tau=\tau_{\rm dec}$. Instead, for
neutrinos with masses of order of 1 eV or smaller, the asymptote is
far from being reached today, except for scales very close to
$k=k_{\rm nr}$. On smaller scales, $ \delta_{\nu}$ remains much
smaller than $\delta_{\rm cdm} \simeq \delta_{\rm b}$.\\

\item
second, for $k<k_{\rm nr}$, {\it i.e.} for modes which are outside the
Hubble radius at the time of the non-relativistic transition, the
density contrast $\delta_{\nu}$ is time-independent and of the same
order as ($\phi$, $\delta_{\rm cdm}$) on super-Hubble scales.  After
horizon crossing, the solution is dominated again by
(\ref{asymp_delta_nu}) and becomes rapidly equal to $\delta_{\rm
cdm}$.
\end{itemize}
The conclusion of this section is that if we could picture the three
density contrasts at a fixed time near the end of matter domination,
we would expect $\delta_{\nu} = \delta_{\rm cdm} =\delta_{\rm b}$ for
$a_0 H_0 < k < k_{\rm nr}$, while for $k> k_{\rm nr}$ the ratio
$\delta_{\nu} / \delta_{\rm cdm} = \delta_{\nu} / \delta_{\rm b}$
would be smaller than one, and tending towards zero as $k$ goes to
infinity.  The recent stage of $\Lambda$ domination does not change
this conclusion: for $a_0 H_0 < k < k_{\rm nr}$, the three density
contrasts are damped by the same amount, and on smaller scales the
ratio $\delta_{\nu} / \delta_{\rm cdm}$ does not change
significantly. All these features can be checked in the numerical
simulations presented in Sec.\ \ref{num_res_for_nu} (see Fig.\
\ref{fig_deltas}).  Therefore, the matter power spectrum today (or at
any time after the non-relativistic transition of the heaviest
neutrinos) is given by
\begin{eqnarray}
P(k) &=& \left\langle 
\left(\frac{\delta \rho_{\rm cdm} + \delta \rho_{\rm b} + \delta \rho_{\nu}}
{\rho_{\rm cdm} + \rho_{\rm b} + \rho_{\nu}}\right)^2 \right\rangle \nonumber\\
&=& \left\langle \left( 
\frac{\Omega_{\rm cdm} \, \delta_{\rm cdm} 
+ \Omega_{\rm b} \, \delta_{\rm b} 
+ \Omega_{\nu} \, \delta_{\nu}}{\Omega_{\rm cdm} + \Omega_{\rm b} + \Omega_{\nu}} 
\right)^2
\right\rangle \nonumber\\
&=&
\left\{
\mathrm{
\begin{tabular}{lcr}
$\langle \delta_{\rm cdm}^2 \rangle$ & ~~~~~~~~for & $k < k_{\rm nr}$,
\\
$
[1-\Omega_{\nu}/\Omega_{\rm m}]^2 \,
\langle \delta_{\rm cdm}^2 \rangle$ & ~~~~~~~~for & $k \gg k_{\rm nr}$,
\end{tabular}
}
\right.
\label{Pk_generic_LMDM}
\end{eqnarray}
with $\Omega_{\rm m} \equiv \Omega_{\rm cdm} + \Omega_{\rm b} +
\Omega_{\nu}$.  If the neutrinos did not induce a gravitational
backreaction effect on the evolution of the metric and of other
perturbations, then $\langle \delta_{\rm cdm}^2 \rangle$ would be the
same with and without neutrinos and the role of the neutrino masses
would be simply to cut the power spectrum by a factor
$[1-\Omega_{\nu}/\Omega_{\rm m}]^2$ for $k\gg k_{\rm nr}$. However, we
will see in the next subsection that the presence of neutrinos
actually modifies the evolution of the CDM and baryon density
contrasts in such way that the suppression factor is greatly enhanced,
more or less by a factor four.

\subsubsection{Modified evolution of matter perturbations}
\label{subsubsec:mod_ev}

{\bf Neutrino backreaction effects: generalities.}\\ In principle, the
impact of neutrinos on the evolution of other perturbations (baryons,
cold dark matter, photons) can be divided in two categories:
\begin{enumerate}
\item
effects from the homogeneous density and pressure of neutrinos, which
affect the Friedmann law. Since the equations of evolution of all
perturbations contain the expansion rate, the solutions can be
affected.\\

\item
direct gravitational back-reaction effect at the level of
perturbations. The neutrino density, pressure and velocity
perturbations provide a significant contribution to the total $\delta
T_{\mu}^{\nu}$, essentially on scales larger than the free-streaming
scale. So, they can change the evolution of metric perturbations, in a
similar way as an increased photon density. In addition, the neutrino
shear induces a non-zero difference $(\phi-\psi)$, which is constant
on super-horizon scale during radiation domination and becomes
progressively negligible on all scales during matter domination.
\end{enumerate}
Some of these effects are independent of the neutrino mass, and their
combination results in a very significant difference between the CMB
and LSS power spectra in absence of neutrinos, and those in presence
of three families of ultra-relativistic neutrinos (see Fig.\
\ref{fig_powspec_LMDM}).  These effects are investigated in details in
\cite{Bashinsky:2003tk}, and are extremely relevant from the
observational point of view. For instance, they give an opportunity to
measure the effective neutrino number $N_{\rm eff}$ (see Sec.\
\ref{subsec:neff}). Also, it was shown in Ref.\ \cite{Trotta:2004ty}
that the specific gravitational backreaction effect induced by
neutrino shear is actually confirmed by current observations of the
CMB temperature spectrum by WMAP.

However, in this review, we are more interested in the effect of the
neutrino mass, in realistic scenarios for which the non-relativistic
transition of at least two species takes place during matter
domination.  Our goal is to understand how the matter power spectrum
of Eq.\ (\ref{Pk_generic_LMDM}) depends on the neutrino mass in the two
limits $k<k_{\rm nr}$ and $k>k_{\rm nr}$. Therefore, we should study
how the evolution of $\delta_{\rm cdm}=\delta_{\rm b}$ is affected by
non-relativistic neutrinos deep in the matter-dominated regime.

{\bf How neutrinos slow down the growth of matter perturbations.}\\ We
are arriving to the most important point of this section, since we are
going to derive the dominant observable effect of neutrino masses.
Before getting into the equations, let us try to describe the effect
qualitatively. For this purpose, we first need to come back to the
issue of linear clustering of CDM and baryons during matter domination
{\it in absence of neutrinos}.  In Sec.\ \ref{subsec:pert_no_nu},
we stated that $\phi=\psi$ remains constant during matter domination,
basing ourselves on Eq.\ (\ref{sys_phi}) with a zero sound speed
(or equivalently on the perturbed Einstein Eqs.\ (\ref{Einstein3}) and
(\ref{Einstein4}) with zero pressure perturbation and shear on the
right-hand side). Then, we used the Poisson equation
\begin{equation}
-\frac{k^2}{a^2} \psi = 4 \pi G \delta \rho~,
\end{equation}
to show that inside the Hubble radius, the density contrast
$\delta_{\rm cdm}=\delta_{\rm b}$ grows like $[a^2 (\bar{\rho}_{\rm
cdm}+\bar{\rho}_{\rm b})]^{-1}$, i.e.\ like a linear function of
the scale factor. The fact that linear structure formation corresponds
to a constant gravitational potential sounds counter-intuitive at
first sight. Actually, it results from an exact compensation between
clustering and the stretching of spacetime.  During gravitational
collapse, the non-relativistic matter density perturbations $\delta
\rho$ do not grow with time, as one would naively expect from a common
intuition based on a static Universe. Instead, $\delta \rho$
decreases, but only like $a^{-2}$ and not as fast as the dilution
factor $a^{-3}$, so that the density contrasts $\delta
\rho/\bar{\rho}$ actually grows like $a$. The behavior $\delta \rho
\propto a^{-2}$ corresponds precisely to a static potential, since the
gradient $\Delta \psi = - (k/a)^2 \psi$ also decreases like $a^{-2}$
due to the stretching of space-time.

Let us restore now the neutrino background.  During matter domination
and on scales smaller than the free-streaming scale, the neutrino
perturbations $\delta_{\nu}$ do not contribute to gravitational
clustering. Indeed, we have seen in Secs.\ \ref{subsec:r} and
\ref{subsec:nr} that free-streaming leads to $\delta_{\nu} \ll
\delta_{\rm cdm}$, and since $\bar{\rho}_{\nu} < \bar{\rho}_{\rm cdm}$
we see that $\delta \bar{\rho}_{\nu} \ll \delta \bar{\rho}_{\rm cdm}$:
so, neutrinos can be simply omitted from the Poisson equation. On the
other hand, they do contribute to the homogeneous expansion through
Friedmann equation.  Therefore the exact compensation between
clustering and expansion described in the previous paragraph is
slightly shifted: the balance is displaced in favour of the expansion
effect, and we expect $\phi=\psi$ to decay slowly, while $\delta_{\rm
cdm}=\delta_{\rm b}$ should grow not as fast as the scale factor.
This mechanism, first described and quantified in Ref.\
\cite{Bond:1980ha}, belongs to the first of the two categories defined
at the beginning of this subsection: physically, it is a pure
background effect, which leads to a modified evolution of metric and
matter perturbations.

Let us estimate the magnitude of this effect in the same fashion as in
the pioneering paper \cite{Bond:1980ha}.  At any time we can combine
the continuity and Euler equations of CDM perturbations
(\ref{continuity}) and (\ref{Euler}) into
\begin{equation}
\ddot{\delta}_{\rm cdm} + \frac{\dot{a}}{a} \, \dot{\delta}_{\rm cdm}
= - k^2 \phi + 3 (\ddot{\psi} + \frac{\dot{a}}{a} \dot{\psi})~,
\label{evol_cdm}
\end{equation}
but well-inside the Hubble radius the source term is dominated by the
comoving gradient $-k^2 \phi$, which is approximately
equal to $-k^2 \psi$ as soon as the neutrino shear can be neglected.
This gradient is given by the Poisson equation, and we obtain
\begin{equation}
\ddot{\delta}_{\rm cdm} + \frac{\dot{a}}{a} \, \dot{\delta}_{\rm cdm}
= 4 \pi G \, a^2 \, \delta \rho~,
\end{equation}
where $\delta \rho$ is the total density perturbation.  In absence of
neutrinos and deep inside the matter-dominated regime, we would have
$\delta_{\rm cdm}=\delta_{\rm b}$, a total density perturbation
$\delta \rho = (\bar{\rho}_{\rm cdm} + \bar{\rho}_{\rm b}) \,
\delta_{\rm cdm}$, an expansion rate given by $3 \left( {\dot{a}}/{a}
\right)^2 = 8 \pi G a^2 (\bar{\rho}_{\rm cdm} + \bar{\rho}_{\rm b})
\propto a^{-1}$, a scale factor $a \propto \tau^2$, and the equation
would read
\begin{equation}
\ddot{\delta}_{\rm cdm} + \frac{2}{\tau} \, \dot{\delta}_{\rm cdm} - 
\frac{6}{\tau^2} \,
\delta_{\rm cdm} = 0~,
\label{deltacdmMD}
\end{equation}
with two solutions $\delta_{\rm cdm} \propto \tau^2$ and $\delta_{\rm
cdm} \propto \tau^{-3}$. Neglecting the decaying mode we recover the
standard result $\delta_{\rm cdm} \propto a$.  Now, let us consider
instead the case with massive neutrinos, still deep inside the
matter-dominated regime and on scales $k \gg k_{\rm nr}$, so that
$\delta \rho_{\nu}$ does not contribute to the Poisson equation:
$\delta \rho = (\bar{\rho}_{\rm cdm} + \bar{\rho}_{\rm b}) \,
\delta_{\rm cdm}$, while the neutrino background density does
contribute to the expansion rate: $3 \left( {\dot{a}}/{a} \right)^2 =
8 \pi G a^2 (\bar{\rho}_{\rm cdm} + \bar{\rho}_{\rm b} +
\bar{\rho}_{\nu})$.  Let us assume that 
$\bar{\rho}_{\nu}$ is dominated by non-relativistic neutrinos, so
that it decays approximately like $a^{-3}$, and the number
\begin{equation}
f_{\nu} \equiv \frac{\rho_{\nu}}{(\rho_{\rm cdm}+\rho_{\rm b}
+\rho_{\nu})} =
\frac{\Omega_{\nu}}{\Omega_{\rm m}}
\end{equation}
remains approximately constant. Then, the scale factor still evolves like
$\tau^2$ and the equation of evolution reads
\begin{equation}
\ddot{\delta}_{\rm cdm} + \frac{2}{\tau} \, \dot{\delta}_{\rm cdm} - 
\frac{6}{\tau^2} (1 - f_{\nu}) \,
\delta_{\rm cdm} = 0~.
\label{evol_cdm_massive_nu}
\end{equation}
Looking for solutions in 
$\delta_{\rm cdm} \propto \tau^{2p}$, we find two roots
\begin{equation}
p_{\pm} = \frac{-1 \pm \sqrt{1+24 (1 - f_{\nu})}}{4}~,
\end{equation}
and we conclude that the growing solution for the CDM density contrast
reads
\begin{equation}
\delta_{\rm cdm} \propto a^{p_+} \simeq a^{1-\frac{3}{5}f_{\nu}}~,
\end{equation}
where in the last step we assumed $f_{\nu} \ll 1$.  As expected, the
growth of $\delta_{\rm cdm}$ is reduced due to the fact that one of
the component in the Universe contributes to the homogeneous expansion
rate but not to the gravitational clustering.  The Poisson equation
gives
\begin{equation}
- k^2 \psi \propto a^{p_+-1} \simeq a^{-\frac{3}{5}f_{\nu}}~,
\label{Poisson_massive_nu}
\end{equation}
showing that for the same reason the gravitational potential
slowly decays during matter domination.

At the end of matter domination and during $\Lambda$ domination, we
have already seen that in absence of neutrinos $\phi$ decays like
$g(a)$ and $\delta_{\rm cdm}$ grows like $a \, g(a)$ (we recall that
the damping factor $g(a)$ is normalized to $g=1$ for $a \ll
a_{\Lambda}$).  The combined effect of $\Lambda$ and of neutrinos on
the growth of $\delta_{\rm cdm}$ can be well approximated by
\cite{Hu:1997vi}
\begin{equation}
\delta_{\rm cdm} \propto [a \, g(a)]^{p_+} \simeq 
[a \, g(a)]^{1-\frac{3}{5}f_{\nu}}~.
\label{cdm_growth_nu_lambda}
\end{equation}

{\bf Matter power spectrum for massive versus massless neutrinos.}\\
Let us try to predict analytically the difference between the power
spectrum of two cosmological models with respectively massive and
massless neutrinos ($\Lambda$MDM versus $\Lambda$CDM) assuming that they
share the same values of 
($\omega_{\rm m}$, $\omega_{\rm b}$, $\Omega_{\Lambda}$,
$A$, $n_s$, $\tau$).  The difference
between the two models lies in the values of $\omega_{\nu}$ and
$\omega_{\rm cdm}=\omega_{\rm m}-\omega_{\rm b}-\omega_{\nu}$, and can be
conveniently parametrized by $f_{\nu}$.
\begin{enumerate}
\item
for very large scales $k<k_{\rm nr}$, neutrino perturbations
were never affected by free streaming, and in the non-relativistic regime
they become indistinguishable from CDM perturbations. In particular,
inside the Hubble radius, neutrinos contribute both to
the Poisson equation with $\delta_{\nu}=\delta_{\rm cdm}$ and
to the background evolution, so the density contrast grows like $a$.
So, for fixed $\omega_{\rm m}$, this branch of the spectrum
is completely insensitive to the value of $f_{\nu}$ and of the neutrino mass.\\

\item
for very small scales with $k\gg k_{\rm nr}$ and
$k\gg k_{\rm eq}$, the two spectra will differ for various
reasons. First, equality does not take place for the same value of the
scale factor in the two models, since one always has $(a_{\rm
eq}/a_0)=\omega_{\rm r}/(\omega_{\rm b}+\omega_{\rm
cdm})=(1-f_{\nu})^{-1} \omega_{\rm r}/\omega_{\rm m}$, where
$\omega_{\rm r}$ includes the density of photons and of three
effectively massless neutrinos. Since the two models share the same
value of $\omega_{\rm r}$ and $\omega_{\rm m}$, the ratio between the
two values of the scale factor at equality is $a_{\rm
eq}^{f_{\nu}}/a_{\rm eq}^{f_{\nu}=0}=(1-f_{\nu})^{-1}$.  At equality,
and more generally at any time before the non-relativistic transition,
the two models are rigorously equivalent, modulo a shift in the scale
factor by the above factor. So, the identity
\begin{equation}
\delta_{\rm cdm}^{f_{\nu}}[a]=\delta_{\rm cdm}^{f_{\nu}=0}[(1-f_{\nu})\,a]
\label{a-identity}
\end{equation}
is valid as long as $a \leq a_{\rm nr}$. After $a=a_{\rm nr}$ and in
the massive model, the mechanism described in
Eqs.\ (\ref{evol_cdm_massive_nu})-(\ref{Poisson_massive_nu}) will
take place.  So, in the approximation of a time-independent fraction
$f_{\nu}$ for $a \geq a_{\rm nr}$ (which is not strictly valid in the
case of several massive non-degenerate neutrinos), the growth of CDM
perturbations is given by Eq.\ (\ref{cdm_growth_nu_lambda}), i.e. by a
constant logarithmic slope $(1-\frac{3}{5} f_{\nu})$ for the scale
factor (or for the product $[a \, g(a)]$ during dark energy
domination),
\begin{equation}
\delta_{\rm cdm}^{f_{\nu}}[a_0]=
\left(\frac{a_0 \, g(a_0)}{a_{\rm nr}}\right)^{1-\frac{3}{5}f_{\nu}}
\,
\delta_{\rm cdm}^{f_{\nu}}[a_{\rm nr}]~.
\end{equation}
In the massless case, everything is identical to the massive model
until the time at which $a = (1-f_{\nu}) \, a_{\rm nr}$ (i.e.\ until
the value of the scale factor for which neutrinos become
non-relativistic in the massive case ``shifted back'' to the massless
case according to Eq.(\ref{a-identity})).  Therefore, at this time,
the logarithmic slope of $\delta_{\rm cdm}^{f_{\nu}=0}$ must also be
equal to $(1-\frac{3}{5} f_{\nu})$, but then it will increase
progressively until unity, so that $\delta_{\rm cdm}^{f_{\nu}=0}$
becomes a linear function of $a$ or $[a \, g(a)]$. In a crude
approximation, we can write
\begin{equation}
\delta_{\rm cdm}^{f_{\nu}=0}[a_0] \simeq
\left(\frac{a_0\, g(a_0)}{(1-f_{\nu})\,a_{\rm nr}}\right) \,
\delta_{\rm cdm}^{f_{\nu}=0}[(1-f_{\nu})\,a_{\rm nr}]~,
\end{equation}
but this tends to overestimate the growth of perturbations
in the massless case: it assumes that right after $a=a_{\rm nr}$ the
logarithmic slope is equal to one, which is not true immediately. 
Indeed, a comparison with numerical results shows that the
total growth factor is a bit smaller,
\begin{equation}
\delta_{\rm cdm}^{f_{\nu}=0}[a_0] \simeq
\left(\frac{a_0\, g(a_0)}{(1-f_{\nu})^{1/2} a_{\rm nr}}\right) \,
\delta_{\rm cdm}^{f_{\nu}=0}[(1-f_{\nu}) \, a_{\rm nr}]~.
\end{equation}
Using this semi-analytic result, we find that the ratio between the
present value of $\delta_{\rm cdm}$ in the two models reads
\begin{equation}
\frac{\delta_{\rm cdm}^{f_{\nu}}[a_0]}{\delta_{\rm cdm}^{f_{\nu}=0}[a_0]}
= (1-f_{\nu})^{1/2}
\left(\frac{a_0\, g(a_0)}{a_{\rm nr}}\right)^{-\frac{3}{5} f_{\nu}}.
\end{equation}
According to Eq.\ (\ref{Pk_generic_LMDM}) this means that the total
matter power spectrum is reduced by
\begin{equation}
\frac{P(k)^{f_{\nu}}}{P(k)^{f_{\nu}=0}}
= \left(1-f_{\nu}\right)^3
\left(\frac{a_0 \, g(a_0)}{a_{\rm nr}}\right)^{-\frac{6}{5}f_{\nu}}~.
\end{equation}
Finally, we can replace $(a_0/a_{\rm nr})$ by $ 2000 \, m_{\nu}/(1\,
{\rm eV})$ and, assuming that the mass $m_{\nu}$ is shared by a number
$N_{\nu}$ of families, we can use $m_{\nu}= (\omega_{\nu}/N_{\nu}) \,
93.2 \, {\rm eV}$.  We obtain an expression that depends only on
($f_{\nu}$, $N_{\nu}$, $\omega_{\rm m}$, $\Omega_{\Lambda}$)
\begin{equation}
\frac{P(k)^{f_{\nu}}}{P(k)^{f_{\nu}=0}}
= \left(1-f_{\nu}\right)^3 \left[
1.9\times10^{5} \, g(a_0) \, \omega_{\rm m} \, f_{\nu} / N_{\nu} \,
\right]^{-\frac{6}{5}f_{\nu}}~.
\label{suppression}
\end{equation}
We show in Fig.\ \ref{fig_tk_vs_fnu} that this semi-analytic
expression is a very good approximation of the exact numerical result,
and also that for plausible values of $(\omega_{\rm m}, N_{\nu},
\Omega_{\Lambda})$ and for $f_{\nu}<0.07$, it can be approximated by
the well-known linear expression~\cite{Hu:1997mj}
\begin{equation}
\frac{P(k)^{f_{\nu}}}{P(k)^{f_{\nu}=0}} \simeq -8 \, f_{\nu}~.
\end{equation}
\end{enumerate}
In conclusion --and as confirmed by the numerical simulations in the
next subsection-- the ratio $P(k)^{f_{\nu}}/\,P(k)^{f_{\nu}=0}$
computed for fixed parameters ($\omega_{\rm m}$, $\Omega_{\Lambda}$)
smoothly interpolates between one for $k<k_{\rm nr}$ and a plateau
given by Eq.\ ({\ref{suppression}) for $k\gg k_{\rm eq}$.

\subsubsection{Numerical results}
\label{num_res_for_nu}

\begin{figure}[t]
\begin{center}
\includegraphics[width=0.49\textwidth]{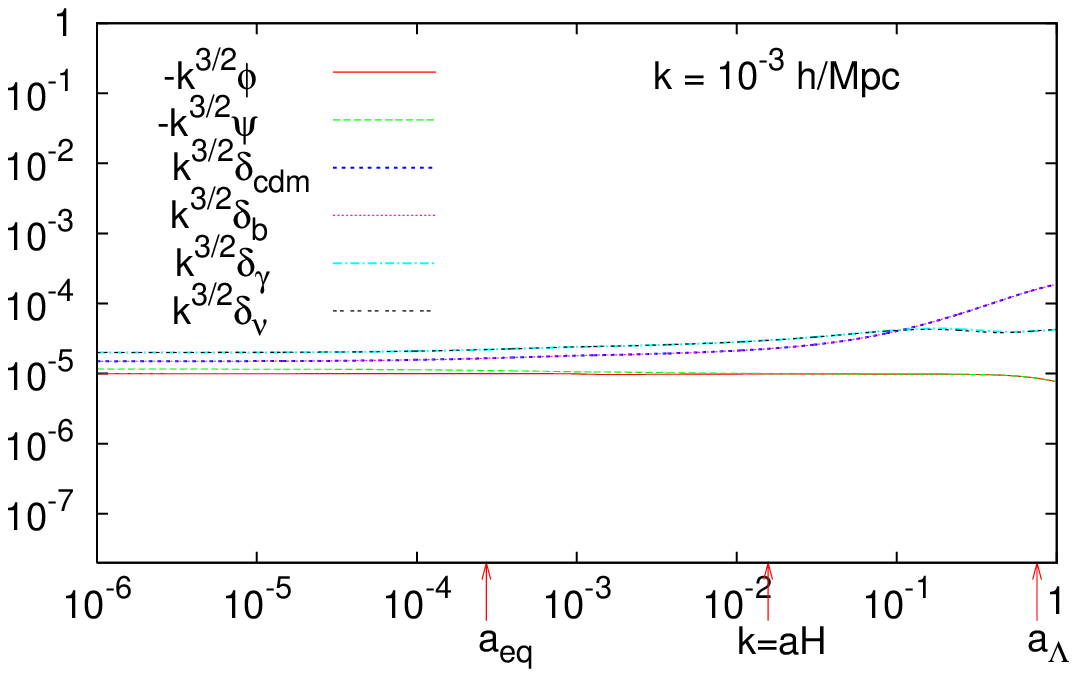}
\includegraphics[width=0.49\textwidth]{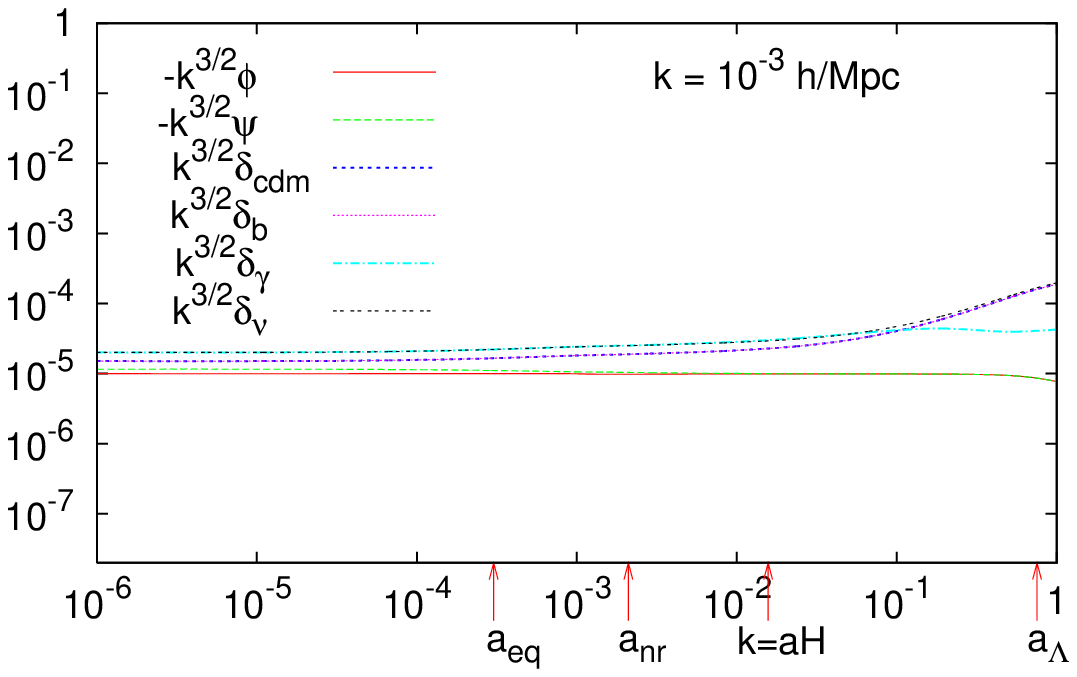}\\
\includegraphics[width=0.49\textwidth]{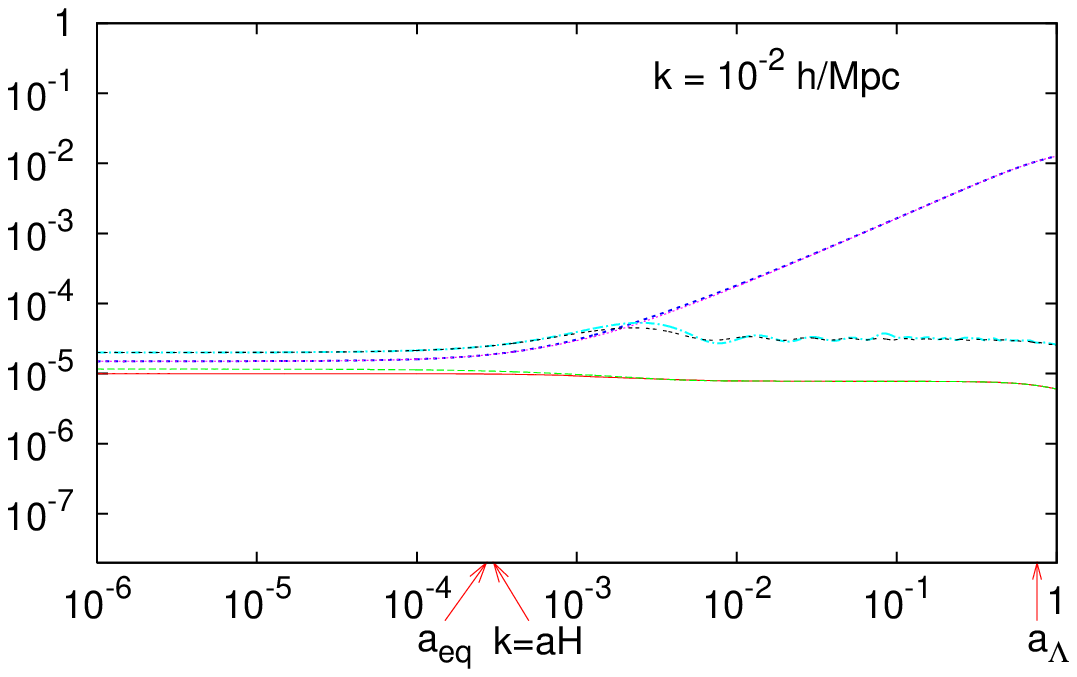}
\includegraphics[width=0.49\textwidth]{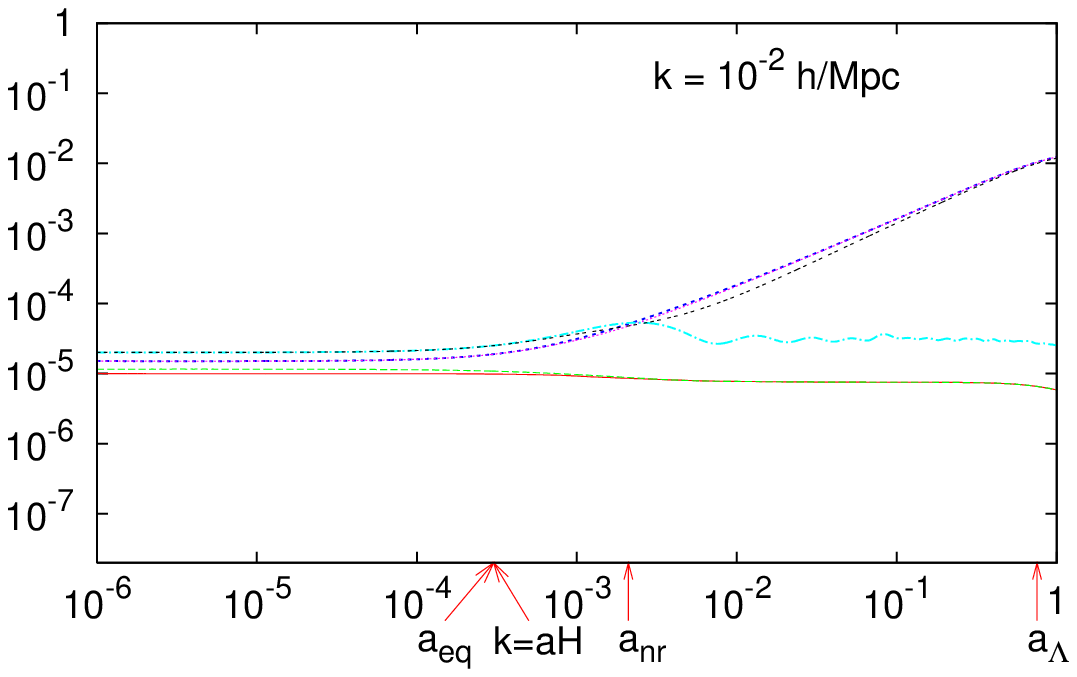}\\
\includegraphics[width=0.49\textwidth]{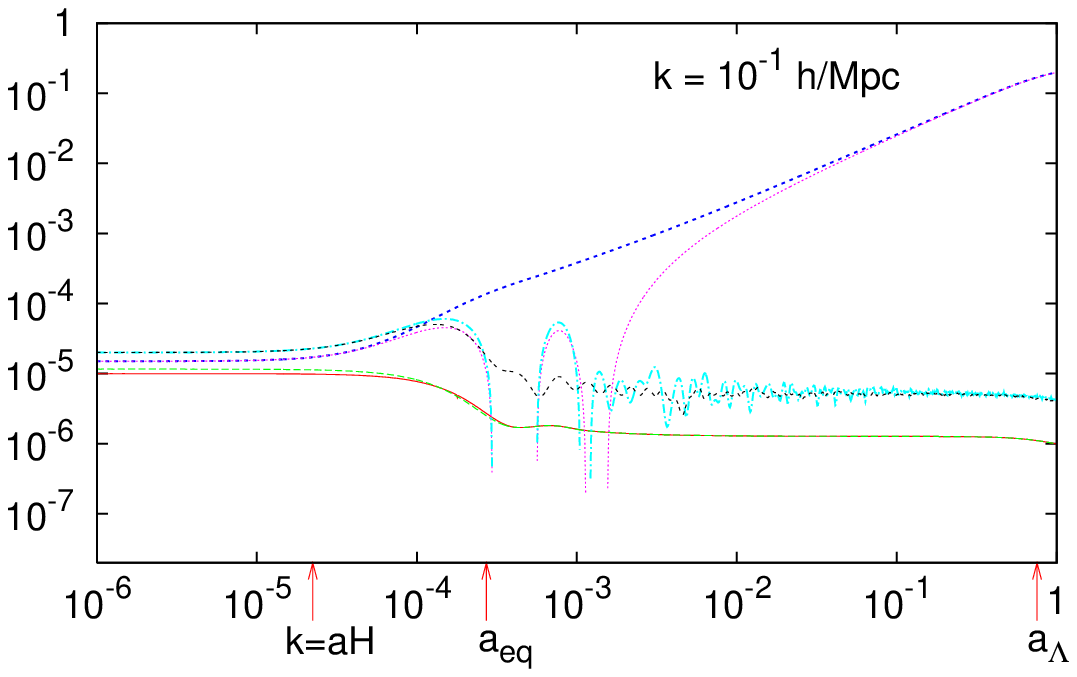}
\includegraphics[width=0.49\textwidth]{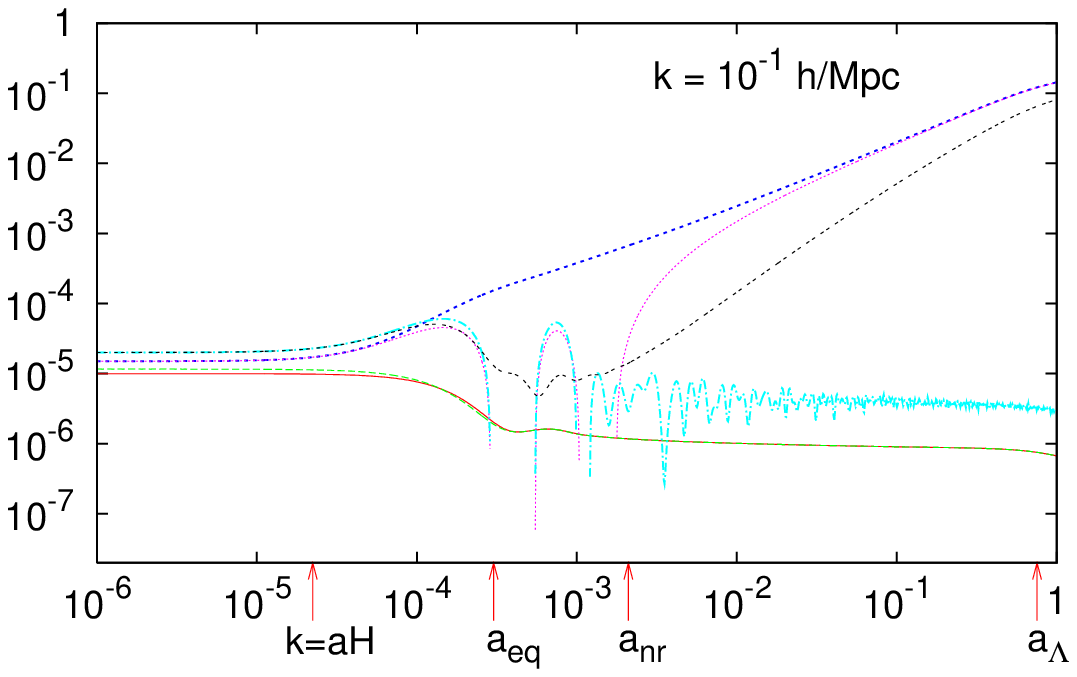}\\
\includegraphics[width=0.49\textwidth]{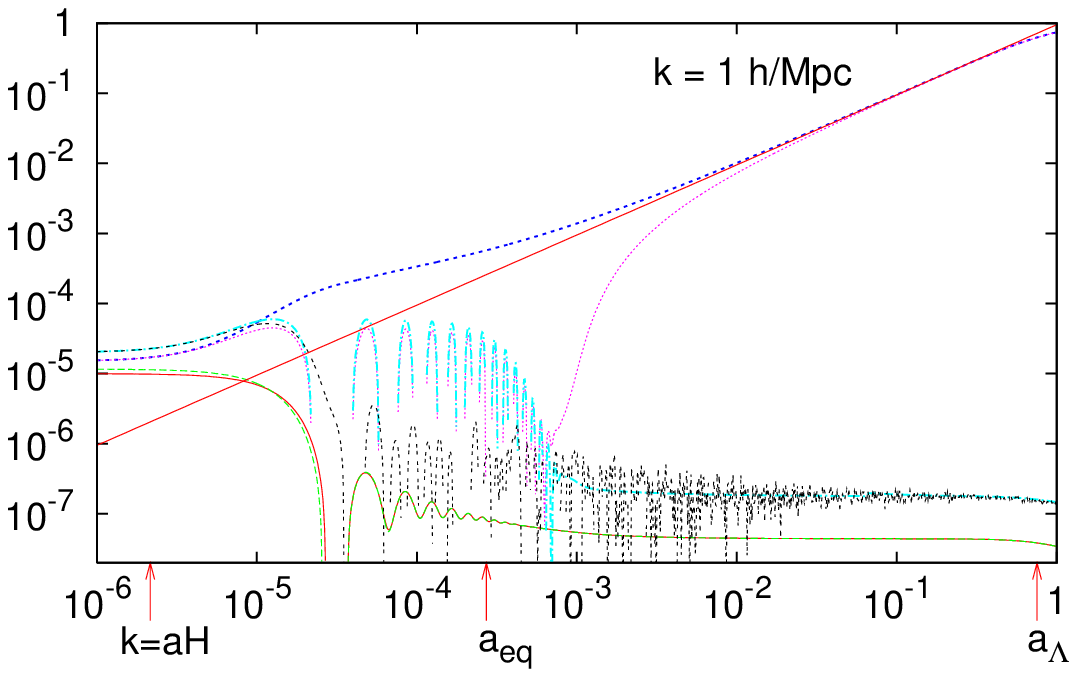}
\includegraphics[width=0.49\textwidth]{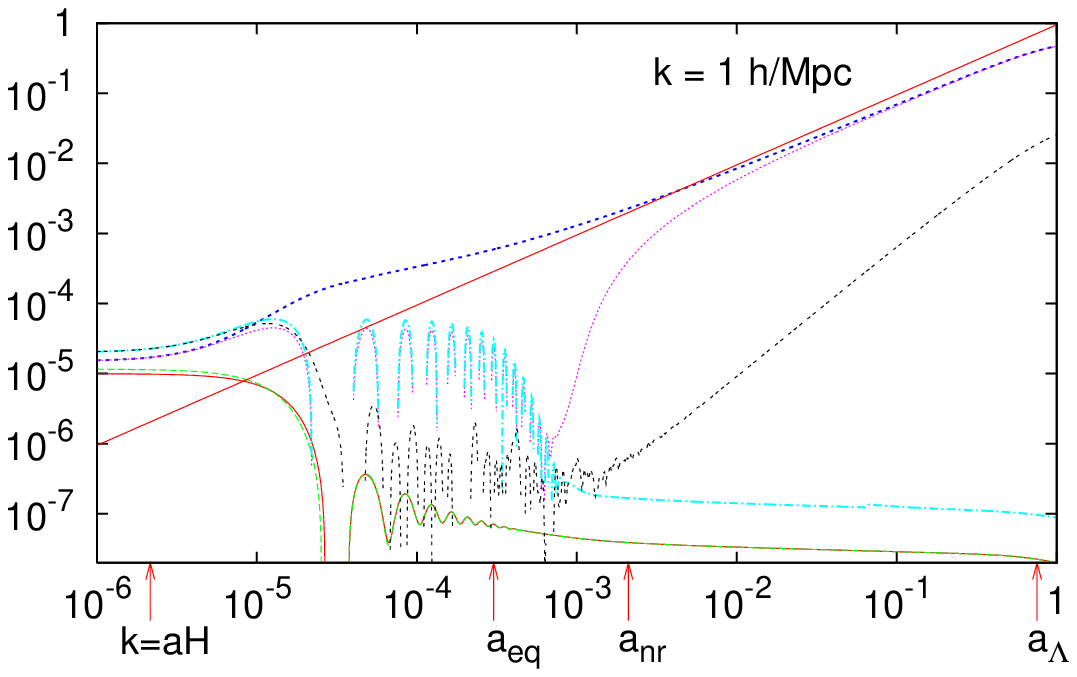}
\caption{\label{fig_deltas} Evolution of the metric and density
perturbations as a function of the scale factor (normalized to $a_0=1$
today), in the longitudinal gauge, for modes $10^{-3}\,h\,$Mpc$^{-1} <
k < 1 \,h\,$Mpc$^{-1}$ (from top to bottom), and for two cosmological
models: $\Lambda$CDM (left) and $\Lambda$MDM (right), both with
$\omega_{\rm m}=0.147$ and $\Omega_{\Lambda}=0.7$. The integration has
been performed with the code {\sc cmbfast} starting from the initial
condition $k^{3/2}\phi=-10^{-5}$.  The $\Lambda$MDM model has three
degenerate neutrinos with $m_{\nu}=0.46$ eV, corresponding to
$f_{\nu}=0.1$.}
\end{center}
\vspace{0.5cm}
\end{figure}

In Fig.\ \ref{fig_deltas}, we tried to summarize all the features
described in this section by showing the numerical solutions for the
metric and density perturbations ($\phi$, $\psi$, $\delta_{\rm cdm}$,
$\delta_{\rm b}$, $\delta_{\gamma}$, $\delta_{\nu}$) at different
wavenumbers, for a $\Lambda$CDM (left) and $\Lambda$MDM (right) a
model. The two models share common values of ($\omega_{\rm m}$,
$\omega_b$, $\Omega_\Lambda$, $\tau$) and the initial amplitudes are
the same. In the $\Lambda$CDM model, the three neutrino families are
massless, while in the $\Lambda$MDM model they share a common value of
the mass corresponding to $f_{\nu}=0.1$.  At the bottom of each plot,
we show the value of the scale factor at the time of Hubble entry
($k=aH$), radiation/matter equality ($a_{\rm eq}$) and
matter/$\Lambda$ equality ($a_{\rm \Lambda}$). Decoupling takes place
soon after equality, around $a=9 \times 10^{-4}$.  These solutions
have been computed with the public code {\sc cmbfast}
\cite{Seljak:1996is} in the synchronous gauge, and then translated
into the longitudinal gauge. Let us briefly summarize the behavior of
each quantity.

For a $\Lambda$CDM model (left plots in Fig.\ \ref{fig_deltas})}, one
has that
\begin{itemize}
\item
the metric perturbations are essentially constant on large scales $k <
k_{\rm eq}$ during radiation and matter domination. Instead, on small
scales $k > k_{\rm eq}$, they decay between Hubble entry and
decoupling, damped by the propagation of acoustic waves. On all
scales, they are also damped during $\Lambda$ domination. Initially
the difference $\psi-\phi$ is constant and non-zero, due to neutrino
shear. After equality, the difference becomes negligible.\\

\item the photon perturbations undergo acoustic oscillations between
Hubble entry and decoupling. Then, during matter domination, they
free-stream, maintaining and average density contrast
$\delta_{\gamma}=-4\phi$, corresponding (as we have seen) to gravity
vs. pressure equilibrium.\\

\item the relativistic neutrino perturbations are damped during
radiation domination after Hubble crossing, because of
free-streaming. During matter domination free-streaming goes on but
$\delta_{\nu}$ maintains itself around $-4\phi$ like for photons.\\

\item the CDM perturbations always grow inside the Hubble radius.
During radiation domination this growth starts abruptly,
but then becomes only logarithmic; during matter domination it is
linear; finally during $\Lambda$ domination it slows down. In the
bottom left plot for $k=1\,h\,$Mpc$^{-1}$, the solid red line is a
linear function of $a$ normalized in order to fit $\delta_{\rm cdm}$
during matter domination.\\

\item the baryon perturbations follow closely photon perturbations
until decoupling, experiencing the same acoustic oscillations. After
decoupling, they grow very quickly towards $\delta_{\rm cdm}$, and
after $a \sim 0.01$ the baryon perturbations remain equal to the CDM
ones.
\end{itemize}

For a $\Lambda$MDM model (right plots in Fig.\ \ref{fig_deltas}), the
main differences with respect to the previous case are 

(1) the fact that equality takes place a bit later, by a fraction
$(1-f_{\nu})^{-1}=1.1$.

(2) the behavior of $\delta_{\nu}$ after the time of the
non-relativistic transition, which is shown at the bottom of the plots
by the label ``$a_{\rm nr}$''.  For $a>a_{\rm nr}$, the neutrino
density perturbations track $\delta_{\rm cdm}=\delta_{\rm b}$ instead
of $\delta_{\gamma}$.  For large scales with $k<k_{\rm nr}$ (upper
plot), the non-relativistic transition takes place when the wavelength
is larger than the Hubble radius, so $\delta_{\rm cdm}$ and
$\delta_{\nu}$ are not very different when $a=a_{\rm nr}$, and soon
after $\delta_{\nu}$ reaches $\delta_{\rm cdm}$. For slightly smaller
scales like $k = 10^{-2} h\,$Mpc$^{-1}$ (right plot in second row),
the neutrino perturbations are much smaller than the CDM ones when
$a=a_{\rm nr}$, but then they have time to grow and reach $\delta_{\rm
cdm}$ before the present time.  For $k \geq 10^{-1} h\,$Mpc$^{-1}$
(lower plots), they do not have time to do so and remain subdominant
today.

(3) the growth of CDM perturbations is suppressed in the
$\Lambda$MDM case for two reasons: first because equality takes place
a bit later, and more importantly because $\delta_{\rm cdm}$ does not
grow as fast as the scale factor for $a>a_{\rm nr}$.  This can be seen
very clearly on the lower right plot, where we show exactly the same
linear function as in the lower left one.  The change in the time of
equality shifts the $\delta_{\rm cdm}$ normalization down, while the
effect of non-relativistic neutrinos lowers the value of the slope as
predicted analytically in the last subsection.

\begin{figure}[t]
\begin{center}
\includegraphics[width=0.73\textwidth]{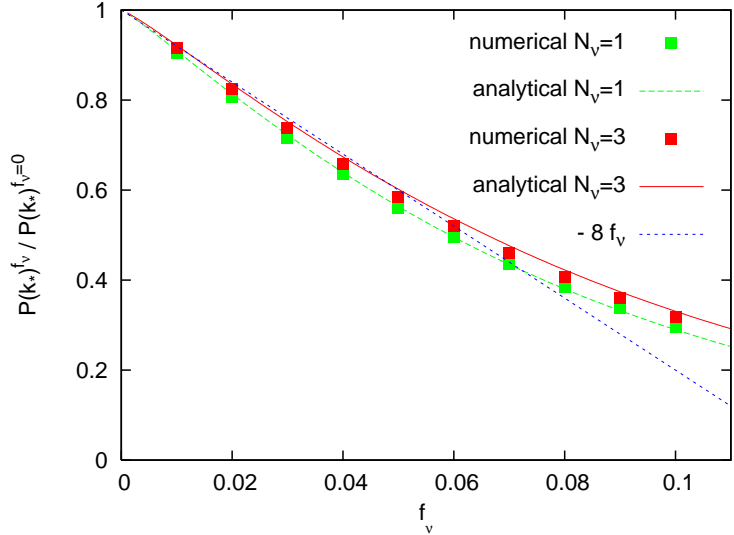}\\
\caption{\label{fig_tk_vs_fnu} Ratio of the matter power spectrum
including $N_{\nu}=1$ or 3 degenerate massive neutrinos to that with
three massless neutrinos, computed at the scale
$k_*=5\,h\,$Mpc$^{-1}$ for fixed parameters $(\omega_{\rm m}, \,
\Omega_{\Lambda})=(0.147,0.70)$, and plotted as a function of the
density fraction $f_{\nu}$.  The numerical result is compared with the
semi-analytical approximation of Eq.\ (\ref{suppression}) and with the
linear approximation $-8\,f_{\nu}$.}
\end{center}
\vspace{0.5cm}
\end{figure}
\begin{figure}[t]
\begin{center}
\includegraphics[width=0.73\textwidth]{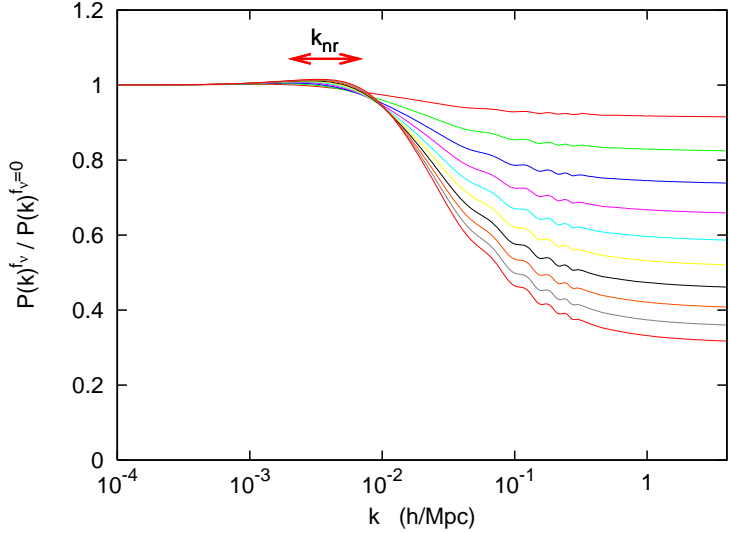}\\
\caption{\label{fig_tk} Ratio of the matter power spectrum including
three degenerate massive neutrinos with density fraction $f_{\nu}$ to
that with three massless neutrinos.  The parameters $(\omega_{\rm m},
\, \Omega_{\Lambda})=(0.147,0.70)$\ are kept fixed, and from top to
bottom the curves correspond to $f_{\nu}=0.01, 0.02, 0.03,\ldots,0.10$.
The individual masses $m_{\nu}$ range from $0.046$ eV to $0.46$ eV,
and the scale $k_{\rm nr}$ from $2.1\times10^{-3}h\,$Mpc$^{-1}$ to
$6.7\times10^{-3}h\,$Mpc$^{-1}$ as shown on the top of the
figure. $k_{\rm eq}$ is approximately equal to
$1.5\times10^{-2}h\,$Mpc$^{-1}$.}
\end{center}
\vspace{0.5cm}
\end{figure}
Looking now at all wavenumbers, we plot in Fig.\ \ref{fig_tk_vs_fnu}
the ratio of the matter power spectrum for $\Lambda$MDM over that of
$\Lambda$CDM, for different values of $f_{\nu}$, but for fixed
parameters $(\omega_{\rm m}, \, \Omega_{\Lambda})$. Here again, the
$\Lambda$MDM model has three degenerate massive neutrinos.  As
expected from the analytical results, this ratio is a step-like
function, equal to one for $k<k_{\rm nr}$ and to a constant for $k\gg
k_{\rm eq}$.  The value of the small-scale suppression factor is
plotted in Fig.\ \ref{fig_tk} as a function of $f_{\nu}$ and of the
number $N_{\nu}$ of degenerate massive neutrinos, still for fixed
$(\omega_{\rm m}, \, \Omega_{\Lambda})$.  The numerical result is
found to be in excellent agreement with the analytical prediction of
Eq.\ (\ref{suppression}). For simplicity, the growth factor
$g(a_0)\simeq0.8$ can even be replaced by one in
Eq.\ (\ref{suppression}) without changing the result significantly.
The well-known formula $P(k)^{f_{\nu}}/P(k)^{f_{\nu}=0} \simeq -8 \,
f_{\nu}$ is a reasonable first-order approximation for $0 < f_{\nu} <
0.07$.

\subsection{Summary of the neutrino mass effects}
\label{subsec:sum_mass-effect}

\subsubsection{Effects on CMB and LSS power spectra for fixed 
$(\omega_{\rm m}, \, \Omega_{\Lambda})$ and degenerate masses}

\begin{figure}[t]
\begin{center}
\includegraphics[width=0.48\textwidth]{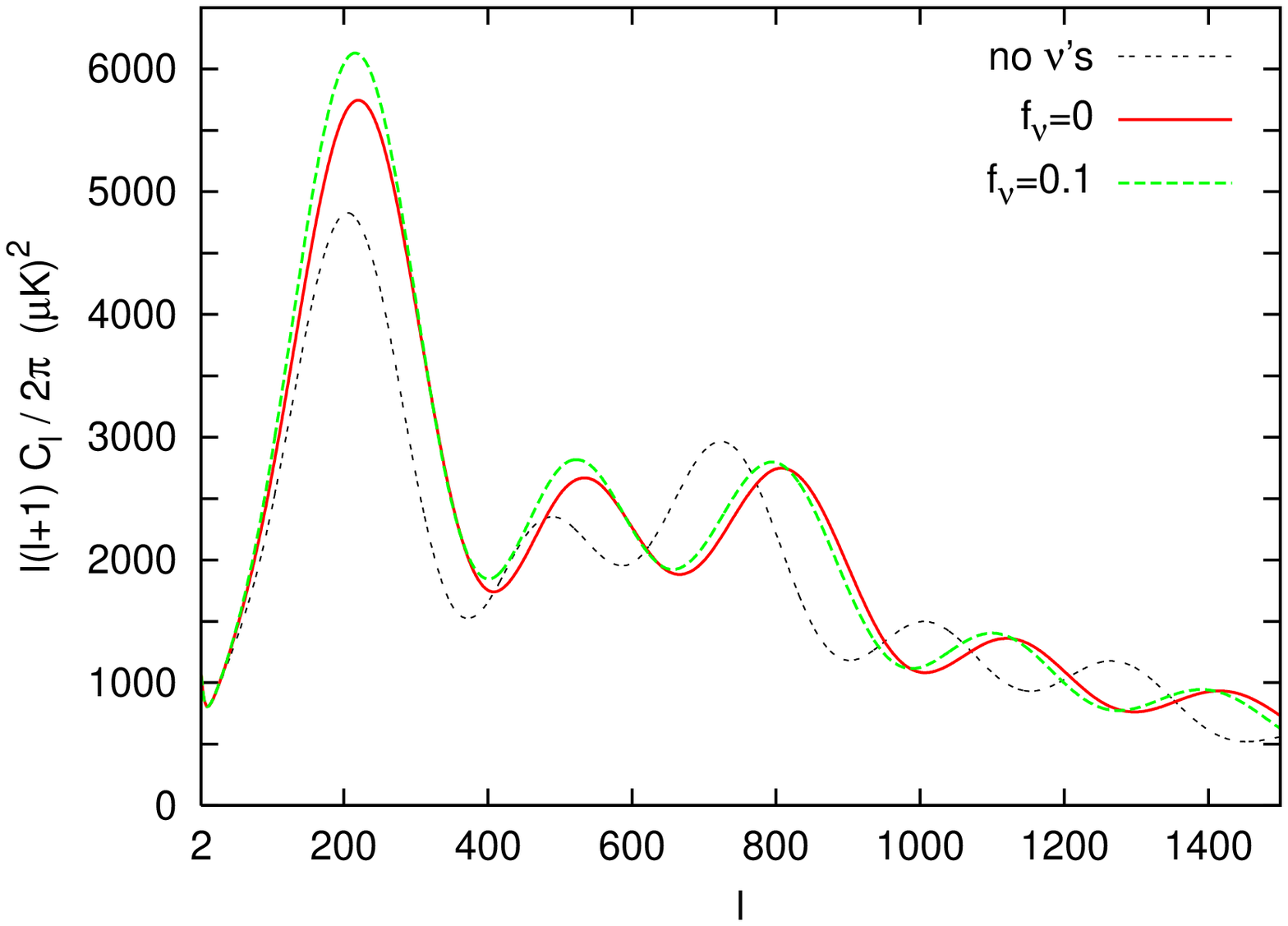}
\includegraphics[width=0.48\textwidth]{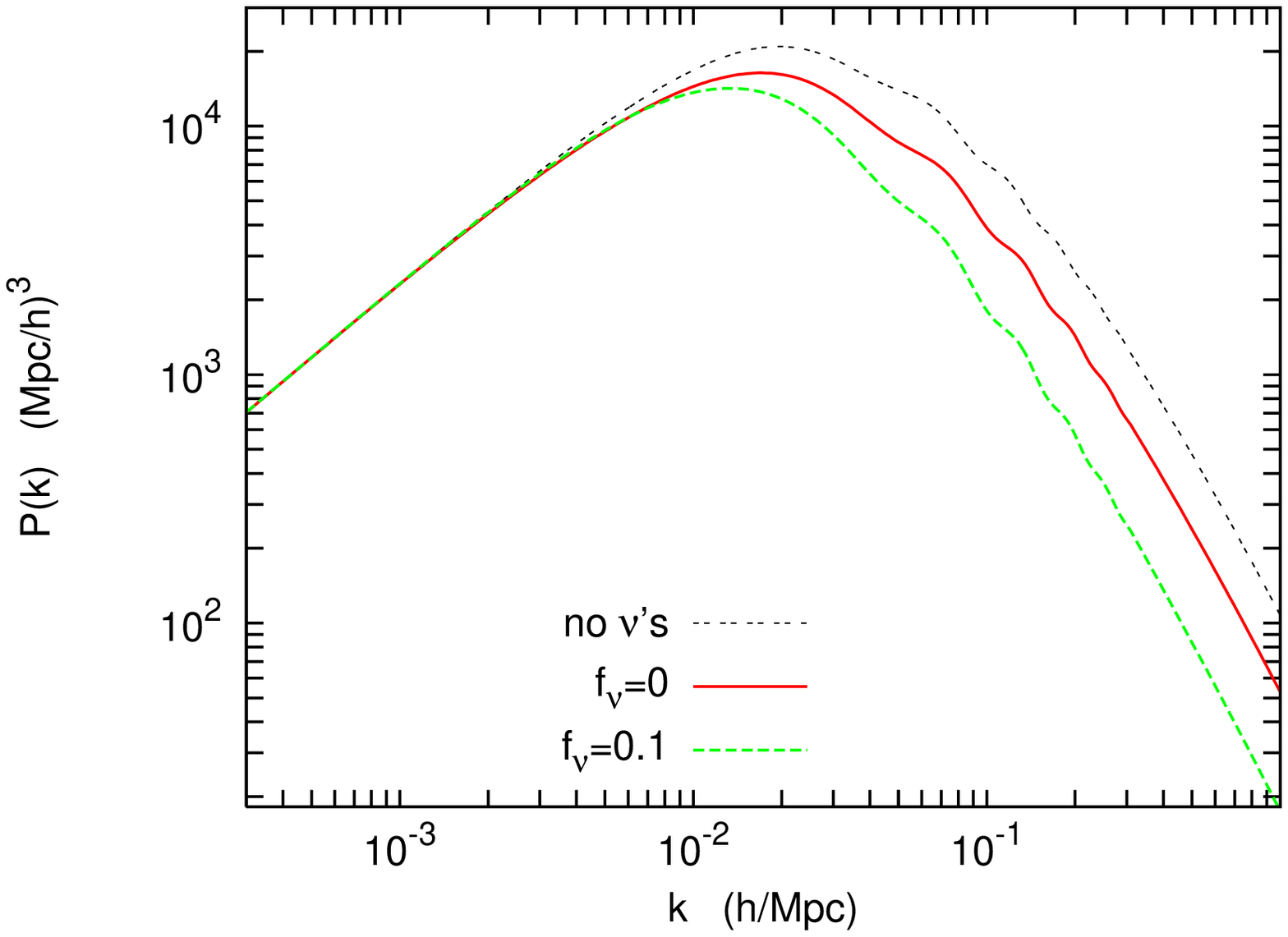}
\caption{\label{fig_powspec_LMDM} CMB temperature anisotropy spectrum
$C_l^T$ and matter power spectrum $P(k)$ for three models: the
neutrinoless $\Lambda$CDM model of section \ref{sec:num_res}, a more
realistic $\Lambda$CDM model with three massless neutrinos ($f_{\nu}
\simeq 0$), and finally a $\Lambda$MDM model with three massive degenerate 
neutrinos and a total density fraction $f_{\nu}=0.1$. In
all models, the values of ($\omega_{\rm b}$, $\omega_{\rm m}$,
$\Omega_{\Lambda}$, $A_s$, $n$, $\tau$) have been kept fixed.}
 \end{center}
\vspace{0.5cm}
\end{figure}

In Fig.\ \ref{fig_powspec_LMDM}, we show $C_l^T$ and $P(k)$ for two
models: $\Lambda$CDM with $f_{\nu}=0$ and $\Lambda$MDM with
$N_{\nu}=3$ massive neutrinos and a total density fraction
$f_{\nu}=0.1$. We also display for comparison the neutrinoless model
of Sec.\ \ref{sec:num_res}.  In all models, the values of
($\omega_{\rm b}$, $\omega_{\rm m}$, $\Omega_{\Lambda}$, $A_s$, $n$,
$\tau$) have been kept fixed, with the increase in $\omega_{\nu}$
being compensated by a decrease in $\omega_{\rm cdm}$.  There is a
clear difference between the neutrinoless and massless neutrino cases,
caused by a large change in the time of equality and by the role of
the neutrino energy-momentum fluctuations in the perturbed Einstein
equation \cite{Bashinsky:2003tk}. However our purpose is to focus on
the impact of the mass, i.e. on the difference between the solid (red)
and thick dashed (green) curves in Fig.\ \ref{fig_powspec_LMDM}.
\begin{figure}[t]
\begin{center}
\includegraphics[width=0.49\textwidth]{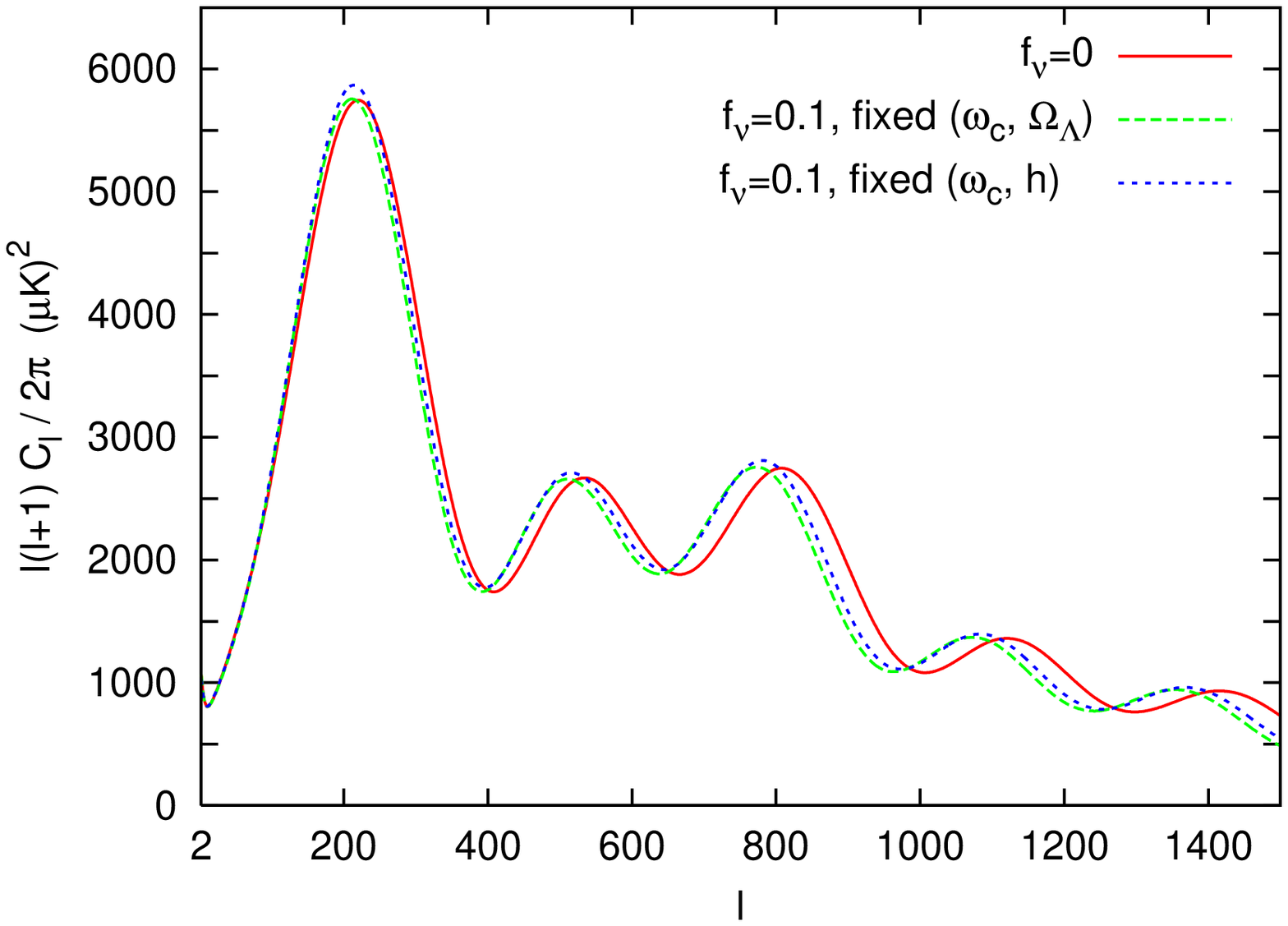}
\includegraphics[width=0.49\textwidth]{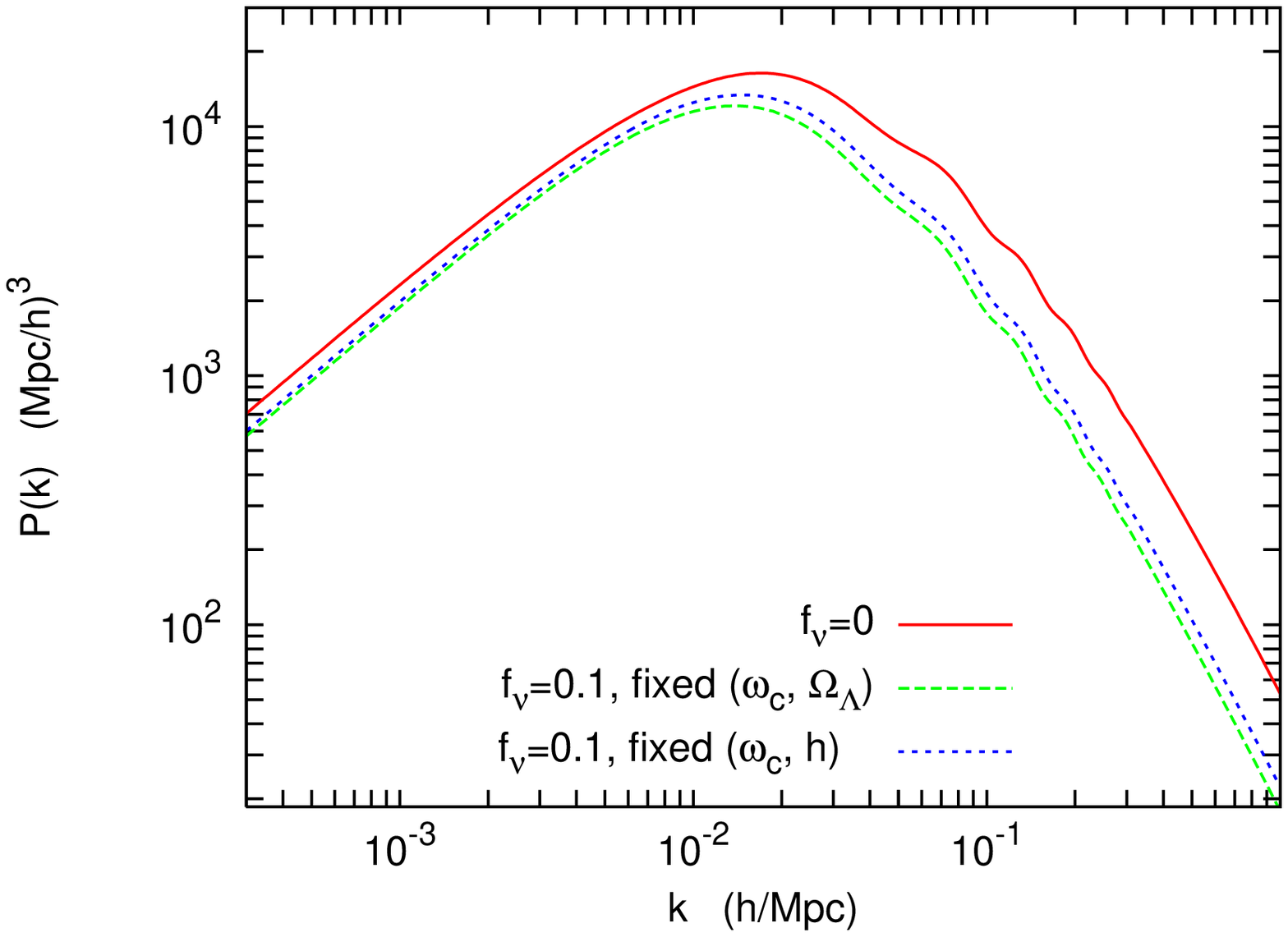}
\caption{\label{fig_powspec_deg} 
CMB temperature anisotropy spectrum
$C_l^T$ and matter power spectrum $P(k)$ for three models: 
the same $\Lambda$CDM model as in the previous figure,
with three massless neutrinos ($f_{\nu}
\simeq 0$); and two models with three massive degenerate 
neutrinos and a total density fraction $f_{\nu}=0.1$, sharing the same value
of $\omega_{\rm b}$ and $\omega_{\rm cdm}$ as the massless model, which
implies a shift either in $h$ (green dashed) or in $\Omega_{\Lambda}$
(blue dotted).}
\end{center}
\vspace{0.5cm}
\end{figure}

{\bf Impact on the CMB temperature spectrum.}  For $f_{\nu} \leq 0.1$,
the three neutrino species are still relativistic at the time of
decoupling, and the direct effect of free-streaming neutrinos on the
evolution of the baryon-photon acoustic oscillations is the same in
the $\Lambda$CDM and $\Lambda$MDM cases. Therefore, the effect of the
mass is indirect, appearing only at the level of the background
evolution: the fact that the neutrinos account today for a fraction
$\Omega_{\nu}$ of the critical density implies some change either in
the present value of the spatial curvature, or in the relative density
of other species. In this section, we choose to maintain a flat
Universe with fixed $(\omega_{\rm m}, \omega_{\rm b},
\Omega_{\Lambda})$ --other options would be possible, as discussed in
the next section.  Thus, while $\Omega_{\rm b}$ and $\Omega_{\Lambda}$
are constant, $\Omega_{\rm cdm}$ is constrained to decrease as
$\Omega_{\nu}$ increases. The main effect on the CMB anisotropy
spectrum results from a change in the time of equality. Since
neutrinos are still relativistic at decoupling, they should be counted
as radiation instead of matter around the time of equality, which is
found by solving $\rho_{\rm b}+\rho_{\rm
cdm}=\rho_{\gamma}+\rho_{\nu}$. This gives
\begin{equation}
a_{\rm eq}=\frac{\omega_{\rm r}}{\omega_{\rm b}+\omega_{\rm cdm}}~,
\end{equation}
where $\omega_{\rm r}$ stands for the radiation density extrapolated
until today {\it assuming that all neutrinos would remain massless},
given by Eq.\ (\ref{neff}) with $N_{\rm eff}=3.04$. So, when $f_{\nu}$
increases, $a_{\rm eq}$ increases proportionally to
$[1-f_{\nu}]^{-1}$: equality is postponed. We already saw in
Sec.\ \ref{subsec:params} that this produces an enhancement of
small-scale perturbations, especially near the first acoustic peak.
Also, postponing the time of equality increases slightly the size of
the sound horizon at recombination. These two features explain why in
Fig.\ \ref{fig_powspec_LMDM} the acoustic peaks are slightly enhanced
and shifted to the left in the $\Lambda$MDM case.

If neutrinos were heavier than a few eV, they would already be
non-relativistic at decoupling. This case would have more complicated
consequences for the CMB, as described in \cite{Dodelson:1995es}.
However, we will see in section \ref{sec:present} that this situation
is disfavoured by current upper bounds on the neutrino mass.

{\bf Impact on the matter power spectrum.}  This issue has been
discussed in details in the previous subsection: the combined effect
of the shift in the time of equality and of the reduced CDM
fluctuation growth during matter domination produces an attenuation of
small-scale perturbations for $k>k_{\rm nr}$, obeying to
Eq.\ (\ref{suppression}) in the large $k$ limit.  Various studies,
including
Refs.\
\cite{Hu:1997vi,Holtzman:1989ki,PSY95,Ma:1996za,Eisenstein:1997jh,Novosyadlyj:1998bw},
have derived some analytical approximation to the full MDM or
$\Lambda$MDM matter power spectrum, valid for arbitrary scales and
redshifts. However, the $\Lambda \neq 0$ case is only considered in
Refs.\ \cite{Hu:1997vi,Eisenstein:1997jh}.

\subsubsection{Degeneracy with other cosmological parameters}

{\bf Possible parameter degeneracy in the CMB temperature spectrum}.
Since the effect of the neutrino mass on CMB fluctuations is indirect
and appears only at the background level, one could think that by
changing the value of other cosmological parameters it would be
possible to cancel exactly this effect. For the simplest $\Lambda$MDM
model with only seven cosmological parameters, this is impossible: one
cannot vary the neutrino mass while keeping fixed $a_{\rm eq}$ and all
other quantities governing the CMB spectrum, that we already listed in
Sec.\ \ref{subsec:params}.  Let us illustrate this with a
particular example. Since the time of equality is related to
$(\omega_{\rm b}+\omega_{\rm cdm})$ rather than $\omega_{\rm m}$, we
could try to maintain this sum fixed. In this case, it is now
$\omega_{\rm m}$ that will increase like $[1-f_{\nu}]^{-1}$. We still
have some freedom concerning other parameters: we could choose either
to maintain a constant $\Omega_{\Lambda}$, in which case $h$ will
increase as $[1-f_{\nu}]^{-1/2}$; or to maintain a constant $h$, in
which case $(1-\Omega_{\Lambda})$ will increase as
$[1-f_{\nu}]^{-1}$. The result of these two possibilities is displayed
on Fig.\ \ref{fig_powspec_deg}.  Since the value of the scale factor
at equality is now more or less fixed, the change in peak amplitude is
smaller than in Fig.\ \ref{fig_powspec_LMDM}.  The counterpart is a
larger shift in peak location: since we now have variations of either
$h$ or $\Omega_{\Lambda}$, the angular scale of the sound horizon at
decoupling cannot be kept constant.

We conclude that for a seven parameter $\Lambda$MDM model, there is no
parameter degeneracy in the CMB spectrum. Therefore, it is possible to
constrain the neutrino mass using CMB experiments alone (although
neutrinos are still relativistic at decoupling).  We will see in the
next sections how this applies to current and future experiments.

This conclusion can be altered in more complicated models with extra
cosmological parameters. For instance, allowing for an open Universe,
it would be possible to increase $f_{\nu}$ with fixed ($\omega_{\rm
cdm}$, $\omega_{\rm b}$, $\Omega_{\Lambda}$), and to tune the spatial
curvature in order to shift back the peaks to their original
location. One could also vary the number of relativistic degrees of
freedom in order to play with the time of equality (see e.g.\
\cite{Crotty:2004gm}). In such extended models the CMB alone is not
sufficient for constraining the mass, but fortunately the LSS power
spectrum can lift the degeneracy.

{\bf Possible parameter degeneracy in the matter power spectrum}.  The
possibility of a parameter degeneracy in the matter power spectrum
depends on the interval $[k_{\rm min},k_{\rm max}]$ in which it can be
accurately measured. Ideally, if we could have $k_{\rm min} \leq
10^{-2}h\,$Mpc$^{-1}$ and $k_{\rm max} \geq 1\,h\,$Mpc$^{-1}$, the
effect of the neutrino mass would be non-degenerate, because of its
very characteristic step-like effect. In contrast, parameters like the
scalar tilt or the tilt running change the spectrum slope on all
scales, while the cosmological parameters which determine the shape of
the matter power spectrum below the scale of equality change the slope
for all wavenumbers $k \geq k_{\rm eq}$.

The problem is that usually, the matter power spectrum can only be
accurately measured in the intermediate region where the mass effect
is neither null nor maximal: in other words, many experiments only
have access to the transition region in the step-like transfer
function. In this region, the neutrino mass affects the slope of the
matter power spectrum in a way which can be easily confused with the
effect of the scalar tilt, tilt running, or many other cosmological
parameters. Because of these parameter degeneracies, the LSS data
alone cannot provide significant constraints on the neutrino mass, and
it is necessary to combine them with CMB data. The latter usually lift
all degeneracies. For instance, the CMB probes primordial fluctuations
on a very broad range of scales and makes a very good determination of
the tilt and tilt running: so, degeneracies with $n_s$ or $d n_s / d
k$ are usually not a problem. We will see however in 
Sec.\ \ref{subsec:extraparam}
that when exotic models with e.g.\ extra relativistic degrees of
freedom, a constant equation-of-state parameter of the dark energy
different from $-1$ or a non-power-law primordial spectrum are
introduced, the neutrino mass bound can become significantly weaker.

\subsubsection{Effects of non-degenerate neutrino masses}

It is often said that cosmological perturbations only probe the total
neutrino mass $\sum_i m_i=m_1+m_2+m_3$ (from now on denoted as
$M_\nu$).  Let us scrutinize this point.

We saw in the previous subsection that small neutrino masses only have
an indirect background effect on the CMB anisotropy spectrum, related
to the current density $\omega_{\nu}$, and therefore to the total mass
$M_{\nu}$ (see Eq.\ \ref{omeganu}).  So, in very good approximation,
CMB anisotropies probe the total mass with no information on the
splitting between different mass eigenstates.

As far as LSS is concerned, the situation is a bit more subtle. In
Sec.\ \ref{subsec:pert_nu}, we always assumed for simplicity that
there was one, two or three massive neutrino species with a common
mass.  In this case, we have seen that the position of the break in
the power spectrum $P(k)$ is related to $k_{\rm nr}$, i.e.\ to the
time of the non-relativistic transition which depends on the
individual mass $m_{\nu}$.  Instead, the amplitude of the break is
related both to $f_{\nu}$ and $a_{\rm nr}$, which depend respectively
on $M_{\nu}$ and $m_{\nu}$.  However, the results in Fig.\
\ref{fig_tk_vs_fnu} present a much stronger dependence on $f_{\nu}$
than on $N_{\nu}$: thus like for the CMB, the dominant effect is that
of the total mass, but the individual masses are not completely
irrelevant.  In conclusion, we expect various scenarios with the same
total mass $M_{\nu}$ and different numbers of massive species
$N_{\nu}=1,2,3$ to share the same CMB spectrum, but not the same
matter power spectrum, with small variations in the scale and
amplitude of the damping caused by neutrino free-streaming.  For the
mass schemes shown in Fig.\ \ref{fig:nuschemes}, the normal hierarchy
(NH) and the inverted hierarchy (IH), the differences are expected to
be even more subtle, because there are three distinct values of
$m_{\nu}$, $a_{\rm nr}$ and $k_{\rm nr}$.

\begin{figure}
\includegraphics[width=.48\textwidth]{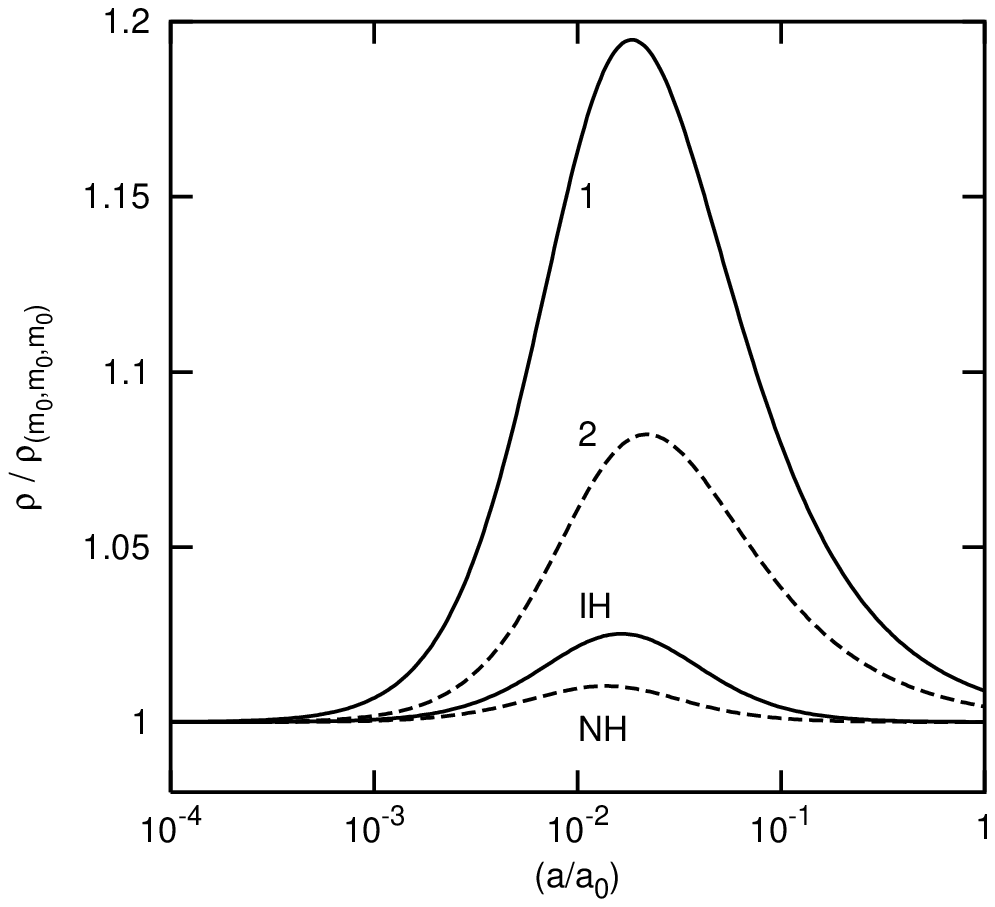}
\includegraphics[width=.48\textwidth]{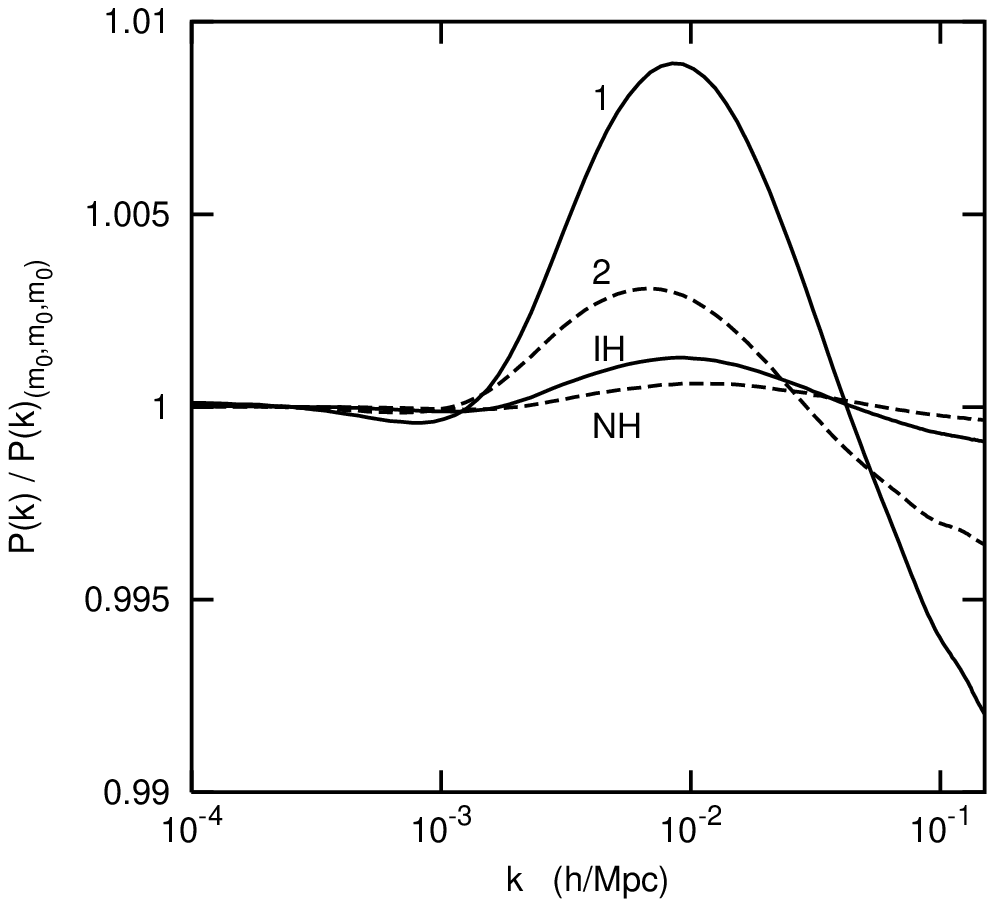}
\caption{\label{splitting} (Left) Evolution of the total
neutrino energy density as a function of the scale factor of the
Universe for models where the same total mass $M_\nu = 0.12$ eV
is distributed among the three species in various ways. Each line
corresponds to the energy density in one of four different cases (only
one massive eigenstates (1), two degenerate massive states (2), normal
hierarchy (NH), inverted hierarchy (IH)) normalized to the case with
three degenerate massive states.  (Right) Comparison of the
matter power spectrum obtained for the same models, divided each time
by that with three degenerate massive states. Differences in the mass
splitting affect the position and amplitude of the break in the power
spectrum.  }
\end{figure}
This problem has been studied numerically in Ref.\
\cite{Lesgourgues:2004ps}, using a modified version of {\sc cmbfast}
with three independent neutrino sectors\footnote{See also the recent
discussion in \cite{Takada:2005si}, as well as the study in
\cite{Slosar:2006xb}, where {\sc camb} was modified in order
to perform calculations with two different neutrino masses.}. Fig.\
\ref{splitting} is taken from this reference. First, for a fixed total
neutrino mass $M_{\nu}=0.12$ eV, we see on the left plot the ratio of
the total neutrino density $\bar{\rho}_{\nu}(a)$ in the NH, IH,
$N_{\nu}=1$ or degenerate $N_{\nu}=2$ scenario over that of the
degenerate $N_{\nu}=3$ scenario. The bump, more or less pronounced
depending on the case, is related to the different epochs of
transition to the non-relativistic regime.  In the right plot, we can
see the ratio between the matter power spectrum $P(k)$ in the same
cases over that of the degenerate $N_{\nu}=3$ scenario.  The bump
around $k=10^{-2}\,h\,$Mpc$^{-1}$ is caused by the shift in the break
scale, and the large-$k$ asymptotic values differ from one because of
the different break amplitudes.  Generally speaking, the NH scenario
interpolates between the $N_{\nu}=1$ and the degenerate $N_{\nu}=3$
cases, depending on the value of the lightest neutrino; the IH case
interpolates between the degenerate $N_{\nu}=2$ and $N_{\nu}=3$
cases. But the difference between the NH and IH power spectra always
remain very small: for instance, in the case shown in Fig.\
\ref{splitting} with $M_{\nu}=0.12$ eV, the two $P(k)$ differ at most
by $0.1\%$. It is very unlikely that this effect can be measured one
day (see however section \ref{sec:future}).

\subsubsection{Massive neutrinos and redshift dependence of the 
matter power spectrum}
\label{subsubsec:Pkz}

In absence of neutrinos, we have already seen that the shape of the
power spectrum during matter and $\Lambda$ domination is
redshift-independent for modes well inside the Hubble radius, while
the global power spectrum amplitude grows like $[a \, g(a)]^2$, where
$g(a)$ is a damping factor normalized to $g=1$ for $a\ll a_{\Lambda}$
(here $a_{\Lambda}$ is the scale factor at matter/$\Lambda$ equality).
In terms of redshift evolution, this gives
\begin{equation}
P(k,z)^{f_{\nu}=0} = \left(\frac{g(z)}{(1+z)\,g(0)}\right)^2
P(k,0)^{f_{\nu}=0} ~, \qquad k \gg aH~.
\end{equation}
The reason for this shape invariance is simply that the matter
fluctuations obey to a scale-independent equation of evolution (see
for instance Eq.\ (\ref{deltacdmMD}), which can easily be generalized
to the case of $\Lambda$ domination).
\begin{figure}[t]
\begin{center}
\includegraphics[width=.73\textwidth]{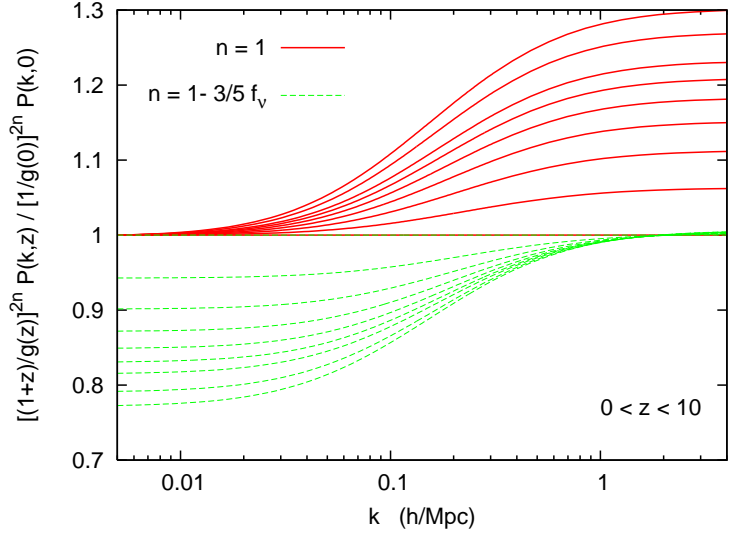}
\end{center}
\caption{\label{fig_tk_vs_z} Shape distortion of the matter power
spectrum $P(k,z)$ at different redshifts. {\it Upper solid curves:}
the power spectra are divided by $[a \, g(a)]^2= [g(z)/(1+z)]^2$,
which accounts for the evolution on large scales. {}From top to
bottom, the curves correspond to $z=10, 8, 6, 5, 4, 3, 2, 1, 0$.  {\it
Lower dashed curves:} the power spectra are divided instead by $[a \,
g(a)]^{2- 6/5\,f_{\nu}}= [g(z)/(1+z)]^{2- 6/5\,f_{\nu}}$, which
approximates the power spectrum evolution on small scales.  {}From top
to bottom, the curves correspond to $z=0, 1, 2, 3, 4, 5, 6, 8, 10$.
All cosmological parameters are kept fixed and the neutrino density
fraction is $f_{\nu}=0.1$.  }
\end{figure}

In presence of massive neutrinos, the shape of the power
spectrum is not invariant with redshift. Indeed, we are effectively
dealing with a two-component system (baryons and CDM on the one hand,
neutrinos on the other) in which the largest scales satisfy
$\delta_{\nu}=\delta_{\rm cdm} \propto [a \, g(a)]$, while for the
smallest scales $\delta_{\nu} \ll \delta_{\rm cdm} \propto [a \,
g(a)]^{1- 3/5\, f_{\nu}}$, as explained in Sec.\
\ref{subsubsec:mod_ev}. Therefore, the power spectrum has a different
redshift dependence on large and small scales, obeying
\begin{equation}
P(k,z) = \left\{
{\rm \begin{tabular}{llr}
$\displaystyle \left(\frac{g(z)}{(1+z)g(0)}\right)^2 \,\,\,P(k,0)$
&~~~for&
$aH < k < k_{\rm nr}$~, \\
$\displaystyle \left(\frac{g(z)}{(1+z)g(0)}\right)^{2- 6/5 \, f_{\nu}} 
\!\!\!P(k,0)$ 
&~~~for&
$k \gg k_{\rm nr}$ ~.
\end{tabular}}
\right.
\end{equation}
We illustrate this behavior in Fig.\ \ref{fig_tk_vs_z}. Had we plotted
the ratio $P(k,z)/P(k,0)$, we would have seen both the shape and
amplitude evolution of the matter power spectrum.  Instead, we want to
visualize only the shape distortion. So, we divide all power spectra
either by $[g(z)/(1+z)]^2$ or $[g(z)/(1+z)]^{2-6/5 \, f_{\nu}}$. As
expected, in the first case, the largest scales are corrected for
their time evolution, and the ratio remains equal to one for $k <
0.01\,h\,$Mpc$^{-1}$; in the second case, the smallest scales are
corrected for their time-evolution and the ratio is equal to one for
$k > 2\,h\,$Mpc$^{-1}$. Between these two limits, the neutrino
perturbations are just about to reach the $\delta_{\nu}
\longrightarrow \delta_{\rm cdm}$ asymptotic behavior around the
present time (see Sec.\ \ref{num_res_for_nu}).  For $f_{\nu}=0.1$, the
maximal shape distortion of the matter power spectrum is as large as
6\% for $z=1$ and 15\% for $z=3$.

\section{Current observations and bounds}
\label{sec:present}

In this section we review how the available cosmological data can be
used to get information on the absolute scale of neutrino masses,
complementary to laboratory experiments. After a brief introduction to
the statistical methods used in cosmology (in particular, Bayesian
inference), we describe the importance of each set or combination of
cosmological data with respect to neutrino masses, and explain how the
bounds depend on the data used and the underlying cosmological model.
We summarize the current limits in Tables
\ref{table:mass_CMB}--\ref{table:mass_LSS_2}, where the upper bounds
on neutrino masses are given at 95\% C.L.\ after marginalization over
all free cosmological parameters. We briefly discuss how the current
bounds on neutrino masses can be relaxed or even avoided when
neutrinos possess non-standard properties.

\subsection{Statistical methods for parameter inference from 
cosmological data}
\label{sec:stat}

The theory of cosmological perturbations is a stochastic theory. For a
given model, corresponding to some parameter values $\theta_i$, the
theory predicts the probability of any particular realization, i.e.\
of any Universe with given perturbations. Therefore, it can also predict
the probability of obtaining a given data set. The likelihood function
$\mathcal{L}(x_n|\theta_i)$ is defined as the probability of
observing the data $x_n$ given the model parameters $\theta_i$. Such a
function can be built for any data which has been properly modeled,
i.e. for which the instrumental noise properties have been fully
understood, and the level of correlation between different data points
has been estimated. 

The likelihood function can be used as a tool for solving the inverse
problem, i.e.\ measuring the goodness-of-fit of a model given the
data, or deriving confidence limits on the model parameters
$\theta_i$. This highly non-trivial task can be performed in different
ways; most particle physicists follow the so-called frequentist
approach, while until now, cosmologists usually prefer the Bayesian
approach (with very few exceptions, see
e.g.\ \cite{Trotta:2002iz,Upadhye:2004hh,Yeche:2005wn}). None of these
methods is better in absolute; they are not directly comparable,
because they address different questions. The selection of one method
rather than the other should depend of the particular goal of the
analysis, on how much the theory is constrained and understood, and on
how much we trust the data modeling (for a detailed comparison of
the two methods, see \cite{Zech:2001eh} or \cite{Trotta:2004qj} and
references therein).

The neutrino mass bounds derived in the next subsections are all based
on the Bayesian inference method, which has presumably no secrets for
readers with some experience in cosmology. On the other hand, many
particle physicists are so familiar with the frequentist philosophy
that cosmological bounds leave them perplex. In the rest of this
subsection, we briefly summarize the method used for deriving Bayesian
confidence limits. Then, we emphasize some properties of Bayesian
inference which are radically different from their frequentist
counterpart.

In any Bayesian parameter extraction, the first step is to assume a
theoretical model, with a set of $N$ free parameters $\theta_i$, and
$N$ probability distributions $\Pi(\theta_i)$ representing our
knowledge on each parameter {\it a priori}, before using the data.
These priors can be simply flat distributions over the interval in
which the parameter is meaningful. For instance, in the case of the
neutrino fraction $f_{\nu}$, it seems reasonable to define
$\Pi(f_\nu)=1$ for $0<f_{\nu}<1$ and $\Pi(f_\nu)=0$ otherwise: such a
prior is called a top-hat. Priors can also contain some information
from external data sets. For instance, if one of the free parameters
is the current expansion rate $H_0$, one can choose to derive its
prior from the results of the Hubble Space Telescope (HST) Key
Project, and define $\Pi(H_0)$ as a Gaussian probability with central
value $72 ~ {\rm km}\, {\rm s}^{-1}\, {\rm Mpc}^{-1}$ and standard
deviation $\sigma=8~ {\rm km}\, {\rm s}^{-1}\, {\rm Mpc}^{-1}$
\cite{Freedman:2000cf}.

Given the likelihood $\mathcal{L}(x_n|\theta_i)$ and the priors
$\Pi(\theta_i)$, one can compute the probability distribution of any
parameter $\theta_j$, or the joint probability for any subset of
parameters. Such a distribution is called a posterior, since it
represents the parameter probability {\it a posteriori}, after
including the full data. Posterior distributions are obtained by
marginalizing over the remaining free parameters. In practice, this
amounts in integrating the product of the likelihood times the priors
over these parameters. The result is a function of one or
several parameters $\theta_j$, which can be normalized to one. One
finally obtains a probability distribution $\mathcal{P}(\theta_j)$,
from which it is straightforward to derive confidence limits: the $n
\%$ confidence interval $I$ is defined as the region in which
\begin{enumerate}
\item the cumulative probability is equal to $n/100$: 
$\int_I \! d\theta_j \,\mathcal{P}(\theta_j) = n/100$, and \\

\item the probability is maximal: the value of $\mathcal{P}
(\theta_j)$ for any point inside $I$ is larger than that for any
point outside $I$.
\end{enumerate}
\begin{figure}[t]
\begin{center}
\includegraphics[width=.49\textwidth]{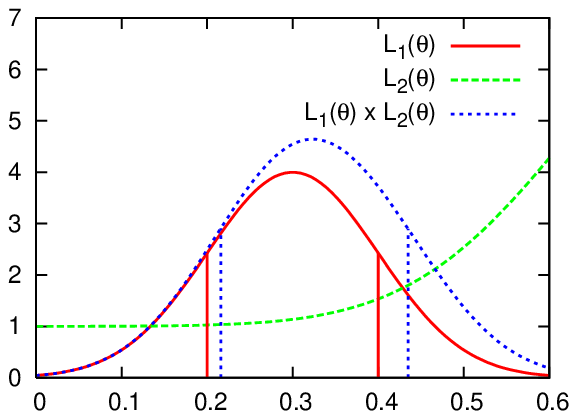}
\includegraphics[width=.49\textwidth]{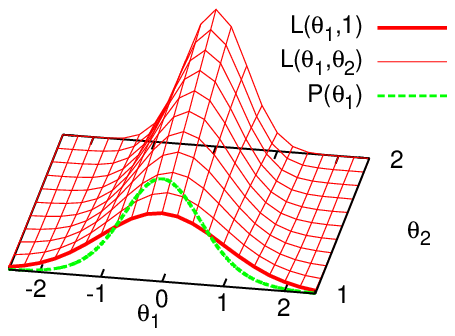}
\end{center}
\caption{\label{fig_stat} Situations which are possible in
Bayesian inference, while they would be paradoxical in a frequentist
analysis. {\it (Left)} Adding more data does not always lead to
stronger bounds. We consider here a model with one free parameter
$\theta$ with flat prior. A first data set has a Gaussian likelihood
$L_1(\theta)$ which predicts $\theta=0.3 \pm 0.1$ at the $68\%$
confidence level (this interval corresponds to the solid vertical
lines).  A second data set has a non-Gaussian likelihood
$L_2(\theta)$. The product $L_1(\theta) \times L_2(\theta)$ leads to a
wider allowed range, corresponding to the dotted vertical lines. {\it
(Right)} Adding more free parameters does not always lead to weaker
bounds.  We study here a two-dimensional model with likelihood
$L(\theta_1,\theta_2)$ and flat priors.  If there are good physical
motivations to fix $\theta_2=1$, one can also study the
one-dimensional model with likelihood $L(\theta_1,1)$ (the thick line
on the plot). We choose the likelihood in such way that $\int d
\theta_1 L(\theta_1,1)=1$: the thick line can be directly
interpreted as the posterior distribution for $\theta_1$ in the
one-dimensional model. For the two-dimensional model, the dashed line
shows the posterior $P(\theta_1)$ obtained by marginalizing over
$\theta_2$, and normalizing to one. Given the very specific form of the
likelihood, the two-dimensional model leads to a narrower posterior
and to stronger bounds on $\theta_1$.}
\end{figure}
The Bayesian inference method is well-defined, efficient, and has
several advantages. However, Bayesian results cannot be interpreted in
the same way as frequentist ones, and there are significant differences
that should be kept in mind:
\begin{itemize}
\item Bayesian inference always produce some confidence intervals,
even if the assumed model is unable to provide good fits to the data.
The results of a Bayesian analysis contain no statements on the
goodness-of-fit of the model.  Therefore, most authors provide some
complementary information, like the maximum value of the likelihood
$\mathcal{L}_{\rm max}$ (or equivalently, the minimum value of the
effective $\chi^2$, which is defined -- even for a non-Gaussian
likelihood -- as $\chi^2_{\rm eff}=-2 \ln \mathcal{L}$). Then,
it is possible to estimate in the frequentist way how
good is the best fit\footnote{However, comparing the merits of
two models is a non-trivial task. A simple comparison of the two
maximum likelihood values is insufficient. Following the Bayesian
philosophy, the most rigorous approach consists in computing the
so-called Bayesian Evidence \cite{Beltran:2005xd,Trotta:2005ar}.}.\\

\item for each parameter, the best-fit value $\theta_i^b$ generally
differs from the most likely value $\theta_i^l$.  Indeed, $\theta_i^b$
corresponds to the best-fit model, while $\theta_i^l$ maximizes the
marginalization integral.  They can be quite different, for instance
if for $\theta_i$ fixed to $\theta_i^b$ there are many bad fits and very
few excellent fits, while for $\theta_i$ fixed to $\theta_i^l$ there are
many good ones. In a future analysis, it would be conceivable that the
best-fit value of the total neutrino mass be zero, while the most
likely value be strictly positive. This is a natural outcome of the Bayesian
approach, which is based on probability distributions in parameter
space, and therefore favours regions in which models are good {\it in
average}.\\

\item the exact meaning of a Bayesian confidence interval is: ``given
a particular underlying model and given the data, it is likely to be
in this region, and unlikely to be outside''. One should be careful
not to over-interpret a Bayesian confidence interval. It is risky to
say that a parameter value sitting outside of the interval is
excluded. First, it is possible that with such a value one can find a
very good fit to the data (in the frequentist sense): the Bayesian
analysis could penalize this point just because the well-fitting
region has a tiny volume in the subspace of parameters to be marginalized
over. Second, one could possibly assume another model (with extra
free parameters) in which the same parameter value would be
comfortably allowed. In summary, Bayesian confidence intervals are
well-defined, but they are attached to a particular underlying
model. It makes no sense to quote a Bayesian upper limit, e.g.\ on the
neutrino masses, without explicitly specifying the model assumed.\\

\item in the frequentist approach, adding more data always lead to
stronger bounds on each parameter, while adding more free parameters
can only enlarge the allowed region. In the Bayesian approach, this is
true in general but not always: it is easy to build counter-examples,
as we did in Fig.\ \ref{fig_stat}.
\end{itemize}
Again, Bayesian inference offers many advantages for
cosmological parameter extraction, and the previous items should
not be regarded as criticisms, but rather as a list of warnings 
for correctly interpreting the results quoted in this section.

\subsection{CMB anisotropies}
\label{sec:present_CMB}

The experimental situation of the measurement of the CMB anisotropies
changed radically with the first year release of WMAP data (WMAP1),
that includes the temperature $\times$ temperature (TT)
\cite{Hinshaw:2003ex} and temperature $\times$ E-polarization (TE)
\cite{Kogut:2003et} correlation functions. The recent third-year
release (WMAP3) improves the TT and TE power spectra
\cite{WMAP3:Hinshaw}, and adds a detection of the E-polarization
self-correlation spectrum (EE) \cite{WMAP3:Page}.  The WMAP experiment
measured with high precision the TT spectrum on large angular scales
($l \lsim 900$), while at similar or smaller scales we have results
from experiments that are either ground-based such as the Arcminute
Cosmology Bolometer Array Receiver (ACBAR) \cite{Kuo:2002ua}, the Very
Small Array (VSA) \cite{Dickinson:2004yr}, the Cosmic Background
Imager (CBI) \cite{Readhead:2004gy} and the Degree Angular Scale
Interferometer (DASI) \cite{Halverson:2001yy}, or balloon-borne such
as ARCHEOPS \cite{Tristram:2004ke}, BOOMERANG (TT \cite{Jones:2005yb},
TE \cite{Piacentini:2005yq} and EE \cite{Montroy:2005yx}) and MAXIMA
\cite{Hanany:2000qf}.  When using data from different CMB experiments,
one should take into account that they overlap in some multipole
region, and not all data are uncorrelated.

Small neutrino masses only have a small influence on the power
spectrum of CMB anisotropies, since for $m_{\nu}$ smaller than
approximately $0.5$ eV the neutrinos are still relativistic at the time
of recombination, so that their direct effect on the CMB perturbations
is identical to that of massless neutrinos.  However, the signature of
a mass smaller than $0.5$ eV does not vanish: we recalled in Sec.\
\ref{subsec:sum_mass-effect} that there is always a background effect,
proportional to $\Omega_{\nu}$, which changes some characteristic
times and scales in the evolution of the Universe, and affects mainly
the amplitude of the first acoustic peak as well as the location of
all the peaks, as shown in Figs.\ \ref{fig_powspec_LMDM} and
\ref{fig_powspec_deg}. Therefore, it is possible to constrain neutrino
masses using CMB experiments only, down to the level at which this
background effect is masked by instrumental noise, or by cosmic
variance, or by parameter degeneracies in the case of some
cosmological models beyond the minimal $\Lambda$ Mixed Dark matter
($\Lambda$MDM) framework.

We list in Table \ref{table:mass_CMB} the bounds on the sum of
neutrino masses that were obtained in analyses of CMB data after the
first WMAP results appeared. In these works, a minimal
$\Lambda$MDM scenario was assumed: the total neutrino mass was the
only additional parameter with respect to a flat $\Lambda$CDM
cosmological model characterized by 6 parameters\footnote{This will be
the case for the bounds reviewed in this section, unless specified
otherwise.}.  However, there were some differences in the analysis
procedure.

The analysis in Ref.\ \cite{Tegmark:2003ud} used a Monte Carlo Markov
Chain (MCMC) method for the minimization, concluding that WMAP data
are consistent with neutrinos making $100\%$ of the dark matter
component (see also \cite{Elgaroy:2003yh}), thus giving the loose
bound on the total neutrino mass $M_\nu < 10.6$ eV. This would mean
that eV masses are basically indistinguishable from CDM at the epoch
of last scattering, and therefore have little impact on the CMB scales
measured by WMAP.

\begin{table*}
\caption{Upper bounds on $M_\nu$ (eV, 95\% CL) from recent analyses of
CMB data only. Note that the results of Refs.\
\cite{Ichikawa:2004zi,Sanchez:2005pi,Rubino-Martin2005,MacTavish:2005yk} are consistent
with each other (see the text for details).
\label{table:mass_CMB}}
\begin{center}
\begin{tabular}{crl}
\hline
Ref.\ & Bound & Data used \\
\hline 
\cite{Tegmark:2003ud}     & 10.6 &  WMAP1 \\
\cite{Ichikawa:2004zi}    & 2.0 & WMAP1 \\
\cite{WMAP3:Spergel} & 2.0 & WMAP3\\
\cite{Sanchez:2005pi}   & 2.1 
& WMAP1, VSA, ACBAR \& CBI\\
\cite{Rubino-Martin2005}  & 1.6 
& WMAP1, VSA, ACBAR \& CBI\\
\cite{MacTavish:2005yk}   & 3.1 
& WMAP1, BOOMERANG03, VSA, ACBAR, \\
{}&{}&CBI, DASI \& MAXIMA\\
\hline
\end{tabular}
\end{center}
\vspace{0.5cm}
\end{table*}
However, a very different result was discussed in ref.\
\cite{Ichikawa:2004zi} (see also \cite{Fukugita:2005sb}), where an
upper limit of $2$ eV was found from the analysis of WMAP1 only using a
grid-based method. The authors of this last analysis ascribe the
difference with respect to \cite{Tegmark:2003ud} to a better
marginalization procedure. However, using the WMAP1 data and an MCMC
method, we find the same result as in \cite{Ichikawa:2004zi}.
Finally, the $2$ eV bound is now confirmed by the analysis of
the WMAP3 data in Ref.\ \cite{WMAP3:Spergel}.

In Table \ref{table:mass_CMB}, we also show the results from two
recent works \cite{Sanchez:2005pi,MacTavish:2005yk} based on the
combination of WMAP1 with other recent CMB experiments. We also show
the result of a very recent analysis \cite{Rubino-Martin2005} in a
cosmological model with 11 parameters (similar to that in Table 4 of
Ref.\ \cite{Rebolo:2004vp}, but in a flat Universe). The $2\sigma$
bound obtained by \cite{Sanchez:2005pi,Rubino-Martin2005} is still
close to 2 eV. The result of \cite{MacTavish:2005yk}, which include
more data from BOOMERANG, DASI and MAXIMA, is larger by 50\%.
We saw in Sec.\ \ref{sec:stat} that in a Bayesian
analysis, it is possible to add more data and obtain weaker bounds. In
the present case, a careful examination of Fig.\ 8 in
\cite{MacTavish:2005yk} shows that their analysis based on WMAP1 only
is in very good agreement with \cite{Ichikawa:2004zi,Sanchez:2005pi};
however, at least one of the DASI or MAXIMA data set tends to push the
bound upwards, because it is better fitted with a non-zero neutrino
mass.

In conclusion, many analyses support the conclusion that a sensible
bound on neutrino masses exists, of order of $2-3$ eV at 2$\sigma$.
This is an important result, since it does not depend on the
uncertainties from LSS data (see Sec.\ \ref{subsec:present_GRS}).
The only way to relax this bound would be to complicate the
cosmological scenario assumed for parameter extraction, but this issue
will be discussed in Secs.\ \ref{subsec:extraparam} and \ref{subsec:nonstand}.

However, these results do not take advantage of the most significant
observable signature of massive neutrinos: the suppression of
small-scale matter fluctuations induced by neutrino free-streaming.
%We will see in Sec.\ \ref{sec:future} that in the future, smaller
%masses could be detected by measuring the effect of weak gravitational
%lensing on the CMB signal. At the present time, 
In order to get
improved bounds on neutrino masses it is crucial to combine CMB data
with independent large-scale structure observations.

\subsection{Galaxy redshift surveys}
\label{subsec:present_GRS}

We have seen in Sec.\ \ref{subsec:sum_mass-effect} that free-streaming
of massive neutrinos produces a direct effect on the formation of
cosmological structures. As shown in Fig.\ \ref{fig_tk}, the presence
of neutrino masses leads to an attenuation of the linear matter power
spectrum on small scales.

In a seminal paper, Hu and collaborators \cite{Hu:1997mj} showed that
an efficient way to probe neutrino masses of order eV was to use data
from large redshift surveys, which measure the distance to a large
number of galaxies, giving us a three-dimensional picture of the
universe (modulo the effect of unknown peculiar velocities on top of
the Hubble flow, which can be corrected only statistically).  At
present, we have data from two large projects: the 2 degree Field
(2dF) galaxy redshift survey \cite{Percival:2001hw,Cole:2005sx}, whose
final results were obtained from more than 220,000 galaxy redshifts,
and the Sloan Digital Sky Survey (SDSS)
\cite{Tegmark:2003uf,Eisenstein:2005su}, which will be completed soon
with data from one million galaxies\footnote{An extension of this
experiment, SDSS-II was recently funded (see the SDSS webpage {\tt
http://www.sdss.org})}.

One of the main goals of galaxy redshift surveys is to reconstruct the
power spectrum of matter fluctuations on very large scales, whose
cosmological evolution is described entirely by linear perturbation
theory. However, the linear power spectrum must be reconstructed from
individual galaxies, i.e.\ from objects which underwent a strongly
non-linear evolution. A simple analytic model of structure formation
suggests that on large scales, the galaxy-galaxy correlation function
should be, not equal, but proportional to the linear matter density
power spectrum, up to a constant factor that is called the
light-to-mass bias.  This parameter, denoted by $b$, can be obtained
from independent methods (e.g.\ high-order statistics beyond the
two-point correlation function). Currently, these methods tend to
confirm that the linear biasing assumption is correct, at least in
first approximation.

A conservative way to use the measurements of galaxy-galaxy
correlations in an analysis of cosmological data is to take the bias
as a free parameter, i.e.\ to consider only the shape of the matter
power spectrum at the corresponding scales and not its amplitude. In
this case, we will refer to the corresponding data of the 2dF and SDSS
surveys as 2dF-gal and SDSS-gal, respectively. Alternatively, one can
add a measurement of $b$, as recently done in three analyses. The
value $b=1.04\pm 0.11 (1\sigma)$ [2dF-bias] \cite{Verde:2003ey} from
the 2dF bispectrum analysis was used in \cite{Spergel:2003cb}, while
\cite{Seljak:2004xh} adopted the estimate $b=0.99\pm 0.11 (1\sigma)$
[SDSS-bias] from galaxy-galaxy lensing from SDSS
\cite{Seljak:2004sj}. Finally, in Ref.\ \cite{MacTavish:2005yk} the
intermediate estimate $b=1.0 \pm 0.10 (1\sigma)$ was used.
\begin{table*}
\caption{Upper bounds on $M_\nu$ (eV, 95\% CL) from recent
analyses combining CMB, galaxy redshift
surveys (shape of the matter power spectrum only)
and other data.\label{table:mass_LSS}}
\begin{center}
\begin{tabular}{lcl}
\hline
Ref.\ & Bound &  Data used 
(in addition to WMAP1 except for Ref.\  \cite{WMAP3:Spergel})\\
\hline 
\cite{Hannestad:2003xv} & 1.2 & other CMB (pre-WMAP), 2dF-gal\\
{}                      & 1.0 & previous + HST, SNIa\\
\cite{Tegmark:2003ud} & 1.74 &  SDSS-gal\\
\cite{Barger:2003vs} & 0.75 & other CMB (pre-WMAP), 2dF-gal, SDSS-gal, HST\\
\cite{Crotty:2004gm} & 1.0 & ACBAR, 2dF-gal, SDSS-gal\\
{}                    & 0.6 & previous + HST, SNIa\\
\cite{Rebolo:2004vp}  & 0.96 & VSA, 2dF-gal\\
\cite{Seljak:2004xh}  & 1.54 & SDSS-gal, SNIa\\
\cite{Fogli:2004as} & 1.4 & other CMB
%\footnote{BOOMERANG98, VSA,CBI, DASI \& MAXIMA}
, 2dF-gal, HST, SNIa\\
\cite{MacTavish:2005yk} & 1.2 & 
other CMB
%\footnote{BOOMERANG03, VSA, ACBAR,CBI, DASI \& MAXIMA},
, 2dF-gal, SDSS-gal\\
\cite{Sanchez:2005pi}  & 1.27 & other CMB
%\footnote{VSA,ACBAR \& CBI}
, SDSS-gal\\
{}& 1.16 & other CMB
%\footnote{VSA, ACBAR \& CBI}
, 2dF-gal\\
\cite{WMAP3:Spergel} & 0.87 & WMAP3, 2dF-gal\\
\hline
\end{tabular}
\end{center}
\vspace{0.5cm}
\end{table*}

We summarize in Table \ref{table:mass_LSS} the bounds on neutrino
masses that were obtained in combined analyses of CMB and LSS data,
using galaxy clustering data in the range of wavenumbers
$(0.015-0.02)\,h/$Mpc$\,<k<(0.15-0.20)\,h/$Mpc (including non-linear
corrections for the smallest wavelengths \cite{Smith:2002dz}). The
results in this Table were obtained leaving the bias as a free
parameter. One concludes from Table \ref{table:mass_LSS} that the
addition of LSS data improves the bounds with respect to the numbers
in Table \ref{table:mass_CMB}, with an upper limit on $M_\nu$ between
$0.9$ and $1.7$ eV for SDSS-gal and/or 2dF-gal added
to CMB data. In general, one obtains weaker bounds on neutrino masses
using preliminary SDSS results instead of 2dF data, but this
conclusion is expected to change after the next SDSS releases.  The
improvement with respect to previous results is evident: in 2001,
ref.\ \cite{Wang:2001gy} found $M_\nu < 4.2$ eV (95\% CL) from the
combination of the PSCz survey \cite{Saunders:2000af} with pre-WMAP
CMB data, while the analysis in \cite{Elgaroy:2002bi} used early 2dF
data to obtain $M_\nu < 2.2$ eV with priors on the values of
$\Omega_{\rm m}$ and $h$.

There exist other measurements of cosmological parameters that are
often added to CMB and LSS data, as shown in Table
\ref{table:mass_LSS}. For instance, the present value of the Hubble
parameter was measured by the Key Project of the Hubble Space
Telescope (HST), giving $h=0.72\pm 0.08 \,(1\sigma)$
\cite{Freedman:2000cf}, which excludes low values of $h$ and leads to
a stronger upper bound on the total neutrino mass. In addition, one
can include the constraints on the current density of the dark energy
component deduced from the redshift dependence of type Ia supernovae
(SNIa) luminosity, which measures the late evolution of the expansion
rate of the Universe. For a flat Universe with a cosmological
constant, these constraints can be translated into bounds for the
matter density $\Omega_{\rm m}$, which range from the conservative
value $\Omega_{\rm m} = 0.28 \pm 0.14 \,(1\sigma)$ from
\cite{Perlmutter:1998np} to the more recent $\Omega_{\rm m} =
0.29^{+0.05}_{-0.03}\,(1\sigma)$ \cite{Riess:2004nr} from the analysis
of SNIa with the HST (other recent works include
\cite{Tonry:2003zg,Astier:2005qq}).

As shown in Table \ref{table:mass_LSS_2}, the bounds on neutrino
masses are more stringent when the amplitude of the matter power
spectrum is fixed with a measurement of the bias
\cite{Spergel:2003cb,MacTavish:2005yk,Seljak:2004xh}, instead of
leaving it as a free parameter. The upper limits on $M_\nu$ are
reduced to values of order $0.5-0.9$ eV ($95\%$ CL),
although some analyses also add Lyman-$\alpha$ data (see next
subsection).

\begin{table*}
\caption{Upper bounds on $M_\nu$ (eV, 95\% CL) from recent
analyses combining CMB, LSS (including bias and/or Ly$\alpha$)
and other cosmological data.\label{table:mass_LSS_2}}
\begin{center}
\begin{tabular}{lcl}
\hline
Ref.\ & Bound &  Data used 
(in addition to WMAP1 except for Ref.\  \cite{WMAP3:Spergel})\\
\hline 
\cite{Spergel:2003cb}   & 0.63 & ACBAR, CBI, 2dF-gal, 2dF-bias\\
{}                      & 0.68 & previous + 
                        Ly$\alpha$ \cite{Croft:2000hs,Gnedin:2001wg}\\
%
%Seljak et al
\cite{Seljak:2004xh}  & 0.54 & SDSS-gal, SNIa, SDSS-Ly$\alpha$\\
{}                    & 0.42 & previous + SDSS-bias\\
\cite{Fogli:2004as} & 0.47 & other CMB,
%\footnote{BOOMERANG98, VSA, CBI, DASI \& MAXIMA}
2dF-gal, HST, SNIa, SDSS-Ly$\alpha$\\
\cite{Hannestad:2004bu} & 0.65 & SDSS-gal, HST, 
                       Ly$\alpha$ \cite{Croft:2000hs,Gnedin:2001wg}\\
\cite{MacTavish:2005yk} & 0.48 & 
other CMB, %\footnote{BOOMERANG03, VSA, ACBAR,CBI, DASI \& MAXIMA},
2dF-gal, SDSS-gal, bias $b=1.0 \pm 0.10$\\
\cite{Goobar:2006xz}  & 0.44 & 
other CMB, 2dF-gal, SDSS-gal, HST, SNIa, SDSS-BAO\\
{}                    & 0.30 & previous + SDSS-Ly$\alpha$\\
\cite{WMAP3:Spergel} & 0.91 & 
WMAP3, SDSS-gal, SDSS-bias\\
                     & 0.68 & 
previous + other CMB, 2dF-gal, SNIa\\
\hline
\end{tabular}
\end{center}
\vspace{0.5cm}
\end{table*}

Finally, a galaxy redshift survey performed in a large
volume can also be sensitive to the imprint created by the baryon
acoustic oscillations (BAO) at large scales on the power spectrum of
non-relativistic matter, that we discussed in Sec.\
\ref{subsec:pert_no_nu} (see Fig.\ \ref{fig_powspec}). Since baryons
are only a subdominant component of the non-relativistic matter, the
BAO feature is manifested as a small single peak in the galaxy
correlation function in real space that was recently detected from the
analysis of the SDSS luminous red galaxy (LRG) sample
\cite{Eisenstein:2005su} at a separation of $100\,h^{-1}$ Mpc.  The
observed position of 
this baryon oscillation peak provides a way to
measure the angular diameter distance out to the typical LRG redshift
of $z=0.35$, which in turn can be used to constrain the parameters of
the underlying cosmological model.
The SDSS measurement of the angular diameter distance at $z=0.35$
[SDSS-BAO] was included in the analysis of
Ref.\ \cite{Goobar:2006xz} to get a bound of $0.44$ eV ($95\%$ CL) on
the total neutrino mass $M_\nu$, as shown in Table 
\ref{table:mass_LSS_2}.

\subsection{Lyman-$\alpha$ forest}

The matter power spectrum on small scales can also be inferred from
data on the so-called Lyman-$\alpha$ forest. This corresponds to the
Lyman-$\alpha$ absorption of photons traveling from distant quasars
($z\sim 2-3$) by the neutral hydrogen in the intergalactic medium. As
an effect of the Universe expansion, photons are continuously
red-shifted along the line of sight, and can be absorbed when they
reach a wavelength of 1216 \AA \ in the rest-frame of the intervening
medium.  Therefore, the quasar spectrum contains a series of
absorption lines, whose amplitude as a function of wavelength traces
back the density and temperature fluctuations of neutral hydrogen
along the line of sight.  It is then possible to infer the matter
density fluctuations in the linear or quasi-linear regime (see e.g.\
\cite{bi,Viel:2001hd,Zaldarriaga:2001xs,Hui:1996fh}).

In order to use the Lyman-$\alpha$ forest data, one needs to recover
the matter power spectrum from the spectrum of the transmitted flux, a
task that requires the use of hydro-dynamical simulations for the
corresponding cosmological model. This matter power spectrum is again
sensitive to the suppression of growth of mass fluctuations caused by
massive neutrinos. In 1999, Ref.\ \cite{Croft:1999mm} obtained an
upper bound of $m_\nu <5.5$ eV (for one massive neutrino species) from
the combined analysis of Lyman-$\alpha$ data and other cosmological
measurements.

Given the various systematics involved in the analysis pipeline, the
robustness of Lyman-$\alpha$ forest data is still a subject of intense
discussion between experts. Among these systematics, we can cite the
uncertainty existing on: the mean absorption by the intergalactic
medium (the effective optical depth), the temperature-density relation
within the absorbing gas, the effect of galactic winds, or the proper
use and accuracy of hydro-dynamical simulations in the mildly
non-linear regime. Different authors tend to describe or to deal with
these uncertainties in slightly different ways
\cite{Croft:2000hs,Gnedin:2001wg,Viel:2004bf,McDonald:2004eu}, an
indication that the field might not be fully mature at the moment.  In
addition, the way in which neutrinos are implemented in
hydro-dynamical simulations is generally rather simplistic.

Some of the results shown in Table \ref{table:mass_LSS_2} were
obtained adding Lyman-$\alpha$ data to CMB and other LSS data.  For a
free bias, one finds that Lyman-$\alpha$ data help to reduce the upper
bounds on the total neutrino mass: using the SDSS Lyman-$\alpha$
forest \cite{McDonald:2004eu,McDonald:2004xn}, Ref.\
\cite{Seljak:2004xh} finds $M_\nu < 0.54$ eV (95\% CL), while Ref.\
\cite{Hannestad:2004bu} quotes $0.65$ eV using Lyman-$\alpha$ data
from \cite{Croft:2000hs,Gnedin:2001wg}.  At least two
recent analyses combined results on the bias and the Lyman-$\alpha$
forest: in Ref.\ \cite{Seljak:2004xh} the addition of SDSS-bias
allowed to reduced the upper bound on $M_\nu$ to $0.42$ eV, while the
first year
analysis of WMAP collaboration \cite{Spergel:2003cb} found a slightly
{\it weaker} bound of $0.68$ eV when adding 2dF-bias and
Lyman-$\alpha$ data from \cite{Croft:2000hs,Gnedin:2001wg}.  Finally,
Ref.\ \cite{Goobar:2006xz} found the upper bound $M_\nu < 0.30$ eV
(95\% CL) adding simultaneously SDSS-Ly$\alpha$ and SDSS-BAO data.

Thus, at present, the impact of Lyman-$\alpha$ data on the neutrino
mass bounds is comparable to that of galaxy bias measurements, if not
weaker.  We can conclude that both the Lyman-$\alpha$ data and the
light-to-mass bias determination are powerful tools for getting
information on neutrino masses, but due to the many involved
systematics, the current bounds obtained with them are probably not as
robust as those in Tables \ref{table:mass_CMB} and
\ref{table:mass_LSS}.

\subsection{Summary of current bounds}
\label{subsec:scb}

\begin{figure}[tb]
\begin{center}
\includegraphics[width=.99\textwidth]{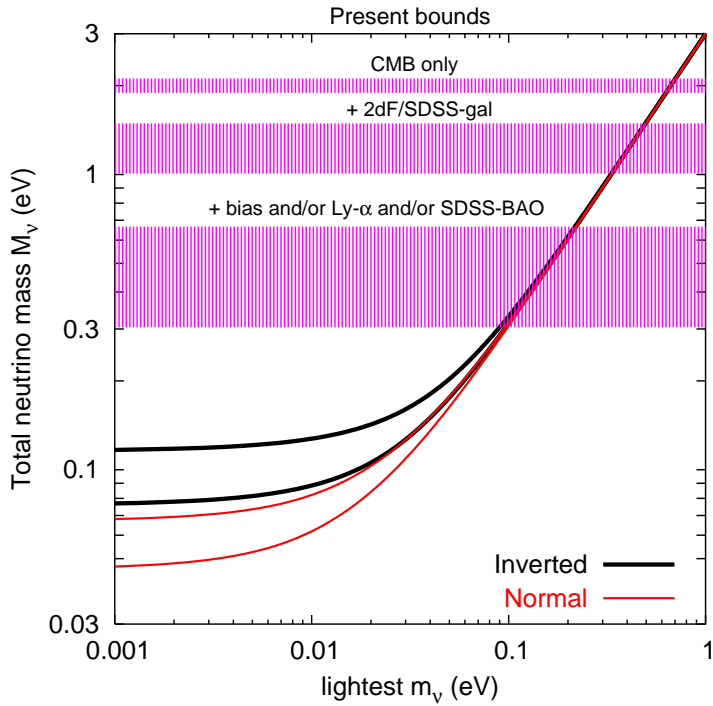}
\end{center}
\caption{\label{fig_current} Current upper bounds (95\%CL) from
cosmological data on the sum of neutrino masses, compared to the
values in agreement at a $3\sigma$ level with neutrino oscillation
data in Eq.\ (\ref{oscpardef}).}
\end{figure}

The results shown in Tables
\ref{table:mass_CMB}-\ref{table:mass_LSS_2} are representative of an
important fact: a single cosmological bound on neutrino masses does
not exist. A graphical summary is presented in Fig.\
\ref{fig_current}, where the cosmological bounds correspond to the
vertical bands, which were grouped according to the ranges quoted in
Tables \ref{table:mass_CMB}, \ref{table:mass_LSS} and
\ref{table:mass_LSS_2}, respectively. The thickness of these bands
roughly describe the spread of values obtained from similar
cosmological data: $2-3$ eV for CMB only, $0.9-1.7$ eV for CMB and
2dF/SDSS-gal or $0.3-0.9$ eV with the inclusion of a
measurement of the bias and/or Lyman-$\alpha$ forest data and/or
the SDSS measurement of the baryon oscillation peak.

One can see from Fig.\ \ref{fig_current} that current cosmological
data probe the region of neutrino masses where the 3 neutrino states
are degenerate, with a mass $M_\nu/3$.  This mass region is
conservatively bounded to values below approximately $1$ eV from CMB
results combined only with galaxy clustering data from 2dF and/or SDSS
(i.e.\ the shape of the matter power spectrum for the relevant
scales). The addition of further data leads to an improvement of the
bounds, which reach the lowest values when data from Lyman-$\alpha$
and/or the SDSS measurement of the baryon oscillation
peak are included or the bias is fixed. In such cases the
contribution of a total neutrino mass of the order
$0.3$-$0.6$ eV seems already disfavoured, in particular covering the
recent positive $0\nu2\beta$ signal from
\cite{Klapdor-Kleingrothaus:2004wj} (of course with the corresponding
caveats, see e.g.\ the discussion in \cite{Fogli:2005cq}).

There exist other measurements of cosmological parameters that were
not included in the analyses we have reviewed so far, although they
could play a role for neutrino mass determination. In particular, we
have seen that any measurement that fixes the amplitude of the matter
power spectrum, such as the bias, would be very important. For
instance, the parameter $\sigma_8$ --which gives the linear rms mass
fluctuations in spheres of radius $8\,h^{-1}$Mpc-- can be measured,
for instance, from the determination of cluster masses through X-ray
observations (see e.g.\ \cite{Pierpaoli:2002rh}) or from weak lensing
(see e.g.\ \cite{Semboloni:2005ct}).  If various methods happened to
converge towards an accurate determination of $\sigma_8$, the
determination of neutrino masses would improve significantly, as
discussed in Ref.\ \cite{Pierpaoli:Mykonos}.  Since $\sigma_8$ probes
the amplitude of matter fluctuations on scales of the order a few Mpc
which are deeply affected by neutrino free-streaming, a large value of
$\sigma_8$ would push down the upper bound on $M_\nu$, while a low
value would potentially bring evidence for non-zero masses.  However,
at the moment, the ensemble of $\sigma_8$ determinations is rather
spread (see Table 5 in \cite{Tegmark:2003ud}), suggesting that not all
systematics errors are properly modelled.  

One of the first analysis of neutrino mass bounds from cosmology used
data on $\sigma_8$ from cluster abundance to obtain an upper bound of
order a few eV \cite{Fukugita:1999as}.  Another possibility was
recently explored in \cite{Kahniashvili:2005sg}, using data on the
redshift-evolution of the number density of massive galaxy clusters to
find the bound $M_\nu < 2.4$ eV (95\% C.L.).
%, accounting for the uncertainties in the measurements.
Ref.\ \cite{Allen:2003pt} used another probe of the small-scale power
spectrum amplitude from X-ray cluster data, and found some marginal
evidence for a non-zero neutrino total mass,
$M_\nu=0.56^{+0.30}_{-0.26}$ eV at 68\% C.L.  This result is related
to the fact that the value of $\sigma_8$ preferred by these X-ray data
%, $\sigma_8=0.74^{+0.12}_{-0.07}$ at 68\% C.L, 
is as small as to contradict most other determinations of this
parameter, for instance from weak lensing.  This example shows that
the sensitivity of cosmological observations to neutrino masses is a
powerful tool, but its implications should not be extracted without
care.

\subsection{Extra parameters}
\label{subsec:extraparam}

Assuming that the relic neutrinos are standard, we have seen that the
limits mainly depend on the cosmological data used in the analysis,
although minor differences on the quoted bounds exist even when using
similar data. These differences arise because not all the analyses in
the Tables used the same set of cosmological parameters and priors
(such as the assumption of a flat Universe).  It is thus important to
test whether the impressive cosmological bounds on neutrino masses
change much if additional cosmological parameters, beyond those
included in the minimal $\Lambda$CDM, are allowed. This could be the
case whenever a new parameter degeneracy with the neutrino masses
arises.

\begin{figure}[t]
\begin{center}
\includegraphics[width=.85\textwidth]{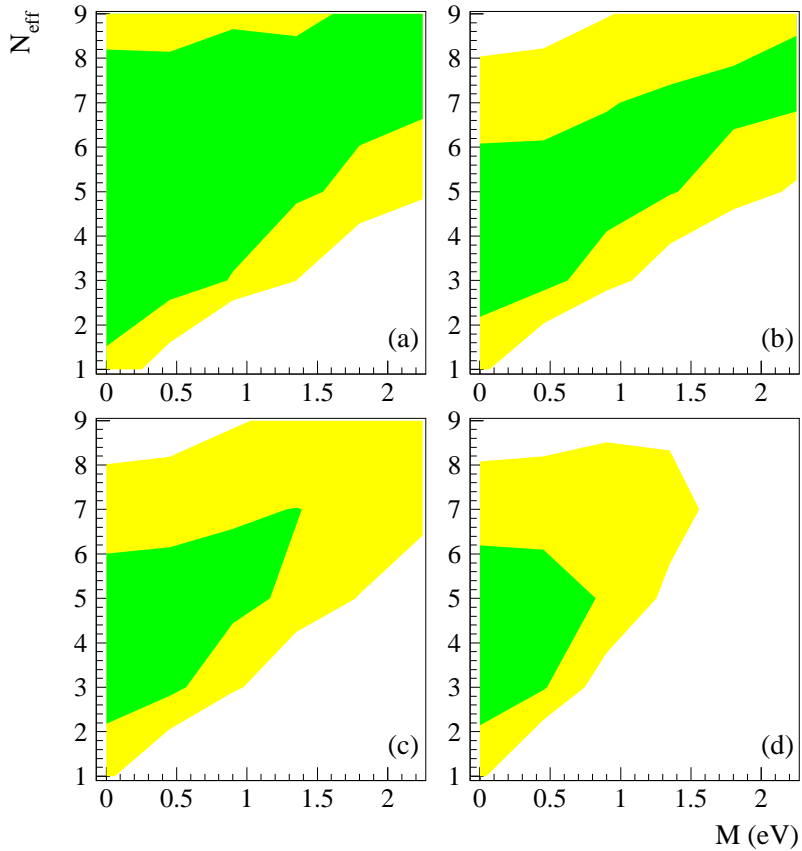}
\caption{\label{fig:Nandm} Two-dimensional likelihood in $(N_{\rm
eff}, M_\nu)$ space, marginalized over the other cosmological
parameters of the model. We plot the 1$\sigma$ (green / dark) and
2$\sigma$ (yellow / light) allowed regions. Here we used CMB 
(WMAP1 \&
ACBAR) and LSS (2dF \& SDSS) data, adding extra priors on $h$ (HST)
and $\Omega_{\rm m}$ (SN99 or SN03) as follows: (a) no priors, (b)
HST, (c) HST+SN99, (d) HST+SN03. For details, see ref.\
\protect\cite{Crotty:2004gm}.}
\end{center}
\end{figure}

An interesting case of degeneracy between cosmological parameters is
that between neutrino masses and the radiation content of the universe
(parametrized via the effective number of neutrinos $N_{\rm
eff}$). The extra radiation partially compensates the effect of
neutrino masses, provided that other cosmological parameters such as
$\Omega_{\rm m}$ and $h$ are varied, leading to a less stringent bound
on $M_\nu$
\cite{Hannestad:2003xv,Crotty:2004gm,Hannestad:2003ye,Lattanzi:2005qq}.
This parameter degeneracy is shown in Fig.\ \ref{fig:Nandm}, taken
from Ref.\ \cite{Crotty:2004gm}. For instance, when one considers a
model with four instead of three species of massive neutrinos, the
upper bound found from CMB and 2dF/SDSS-gal is relaxed from $0.8$ to
$1.2$ eV when one of the states is much heavier than the others. These
results are interesting for the 4-neutrino mass schemes that also
incorporate the results of the LSND experiment
\cite{Aguilar:2001ty}, that we discussed in Sec.\ \ref{sec:numasses}.
At present, the LSND regions in the space of oscillation
parameters are not yet completely disfavoured\footnote{However, a very
recent analysis \cite{Dodelson:2005tp} showed that in a 4-neutrino
scenario, the heavier neutrino mass is bounded to values smaller than
$0.55$ eV ($95\%$ CL) when data from CMB, LSS and Lyman-$\alpha$ are
included.} by cosmological data.

Recently a degeneracy between neutrino masses and the parameter $w$,
that characterizes the equation of state of the dark energy component
$X$ ($p_X= w \rho_X$), was studied in \cite{Hannestad:2005gj}. For
constant values $w <-1$, in the so-called phantom energy regime (which
could describe particular scenarios based on string theory or modified
gravity), it was shown that the neutrino mass bound can be relaxed to
$M_\nu < 1.48$ eV (95\% C.L.), compared with $M_\nu < 0.65$ eV for
fixed $w =-1$. Similar conclusions were obtained by the WMAP team (see
Fig.\ 18 in \cite{WMAP3:Spergel}). Ref.\ \cite{Hannestad:2005gj}
studied the effect of this parameter degeneracy on the analysis of
future cosmological data (see also \cite{Ichikawa:2005hi} for models
with a time-varying equation of state of dark energy). Finally, Ref.\
\cite{Goobar:2006xz} showed that data on baryon acoustic oscillations
(see Sec.\ \ref{subsec:present_GRS}) helps in breaking the $m_\nu-w$
degeneracy.

Finally, the neutrino mass bounds could certainly be relaxed by
assuming that the primordial power spectrum has a non-trivial shape or
Broken Scale Invariance (BSI). This eventuality cannot be excluded,
for instance if a phase transition takes place during inflation. Such
an assumption opens so many degrees of freedom in the parametrization
of cosmological models that a systematic study is almost impossible.
However, some authors \cite{Blanchard:2003du} used this possibility in
combination with a non-zero neutrino mass for building a cosmological
scenario without any form of dark energy, that could fit most of the
data available at the time of publication, except for SNIa data. This
model is now ruled out by the combination of WMAP with the SDSS
galaxy-galaxy correlation function, but it illustrates the fact that
if one is ready to add more degrees of freedom in cosmological
scenarios (in the case of Ref.\ \cite{Blanchard:2003du}, they were
actually not so many), then there is most probably a way to evade the
bounds presented in this review.

\subsection{Non-standard relic neutrinos}
\label{subsec:nonstand}

Finally, let us remind the reader that the cosmological implications
of neutrino masses could be very different if the spectrum or
evolution of the cosmic neutrino background was non-standard. Our aim
here is not to give an exhaustive list of all possibilities
(many results on non-standard neutrinos were reviewed in
\cite{Dolgov:2002wy}), but only to briefly summarize some recent
works.

For instance, the bounds on neutrino masses would be modified if relic
neutrinos presented a momentum spectrum significantly different from the
equilibrium Fermi-Dirac distribution of Eq.\ (\ref{FD}). For
instance, neutrinos could violate the Pauli exclusion principle and
obey Bose-Einstein statistics with important cosmological and
astrophysical implications
\cite{Dolgov:2005qi,Dolgov:2005mi}. Non-equilibrium neutrino spectra
can also be produced in low-reheating scenarios
\cite{Kawasaki:1999na,Kawasaki:2000en,Giudice_reaheat,Adhya:2003tr,Hannestad_reheat,Ichikawa:2005vw},
from the decay of massive neutrinos into relativistic products (see
e.g.\ \cite{Hannestad_nudecay,Kaplinghat_decay}) or as the result of
active-sterile neutrino oscillations after decoupling (see
\cite{Kirilova:2002ss,Abazajian:2004aj} and references therein).  A
recent analysis of cosmological data including large non-thermal
corrections in the
neutrino spectra was done in Ref.\ \cite{Cuoco:2005qr}.

We conclude giving two examples of models where the cosmological
bounds on neutrino masses that we have discussed can be evaded.  In
the first case, massive neutrinos are strongly coupled to a light
scalar field, in such a way that they annihilate as soon as they
become non-relativistic and do not affect the formation of LSS
\cite{Beacom:2004yd}. The viability of this kind of
models depends on the cosmological data included in the analysis and
on the number of neutrino species strongly interacting (see the
discussions in Refs.\
\cite{Hannestad:2004qu,Hannestad:2005ex,Bell:2005dr}). 

The second possibility, described in \cite{Hung:2000yg,Fardon:2003eh}
and many other recent works (see e.g.\ the discussion in
\cite{Copeland:2006wr}), concerns neutrinos coupled to the dark energy
so that the dark energy density is a function of the neutrino mass
(Mass-Varying Neutrinos or MaVaNs). Then, neutrino masses are only
relevant for the very recent stages of the Universe, so that no bounds
can be placed from high-redshift cosmological data. Note, however,
that this is not true for all MaVaNs scenarios, see for instance
\cite{Brookfield:2005bz}.

\section{Future sensitivities and new experimental techniques}
\label{sec:future}

If the characteristics of future experiments are known with some
precision, it is possible to assume a ``fiducial model'', i.e.\ a
cosmological model that would yield the best fit to future data, and
to estimate the error bar on a particular parameter that will be
obtained after marginalizing the hypothetical likelihood distribution
over all the other free parameters (see Sec.\
\ref{sec:stat}). Technically, the simplest way to forecast this error
is to compute a Fisher matrix, as explained in the next
subsection. This technique has been widely used in the literature, for
many different models and hypothetical datasets. Here we will focus on
the results for $\sigma(M_{\nu})$, the forecast $68\%$ CL error on the
total neutrino mass, assuming various combinations of future
observations (CMB anisotropies, galaxy redshift surveys, weak lensing
surveys, \ldots) and various cosmological models with fiducial
parameter values. Many papers derived some predictions for
$\sigma(M_{\nu})$; we will summarize here the main results and see
whether the expected sensitivities will be sufficient for detecting
the total neutrino mass, even if it is close to the minimum values
guaranteed by present data on flavour neutrino oscillations.

\subsection{Fisher matrix}
\label{fisher-matrix}

In a Bayesian analysis, the probability distribution in the space of
cosmological parameters is inferred from the likelihood ${\mathcal
L}(x_n|\theta_i)$ of the data $x_n$ given a theoretical model
described by a set of parameters $\theta_i$ (see Sec.\
\ref{sec:stat}). If we assume that the maximum likelihood ${\mathcal
L}_{\rm max}$ is obtained for a particular model $\theta_i^0$ (the
``fiducial model''), and if we approximate the likelihood function as
a multivariate Gaussian in the vicinity of the maximum, the
symmetrical error on each parameter can be inferred from the Fisher
matrix, defined as
\begin{equation}
F_{ij}\equiv 
\left.
- \frac{\partial^2 \ln {\mathcal L}}{\partial \theta_i \partial \theta_j}
\right|_{\, \theta_i^0}~.
\label{fisher_def}
\end{equation}
The 1$\sigma$ (or 68\% confidence) error bar on a particular parameter
${\theta}_j$ assuming that all other parameters are fixed to the
fiducial values ${\theta}_i^0$ would be given by
$(F_{jj})^{-1/2}$. However, the really interesting quantity is the
error on ${\theta}_j$ assuming that the other parameters are unknown
(and simply marginalized out by integrating over the likelihood). In
that case, the error can be shown to be~\cite{Tegmark:1996bz}
\begin{equation}
\sigma(\theta_j)=(F^{-1})_{jj}^{1/2}~.
\end{equation}
Geometrically, and in the multivariate Gaussian approximation, the
Fisher matrix gives the coefficients of the ellipsoids $F_{ij}
\theta_i \theta_j = f(c)$ corresponding to given confidence levels $c$
in parameter space. Therefore, the study of its eigenvectors and
eigenvalues gives an idea of the linear parameter combinations with
the best and poorest determination (the latter correspond to parameter
degeneracies).

In many cases one refers to data points $x_i$ ($i=1,...,n$) with a
Gaussian probability distribution centered in zero
\begin{equation}
{\mathcal L}\left(\vec{x}|\vec{\theta}\right) = [ (2 \pi)^n |C| ]^{-1/2}
e^{-\frac{1}{2}  \vec{x}^t C \vec{x}}~,
\label{gaussian_likelihood}
\end{equation}
where the ``data covariance matrix'' $C$ contains the variance
expected for the data, $C \equiv \langle \vec{x} \vec{x}^t \rangle$,
after a detailed modelling of all possible contributions. Typically,
the data covariance matrix will be the sum of the theoretical model
prediction (for parameters $\vec{\theta}$) and of the noise variance
estimated from the instrumental characteristics. In general, the data
covariance matrix is non-diagonal and the data points are correlated
with each other.  A purely algebraic calculation shows that the
Gaussian likelihood of Eq.\ (\ref{gaussian_likelihood}) inserted in
Eq.\ (\ref{fisher_def}) leads to the following expression for the
Fisher matrix~\cite{Tegmark:1996bz}
\begin{equation}
F_{ij} = \frac{1}{2} \, \, \mathrm{Trace} 
\left[C^{-1} \frac{\partial C}{\partial \theta_i}
 C^{-1} \frac{\partial C}{\partial \theta_j} \right]~.
\label{fisher_gaussian}
\end{equation}

\subsection{Future CMB experiments}

The quality of the data from the WMAP satellite
\cite{Spergel:2003cb,WMAP3:Spergel}, complemented by the results of
other experiments at smaller angular scales such as those listed in
section \ref{sec:present_CMB}, has shown the importance of CMB data as
a probe of cosmological parameters. There are several other projects
in operational or design stage, aimed mainly at improving the
sensitivity towards small scales and/or polarization anisotropies. The
neutrino mass is not the primary target of these experiments, because
as we have already seen in Sec.\ \ref{subsec:sum_mass-effect}, small
masses only have a modest signature on the CMB. But still, we know
that there is a background effect proportional to $\omega_{\nu}$ which
renders the primary CMB anisotropies sensitive --at least in
principle-- to very small values of the total neutrino mass.  Instead,
typical targets are the scalar tilt and the tensor-to-scalar ratio, in
order to better constrain the predictions of inflationary theories.
However, in order to measure with precision the neutrino masses, it
will be useful to combine some large scale structure observations with
the best possible CMB dataset in order to reduce as much as possible
parameter degeneracies. In this section we review the Fisher matrix
computation and expected sensitivity of future CMB experiments alone;
in the next sections the CMB contribution to the Fisher matrix will be
combined with that from other hypothetical observations.

The CMB temperature maps can be decomposed in multipoles
\begin{equation}
\frac{\Delta T}{T} (\theta,\phi) = \sum_{l,m} a_{lm}^T Y_{lm}(\theta,\phi)
\label{dT} \, ,
\end{equation}
and similarly the polarization map are described by multipoles
$a_{lm}^E$ and $a_{lm}^B$. The data covariance matrix
$C = \langle a_{lm}^X a_{l'm'}^{X'} \rangle$ receives contributions
from: 
\begin{itemize}
\item primary anisotropies, for which the expected variance
is predicted by the cosmological perturbation theory:
$\langle a_{lm}^X a_{l'm'}^{X'} \rangle = \delta_{ll'} \delta_{mm'}
C_l^{XX'}$. For parity reasons the non-vanishing power spectra
are $C_l^{TT}$, $C_l^{EE}$, $C_l^{BB}$ and
$C_l^{TE}$.\\

\item secondary anisotropies imprinted at redshifts $z \ll 1100$, like
the late Integrated Sachs-Wolfe (ISW) effect and the weak lensing
induced by nearby clusters, as well as various foreground
contaminations. The late ISW effect is usually included in the
theoretical power spectrum computation, while the lensing effect is
small enough to give a negligible contribution to $C$ (excepted for
the small-scale B-mode variance). Here we will neglect the lensing
distortions, which will be the object of Sec.\ \ref{cmb-lensing}.
Most of the foreground contamination is expected to be accurately
removed from the raw data using a multi-frequency analysis (since most
foreground signals do not have a blackbody distribution).\\

\item experimental noise, which is usually parametrized by two
quantities: the angular resolution and the sensitivity per pixel
(here, a pixel represents the smallest area in a map on which an
independent measurement is performed). Intuitively, a given angular
resolution implies that the error-bar will grow dramatically below a
given scale, and the finite sensitivity per pixel implies that even in
absence of a primary signal, the covariance matrix would get some
non-zero contribution at a given level. Technically, the angular
resolution is expressed as an angle $\theta_b$, the Full-Width at
Half-Maximum (FWHM) of the instrument beam. The sensitivity (or noise
squared variance) per pixel $\Delta_X^2$ (for $X=T,E,B$) is usually
given in units of a squared temperature, and the dimensionless
sensitivity per pixel $\bar{\Delta}_X^2$ is obtained dividing by the
square of $T_0\simeq2.726$ K.  The solid angle of a pixel can be
approximated by $\theta_b^2$, so that the dimensionless sensitivity
per steradians reads $(\theta_b^2 \bar{\Delta}_X^2)$. For one
instrument, the noise variance can be shown to be
\cite{Tegmark:1997vs}
\begin{equation}
\langle a_{lm}^X a_{l'm'}^{X'} \rangle = \delta_{ll'} \delta_{mm'}
\delta^{X X'} N_l^{XX}~, \qquad
N_l^{XX} \equiv (\theta_b^2 \bar{\Delta}_X^2)B_l^{-2}~,
\label{define_CMB_noise}
\end{equation}
where the Gaussian beam window function
$B_l\equiv\exp[-l(l+1)\theta_b^2 \, / \, (16\ln2)]$ has the effect of
cutting the sensitivity for $l \gg \theta_b^{-1}$. Note that for $X
\neq X'$ the variance vanishes, since the noise in one map (for mode
$T$, $E$ or $B$) is expected to be uncorrelated with the noise in
another map.  Finally, most experiments consist of various detectors,
for which a global sensitivity can be built by summing the inverse
variances
\begin{equation}
\langle a_{lm}^X a_{l'm'}^{X'} \rangle = 
\left( \sum_i \left[ \theta_b^{-2} \bar{\Delta}_X^{-2} B_l^{2} \right]_i 
\right)^{-1}~.
\end{equation}
Detectors are generally grouped in {\it channels}, i.e.\ ensembles of
detectors working at a given frequency. An efficient foreground
extraction requires a large number of channels, at the expense of
decreasing the number of detectors per channel, and hence increasing
the noise per channel.
\end{itemize}
The theoretical and noise variance are usually assumed to be
uncorrelated, which allows to express the covariance matrix in a
rather simple way.  If we represent the data as a vector $\vec{x}={\bf
x}_{lm}=(a^T, a^E, a^B)_{lm}$, where the indexes $(l,m)$ run over all
multipoles, the full data covariance matrix can be decomposed in
diagonal blocks $C_{lm}$ representing the covariance $\langle {\bf
x}_{lm} {\bf x}_{lm}^t \rangle$ for a given $(l,m)$
\begin{equation}
C_{lm}=C_l=\left(
{\rm
\begin{tabular}{ccc}
$C_l^{TT}+N_l^{TT}$ & $C_l^{TE}$ & 0 \\
$C_l^{TE}$ & $C_l^{EE}+N_l^{EE}$ & 0 \\
0 & 0 & $C_l^{BB} +N_l^{BB}$
\end{tabular}
}
\right)~.
\end{equation}
The index $m$ does not appear in this matrix, since all multipoles
$(l,m)$ with the same $l$ share the same variance.  If we consider the
data as Gaussian (which is a very good approximation except for
small-scale $B$-mode multipoles), we can use
Eq.\ (\ref{fisher_gaussian}). The first step of the calculation is to
estimate the number of multipoles $(l,m)$ with a fixed $l$. For
full-sky experiments, $m$ runs from $-l$ to $l$ and this number is
$(2l+1)$. For experiments probing a fraction $f_{\rm sky}$ of the sky,
the number of independent multipole measurement at a given $l$ is
close to $(2l+1)f_{\rm sky}$. This leads to
\begin{equation}
F_{ij}^{\rm CMB} = \frac{1}{2} \, \sum_{l=2}^{l_{\rm max}} (2l+1)f_{\rm sky} 
\mathrm{Trace} 
\left[C_l^{-1} \frac{\partial C_l}{\partial \theta_i}
 C_l^{-1} \frac{\partial C_l}{\partial \theta_j} \right]~,
\end{equation}
which can be further reduced to an expression of the type
\cite{Jungman:1995av}
\begin{equation}
F_{ij}^{\rm CMB} = \sum_{l=2}^{l_{\rm max}} \sum_{\alpha,\beta}
\frac{\partial C_l^{\alpha}}{\partial \theta_i}
({\rm Cov}_l)_{\alpha \beta}^{-1}
\frac{\partial C_l^{\beta}}{\partial \theta_j}~,
\label{gaussian_fishmat}
\end{equation}
where $\alpha$ and $\beta$ run over $TT$, $EE$, $TE$, $BB$ and ${\rm
Cov}_l$ is the ``spectra covariance matrix''. For space reason we do \
not give here the full expression of ${\rm Cov}_l$ (see e.g.\  Ref.\
\cite{Eisenstein:1998hr}). As an example, the first coefficient of the
matrix reads
\begin{equation}
({\rm Cov}_{l})_{TT,TT} = \frac{2}{(2l+1)f_{\rm sky}} (C_l^{TT}+N_l^{TT})^2~.
\end{equation}
The factor $(2l+1)f_{\rm sky}$ which appears in the denominator of
${\rm Cov}_l$ accounts for the effect of {\it cosmic variance}: for
large $l$ values, an experiment makes many independent measurements
(corresponding to several $m$'s) and the variance in absence of noise
is much smaller than the power spectrum $C_l^{XX}$ of a single
multipole $a_{lm}^X$. For small $l$, there are only few independent
measurement and the variance is of the same order of magnitude as
$C_l^{XX}$.

Let us apply this machinery to concrete cases. In Table
\ref{tableexp}, we present the expected sensitivity of a
non-exhaustive list of future CMB experiments.  The design of
projected experiments is subject to constant evolution, and this
compilation should be regarded as indicative only. We did not take the
numbers from some official documentation edited by each experiment,
but from the papers cited below, which worked out some neutrino mass
error forecasts on the basis of the numbers presented here. We refer to
these papers for more details and references concerning the
sensitivity parameters.

\begin{table*}
\caption{Experimental parameters of a few CMB projects, used in the
neutrino mass forecast papers mentioned in this section. Here $f_{\rm
sky}$ is the observed fraction of the sky, $\nu$ is the channel
frequency in GHz, $\theta_b$ measures the FWHM of the beam in
arc-minutes and $\Delta_{T,P}$ is the square root of the sensitivity
per pixel in $\mu$K for temperature and polarization.
\label{tableexp}}
\begin{center}
\begin{tabular}{cccccc}
\hline
Experiment & $f_{\rm sky}$ & $\nu$ & $\theta_b$ & $\Delta_T$ & $\Delta_P$\\
\hline
SPT \cite{Kaplinghat:2003bh} & 0.1 & 217 & 0.9' & 12 & 17\\
\hline
BICEP \cite{LPPP} & 0.03
    & 100 & 60' & 0.33 & 0.47\\
&   & 150 & 42' & 0.35 & 0.49\\
QUaD \cite{LPPP} & 0.025
    & 100 & 6.3' & 3.5 & 5.0\\
&   & 150 & 4.2' & 4.6 & 6.6\\
\hline
BRAIN \cite{LPPP} & 0.03 
    & 100 & 50' & 0.23 & 0.33\\
&   & 150 & 50' & 0.27 & 0.38\\
&   & 220 & 50' & 0.40 & 0.56\\
ClOVER \cite{LPPP} & 0.018
    & 100 & 15' & 0.19 & 0.30\\
&   & 143 & 15' & 0.25 & 0.35\\
&   & 217 & 15' & 0.55 & 0.76\\
\hline
{\sc Planck} \cite{LPPP} & 0.65
    &  30 & 33'  &  4.4 &  6.2\\
&   &  44 & 23'  &  6.5 &  9.2\\
&   &  70 & 14'  &  9.8 & 13.9\\
&   & 100 & 9.5' &  6.8 & 10.9\\
&   & 143 & 7.1' &  6.0 & 11.4\\
&   & 217 & 5.0' & 13.1 & 26.7\\
&   & 353 & 5.0' & 40.1 & 81.2\\
&   & 545 & 5.0' & 401  & $\infty$\\
&   & 857 & 5.0' & 18300 & $\infty$\\
\hline
SAMPAN \cite{LPPP} & 0.65
  & 100 & 42' & 0.13 & 0.18\\
& & 143 & 30' & 0.16 & 0.22\\
& & 217 & 20' & 0.26 & 0.37\\
\hline
CMBpol \cite{Kaplinghat:2003bh} & 0.65 & 217 & 3.0' & 1 & 1.4\\
\hline
{\sc Inflation Probe}       & 0.65 &  70 & 6.0' & 0.29 & 0.41 \\
{\em (hypothetical)} \cite{LPPP}  
                      &      & 100 & 4.2' & 0.42 & 0.59 \\
                      &      & 150 & 2.8' & 0.63 & 0.88 \\
                      &      & 220 & 1.9' & 0.92 & 1.30 \\
\hline
\end{tabular}
\end{center}
\end{table*}

After WMAP, we can expect some very interesting results from a large
number of ground-based experiments, mapping the anisotropies only in a
small regions of the sky but usually with excellent sensitivity and
resolution. Most of these experiments are optimized for measuring
E-polarization or even B-polarization anisotropies, which will still
be poorly constrained after the completion of WMAP, and even after
{\sc Planck} in the case of B-polarization.  Among these
collaborations, we can cite the South Pole 
Telescope\footnote{\tt 
http://spt.uchicago.edu/} (SPT, in construction)
\cite{Ruhl}, the Atacama Cosmology Telescope\footnote{\tt
http://www.hep.upenn.edu/$\sim$angelica/act/act.html} (ACT, funded in
January 2004), and two complementary experiments: Background Imaging
of Cosmic Extragalactic Polarization\footnote{\tt
http://www.astro.caltech.edu/$\sim $lgg/bicep$\_$front.html} (BICEP)
\cite{Keating}, designed for large angular scale, and QUest at 
DASI\footnote{\tt 
http://www.astro.cf.ac.uk/groups/instrumentation/projects/quad/}
(QUaD) \cite{Church}, designed for small angular scales.  The second
experiment, which is already collecting data, is composed of the Q and
U Extragalactic Sub-mm Telescope (QUEST) instrument mounted on the
structure of the DASI experiment.  A second set of experiments is
scheduled in Antarctica at the French-Italian Concordia station and in
the Atacama plateau in Chile, for unprecedented precision measurements
of the $B$-mode for $l<1000$ (which is particularly useful for probing
the primordial gravitational waves from inflation): the B-modes
Radiation measurement from Antarctica with a bolometric
INterferometer\footnote{\tt 
http://apc-p7.org/APC$\_$CS/Experiences/Brain/index.phtml}
(BRAIN) \cite{Piat} instrument for measuring large scales, and the Cl
ObserVER\footnote{\tt http://www-astro.physics.ox.ac.uk/$\sim
$act/clover.html} (ClOVER) \cite{Maffei} instrument for intermediate
scales.  ClOVER was approved for funding by PPARC in late 2004 and
could be operational by 2008.  

At that time, a CMB satellite (the third one after COBE and WMAP)
should be already collecting data: the {\sc Planck}\footnote{\tt
http://sci.esa.int/science-e/www/area/index.cfm?fareaid=17} satellite
\cite{Tauber} has already been built and should be launched in late
2007 or early 2008 by the European Space Agency (ESA).  The
temperature sensitivity of {\sc Planck} will be so good that it can be
thought as the ``ultimate'' observation in the sector of temperature
anisotropies. However, it will still be possible to improve the
measurement of E-polarization on small angular scales, and
B-polarization will be poorly constrained by {\sc Planck}.  On
intermediate scales, the ground-based experiments should be quite
efficient, but progress will still be needed on very large scales
(requiring full-sky coverage) and very small scales (requiring both
high resolution and excellent sensitivity).  Beyond {\sc Planck}, at
least two space projects are under investigation. The mini-satellite
SAMPAN (SAtellite to Measure the Polarized ANisotropies)
\cite{Bouchet} is targeted for large scales, and aims at improving the
ClOVER measurement of the $B$-mode for $l<1000$, thanks to its
full-sky coverage and slightly better sensitivity.  The NASA wishes to
launch a more ambitious satellite project in order to make the
``ultimate'' measurement of E-polarization, like {\sc Planck} for
temperature, and a very good measurement of B-polarization on all
scales, with for the first time an instrumental noise smaller than the
B-mode cosmic variance up to $l\sim 1000$ or even maybe $l\sim
1500$. The generic name of the NASA call for projects is {\sc
Inflation Probe}\footnote{\tt
http://universe.gsfc.nasa.gov/program/inflation.html}. Its design is
still very uncertain. A very preliminary project was called CMBpol;
this name is mentioned in some works presented thereafter
\cite{Kaplinghat:2003bh,Song:2003gg,Lesgourgues:2004ps}, with sketchy
characteristics mentioned in Table \ref{tableexp}. There are now more
advanced projects under investigation. Nobody knows which project and
which design will be eventually approved.  We mention in the Table
some {\it hypothetical} characteristics for {\sc Inflation Probe}
suggested in Ref.\ \cite{LPPP}, assuming a satellite project with a
bolometer array and a passively cooled telescope of 3-4 m of
aperture. This is one of the various possibilities which are being
discussed, and it should be regarded as purely indicative.

Some predictions concerning the sensitivity of these experiments (not
combined with other observations) to the neutrino mass are published
in Refs.\
\cite{Eisenstein:1998hr,Hannestad:2002cn,Lesgourgues:2004ps,LPPP} for
{\sc Planck} and in Ref.\ \cite{LPPP} for the combinations BICEP
+QUaD, BRAIN+ClOVER, as well as SAMPAN alone, {\sc Planck}+SAMPAN and
the hypothetical version of {\sc Inflation probe} (see also
\cite{Lesgourgues:1999ej,Popa:2000hv} for forecasts including a large
neutrino asymmetry, now disfavoured as we saw in Sec.\
\ref{sec:basics}). Like for current bounds, forecast errors are not
straightforward to compare from paper to paper, because different
authors use
\begin{itemize}
\item different fiducial models with different fiducial values for the
cosmological parameters.  In Ref.\ \cite{Eisenstein:1998hr}, the
fiducial model which is closer to the current concordance $\Lambda$CDM
model is the one in Table II of that paper; here, we will only quote
results from this table.  There are eleven free parameters, which are
the usual six of the $\Lambda$CDM model, the mass of a single massive
neutrino, the tensor-to-scale ratio $T/S$, the scalar tilt running
$\alpha$, the curvature fraction $\Omega_k$ and the ionized helium
fraction $Y_{\rm He}$. The fiducial value of $M_{\nu}$ is very close
to zero (so that the fiducial CMB spectra are indistinguishable from
those with massless neutrinos).  In Ref.\ \cite{Hannestad:2002cn}, the
number of free parameters is reduced to eight, by removing ($T/S$,
$\alpha$, $Y_{\rm He}$) which are fixed respectively to
$(0,0,0.24)$. There is a single massive neutrino with fiducial mass
$0.07$ eV. Next, in Table III of Ref.\ \cite{Lesgourgues:2004ps},
there are also eight free parameters, but the last one is $Y_{\rm He}$
instead of $\Omega_k$. There are three massive neutrinos with total
mass $M_{\nu}=0.3$ eV distributed according to the normal hierarchy
scheme.  Finally, in Table II of Ref.\ \cite{LPPP}, there authors
marginalize either over eight parameters (the usual six of
$\Lambda$CDM model, the mass of a single massive neutrino and $Y_{\rm
He}$) or over eleven parameters (the same ones, plus the scalar tilt
running $\alpha$, the parameter $w$ of the equation of state of dark
energy, and the effective number $N_{\rm eff}$ of extra relativistic
degrees of freedom). The fiducial value of the neutrino mass is
$M_{\nu} = 0.1$ eV.\\

\item different experimental sensitivities.  For instance, the
modelling of {\sc Planck} in Refs.\ \cite{Eisenstein:1998hr},
\cite{Lesgourgues:2004ps} and \cite{LPPP} are roughly equivalent,
although in \cite{Lesgourgues:2004ps,LPPP} it was updated to the
current instrument design.  Instead, in Ref.\ \cite{Hannestad:2002cn},
the author replaces the expression of the noise variance $N_l^{TT}$,
$N_l^{EE}$ [see Eq.\ (\ref{define_CMB_noise})] by two step functions
equal to zero below some critical multipoles ($l_{\rm max}^T$, $l_{\rm
max}^E$)=(2500,1500) and to infinity above. This is equivalent to
assuming a slightly over-optimistic version of {\sc Planck}.\\

\item different parameter basis and different finite steps for the
computation of the derivatives $\partial C_l^{XX'} / \partial
\theta_i$.  The likelihood ${\mathcal
L}\left(\vec{x}|\vec{\theta}\right)$ can be close to a multi-variate
Gaussian in one basis, and more complicated in another basis - in
which case Eq.\ (\ref{gaussian_fishmat}) may give only a poor
approximation of the Fisher matrix.  The possible non-gaussianity also
implies that the numerical value of the derivative $\partial C_l^{XX'}
/ \partial \theta_i$ depends on the step $\Delta \theta_i$ used for
its calculation.  However, authors usually optimize their choice of
basis and step size [see Ref.\ \cite{Eisenstein:1998hr})] in order to
obtain robust results.
\end{itemize}
In order to illustrate how these different assumptions are translated
into different forecasts from paper to paper, we summarize in Table
\ref{sigma_mnu_CMB} the predictions for {\sc Planck} published by the
four references
\cite{Eisenstein:1998hr,Hannestad:2002cn,Lesgourgues:2004ps,LPPP}.  In
the case of Ref.\ \cite{LPPP} we quote two numbers, which represent
some optimistic or pessimistic assumptions concerning the subtraction
of astrophysical foregrounds from the observed CMB maps: these two
values, which are not very different from each other, are supposed to
bracket the true realistic 1$\sigma$ error.
\begin{table}
\caption{Forecast error $\sigma(M_{\nu})$ in eV using {\sc Planck}
alone: compared results from the literature.  The third column shows
the free parameters over which the final result is
marginalized (``7'' stands for six usual parameters of the minimal
$\Lambda$CDM model, plus total neutrino mass $M_{\nu}$).  The fourth
column gives the assumed fiducial value of the mass in eV.  In the
last two lines, the two numbers correspond to optimistic or
pessimistic assumptions concerning the foregrounds contamination of
the primary CMB signal.
\label{sigma_mnu_CMB}}
\begin{center}
\begin{tabular}{ccccc}
\hline
Ref. & $\sigma(M_{\nu})$ & parameters & fiducial $M_{\nu}$ & 
{\sc Planck} sensitivity \\
\hline
\cite{Eisenstein:1998hr} & 0.3 & 7+\{$\alpha$, $T/S$, $\Omega_{\rm k}$, 
$Y_{\rm He}$\}
& 0 & slightly optimistic \\
%\hline
\cite{Hannestad:2002cn} & 0.07 & 7+\{$\Omega_{\rm k}$\} 
& 0.07 & very optimistic \\ 
%\hline
\cite{Lesgourgues:2004ps} & 0.3 & 7+\{$Y_{\rm He}$\} & 0.3 & up-to-date \\
%\hline
\cite{LPPP} & 0.45-0.49 & 7+\{$Y_{\rm He}$\} & 0.1 & up-to-date \\
%\hline
\cite{LPPP} & 0.51-0.56 & 7+\{$\alpha$, $w$, $N_{\rm eff}$, 
$Y_{\rm He}$\} & 0.1 & up-to-date\\
\hline
\end{tabular}
\end{center}
\vspace{0.5cm}
\end{table}
\begin{table}
\caption{Forecast error $\sigma(M_{\nu})$ in eV using various CMB
experiments or combinations of them. These numbers are all taken from
Ref.\ \cite{LPPP}, which assumes either an eight parameter model (the
usual six ones of $\Lambda$CDM plus $M_{\nu}$ and $Y_{\rm He}$), or an
eleven parameter model (adding $\alpha$, $w$ and $N_{\rm eff}$).  The
fiducial value of the total mass is taken to be $M_{\nu}=0.1$ eV, and in
each case the two numbers correspond to optimistic or pessimistic
assumptions concerning the foregrounds contamination of the primary
CMB signal.}
\label{sigma_mnu_otherCMB}
\begin{center}
\begin{tabular}{ccccc}
\hline
Experiment & 8 parameters & 11 parameters\\
\hline
BICEP+QUaD   &1.3 - 1.6 & 1.5 - 1.9 \\
BRAIN+ClOVER &1.5 - 1.8 & 1.7 - 2.0 \\
{\sc Planck} &0.45 - 0.49 & 0.51 - 0.56 \\
SAMPAN       &0.34 - 0.40 & 0.37 - 0.44 \\
{\sc Planck}+SAMPAN&0.32 - 0.36&0.34 - 0.40 \\
{\sc Inflation probe}  &0.14 - 0.16&0.25 - 0.26\\
\hline
\end{tabular}
\end{center}
\vspace{0.5cm}
\end{table}
The difference between $\sigma(M_{\nu})=0.3$ eV in
\cite{Eisenstein:1998hr} and $\sigma(M_{\nu})=0.07$ eV in
\cite{Hannestad:2002cn} (taken from Fig.\ 3 of this reference)
probably comes from the sketchy and very optimistic modelling of the
noise in the second reference.  The results of
\cite{Eisenstein:1998hr} and \cite{Lesgourgues:2004ps} seem to be
consistent with each other, since in the latter reference the number
of parameters is smaller but the experimental characteristics are more
realistic.  The difference between $\sigma(M_{\nu})=0.3$ eV in
\cite{Lesgourgues:2004ps} and $\sigma(M_{\nu})\simeq 0.45$ eV in
\cite{LPPP}, for identical cosmological models and experimental
characteristics, comes from variations in the fiducial value of the
total neutrino mass, which decreases from $0.3$ to $0.1$ eV.  Indeed,
Ref.\ \cite{Lesgourgues:2004ps} explores the dependence of
$\sigma(M_{\nu})$ on the fiducial value $M_{\nu}$ (see Fig.\
\ref{figIdeal}), and find that when $M_{\nu}$ decreases from
approximately $0.5$ to $0.1$ eV, $\sigma(M_{\nu})$ is multiplied
roughly by a factor two. Finally, the comparison of the two results
obtained in Ref.\ \cite{LPPP} shows that the determination of the
neutrino mass by {\sc Planck} alone does not suffer significantly from
parameter degeneracies which appear when $\alpha$, $w$ or $N_{\rm
eff}$ are included as extra free parameters. Similarly, the comparison
of \cite{Eisenstein:1998hr} and \cite{Lesgourgues:2004ps} suggests
that this also applies for the possible inclusion of $T/S$ and
$\Omega_{\rm k}$.

For other CMB experiments than {\sc Planck}, not combined with other
type of observations, the only available reference is \cite{LPPP},
whose results are summarized in Table \ref{sigma_mnu_otherCMB}. We see
in this table that ground based experiments alone are not helpful for
constraining the neutrino mass: this is due to that fact that with a
small sky coverage, and in spite of their excellent sensitivity, they
have relatively large error-bars (with respect to {\sc Planck}) for
small and intermediate values of $l$. We will see in Sec.\
\ref{cmb-lensing} how this conclusion changes when one assumes that
the results of these experiments are analyzed with the ``lensing
extraction'' technique.

In summary, the satellites {\sc Planck}, SAMPAN or the combination of
both would reach a 1$\sigma$ error in the range
$\sigma(M_{\nu})=0.3-0.4$ eV, depending on the case. This is of the
same order of magnitude as the current neutrino mass bounds presented
in section \ref{sec:present}, when all available CMB and LSS data are
combined with each other. Finally, if the Inflation probe mission can
actually reach the sensitivity assumed in Table \ref{tableexp}, it
will provide a sensitivity ranging from $\sigma(M_{\nu})=0.15$ eV to
$0.25$ eV depending on the number of marginalized
parameters.

\subsection{Future galaxy redshift surveys}

As already discussed in Sec.\ \ref{subsec:present_GRS}, Galaxy
Redshift Surveys (GRS) allow for a reconstruction of the matter power
spectrum modulo a light-to-mass bias factor. The analytic models for
non-linear structure formation show that the relation between the
reconstructed power spectrum and the total matter linear power
spectrum should be scale-independent, $P_{\rm obs}(k) \equiv b^2
P(k)$, at least up to a wavenumber $k_{\rm max}$ where non-linear
corrections start to induce scale-dependent biasing. At redshift zero
and for the concordance $\Lambda$CDM model, this wavenumber is
expected to be around $k_{\rm max}=0.15 h\,$Mpc$^{-1}$.  The usual
approach is to discard any information for $k>k_{\rm max}$ in order to
avoid many complications related to biasing, non-Gaussian statistics
and uncertainties in the predictions of the non-linear matter power
spectrum.

A Fisher matrix can be derived for future surveys following the lines
of Sec.\ \ref{fisher-matrix}.  It was shown by Tegmark
\cite{Tegmark:1997rp} that the Fisher matrix takes a very simple form
under a couple of assumptions.
\begin{itemize}
\item
For $k \leq k_{\rm max}$, one assumes that the power spectrum
reconstructed from the data, $P_{\rm obs}(k)$, is really a tracer of
the total matter power spectrum, not significantly affected by
non-linear corrections or by redshift-space distortions. We recall
that what is measured is the redshift of each galaxy, not its physical
distance; therefore, strictly speaking, the power spectrum is not
reconstructed in three-dimensional real space but in redshift space;
the redshift-space and real-space power spectra do not coincide
exactly, they are related through a function which depends on the
cosmological parameters, and this complication is ignored here in
first approximation.\\

\item
One assumes that the survey is volume-limited rather than
brightness-limited, i.e.\ the spectrum is reconstructed from
data within a volume $V$ in which the selection function $n(r)$, which
represents the number density of observed galaxies, is approximately
constant (instead of being suppressed near the edges due to the fact
that only the brightest galaxies would be seen). Moreover, one assumes
that the density is large enough so that $1/n(r) \leq P(k)$ for all
relevant wavenumbers $k_{\rm min} < k < k_{\rm max}$, which means
that the galaxy correlation function really probes cosmological
information rather than shot noise due to insufficient sampling.
\end{itemize}
Under these assumptions, the Fisher matrix takes a form which is analogous
to its CMB counterpart of Eq.~(\ref{gaussian_fishmat}):
\begin{equation}
F_{ij}^{\rm GRS}= \frac{1}{2} \int_{k_{\rm min}}^{k_{\rm max}}
\frac{\partial P_{\rm obs} (k)}{\partial \ln \theta_i}
\frac{\partial P_{\rm obs} (k)}{\partial \ln \theta_j} 
\frac{4 \pi w(k)}{P_{\rm obs}^2 (k)} ~d \ln k~,
\label{fisher.matrix2}
\end{equation}
where the weight function $w(k)$ of the survey is defined as
\begin{equation}
w(k) = \frac{V}{\lambda^3} = \frac{V}{(2 \pi / k)^3}~.
\label{weight_function}
\end{equation}
In Eq.\ (\ref{fisher.matrix2}), the factor $4 \pi w(k) d \ln k$
accounts for cosmic (or sampling) variance: it gives the number of
independent measurements with wavenumber in the range $[k, k + dk]$
within the survey volume.  This factor is negligible for wavenumbers
smaller than $k_{\rm min}$, the smallest wavenumber probed by the
survey: so, for practical purposes, the integral can be performed
starting from zero. The result of Ref.\ \cite{Tegmark:1997rp} is
actually more general than above, and includes the case where shot
noise is not negligible and where the selection function is not
constant inside the volume $V$. In these cases, one must replace $V$
in Eqs.\ (\ref{fisher.matrix2}) and (\ref{weight_function}) by an
effective volume $V_{\rm eff}(k)$ defined as
\begin{equation}
V_{\rm eff}(k) = \int d^3{\bf r} 
\left[ \frac{P_{\rm obs}(k)}{1/n({\bf r})+P_{\rm obs}(k)} \right]^2~.
\end{equation}
However, for the future surveys discussed in this section, we can
systematically assume that the power spectrum reconstruction will be
performed inside a volume $V$ chosen in such way that throughout this
volume $1/n({\bf r})$ is smaller than $P_{\rm obs}(k)$ for all $k_{\rm
min} < k < k_{\rm max}$, so that $V_{\rm eff}(k)$ can be replaced by
$V$.

For instance, for the completed SDSS Bright Red Galaxy (BRG) survey,
this volume will include galaxies up to a distance $\lambda \simeq
1\,h^{-1}$Gpc with a sky coverage $f_{\rm sky}=0.25$. So, one can take
$V\sim f_{\rm sky} (4/3) \pi \lambda^3 = 1~ ({\rm Gpc}/h)^3$
\cite{Eisenstein:1998hr}.  Beyond SDSS, there are various plans for
larger surveys, like for instance the SDSS-II LEGACY project, the
Advanced Large, Homogeneous Area Medium Band Redshift Astronomical
survey\footnote{\tt http://alhambra.iaa.es:8080/} (ALHAMBRA)
\cite{Moles:2005dv}, the Dark Energy Survey\footnote{\tt
http://www.darkenergysurvey.org/} (DES) \cite{DES,DES1,DES2}, or the
Kilo-Aperture Optical Spectrograph\footnote{\tt
http://www.noao.edu/kaos} (KAOS).  Some of these surveys will go to
such high redshift that it will be possible, first, to compare the
observed power spectrum with the linear one up to wave-numbers
significantly larger than the usual $k_{\rm max} \sim 0.15 h$
Mpc$^{-1}$ (since the non-linear evolution affects larger and larger
modes as time passes by); and second, to do some tomography, i.e.\ to
reconstruct $P(k)$ in various redshift bins, corresponding to
different times in the evolution of the Universe, in order to measure
the variations of the linear growth factor and to get some good handle
on the dark energy variables. Tomography is also ideal for neutrino
mass extraction, since massive neutrinos induce a very peculiar
redshift-dependence on the matter power spectrum, as we have seen in
Sec.\ \ref{subsubsec:Pkz}.  For instance, KAOS could build two
catalogs centered around redshifts $z= 1$ and $z=3$, corresponding
roughly to $k_{\rm max} \sim 0.2~h$ Mpc$^{-1}$ and $k_{\rm max} \sim
0.48~h$ Mpc$^{-1}$ respectively, instead of $k_{\rm max} \sim 0.1~h$
Mpc$^{-1}$ for current surveys (conservative values).  In both
catalogs, the number density would be such that $1/n \sim P(k_{\rm
max})$, and the effective volume of the two samples close to $V \sim
0.5\, ({\rm Gpc}/h)^3$ and $V \sim 0.6\, ({\rm Gpc}/h)^3$
respectively\footnote{The characteristics of KAOS are taken from the
``Purple Book'' available on-line at {\tt http://www.noao.edu/kaos}.}.

In order to forecast the sensitivity of a galaxy survey combined with
a CMB experiment, one can simply add the two Fisher matrices and
invert their sum.

The sensitivity of large redshift surveys to the neutrino mass was
fist studied in Ref.\ \cite{Hu:1997mj}; however this pioneering paper
was rather crude in the way to combine GRS with CMB experiments. This
work was updated and generalized in \cite{Eisenstein:1998hr}, and
later in \cite{Hannestad:2002cn,Lesgourgues:2004ps,Takada:2005si}. In
Table \ref{sigma_mnu_SDSS}, we quote from these papers the forecast
error $\sigma(M_{\nu})$ obtained assuming that the completed SDSS
survey is combined either with the WMAP or {\sc Planck} data. These
results are based on Table II in \cite{Eisenstein:1998hr}, Figs. 4 and
5 in \cite{Hannestad:2002cn}, and Table III in
\cite{Lesgourgues:2004ps}, for a common conservative value $k_{\rm
max}=0.1h\,$Mpc$^{-1}$. With respect to the results quoted in the
previous subsection, the number of free parameters has increased by
one unit, since there is now an unknown free bias parameter $b=[P_{\rm
obs}(k)/P(k)]^{1/2}$ which is marginalized out.  We see that SDSS
alone gives essentially no information on $M_{\nu}$ given the various
parameter degeneracies. For the SDSS+WMAP combination, there is an
excellent agreement between \cite{Eisenstein:1998hr} and
\cite{Hannestad:2002cn}, and the forecast error $\sigma(M_{\nu})=0.3$
eV is three times smaller than the current error based on incomplete
WMAP+SDSS results $\sigma(M_{\nu})=0.9$ eV \cite{Tegmark:2003uf}.  For
SDSS+{\sc Planck}, it seems again that the {\sc Planck} noise
modelling in \cite{Hannestad:2002cn} is much too optimistic, while
Refs.\ \cite{Eisenstein:1998hr} and \cite{Lesgourgues:2004ps} agree on
$\sigma(M_{\nu})=0.2$ eV. If we compare this with the result obtained
in \cite{Lesgourgues:2004ps} under the same assumptions for {\sc
Planck} alone, $\sigma(M_{\nu})=0.3$ eV, we see that SDSS allows for a
significant improvement.

\begin{table}
\caption{Forecast error $\sigma(M_{\nu})$ in eV for the completed SDSS
survey, with volume $V=1 ({\rm Gpc}/h)^3$, eventually combined with
WMAP or {\sc Planck} data: compared results from the literature,
assuming the conservative value $k_{\rm max}=0.1h\,$Mpc$^{-1}$.
\label{sigma_mnu_SDSS}}
\begin{center}
\begin{tabular}{ccccc}
\hline
Ref. & SDSS & +WMAP & +Planck & assumptions \\
\hline
\cite{Eisenstein:1998hr} & 9 & 0.3 & 0.2 & 12 free parameters, \\
& & & & optimistic {\sc Planck} characteristics \\
\hline
\cite{Hannestad:2002cn} & -- & 0.3 & 0.06 & 9 free parameters, \\ 
& & & & very optimistic WMAP/Planck charact. \\
\hline
\cite{Lesgourgues:2004ps} & 7 & -- & 0.2 & 9 (other) free parameters,\\
& & & & up-to-date {\sc Planck} characteristics\\
\hline
\end{tabular}
\end{center}
\vspace{0.5cm}
\end{table}

Concerning experiments beyond SDSS, Ref.\ \cite{Hannestad:2002cn} and
later Ref.\ \cite{Lesgourgues:2004ps} present forecast errors in a
very general way, by simply showing $\sigma(M_{\nu})$ as a function of
two free parameters $V$ and $k_{\rm max}$. Instead Ref.\
\cite{Takada:2005si} assumes some detailed characteristics for three
high-redshift surveys, each of them operating in three redshift bins.
Here we will summarize the results of Ref.\ \cite{Lesgourgues:2004ps},
assuming that an hypothetical future redshift survey with given $(V,
k_{\rm max})$ is combined with a CMB experiment that could be SPTpol,
{\sc Planck}, CMBpol (see Table \ref{tableexp} for the value of
assumed sensitivity parameters), or an ``ideal CMB experiment'' that
would be limited only by cosmic variance up to $l=2500$ (both for
temperature and polarization). The main difficulty for reaching this
goal would be to subtract accurately small-scale foregrounds
(point-like sources, dusty galaxies, etc.)  but even with current
technology such an ideal experiment is not unconceivable.
\begin{figure}
\includegraphics[width=.85\textwidth]{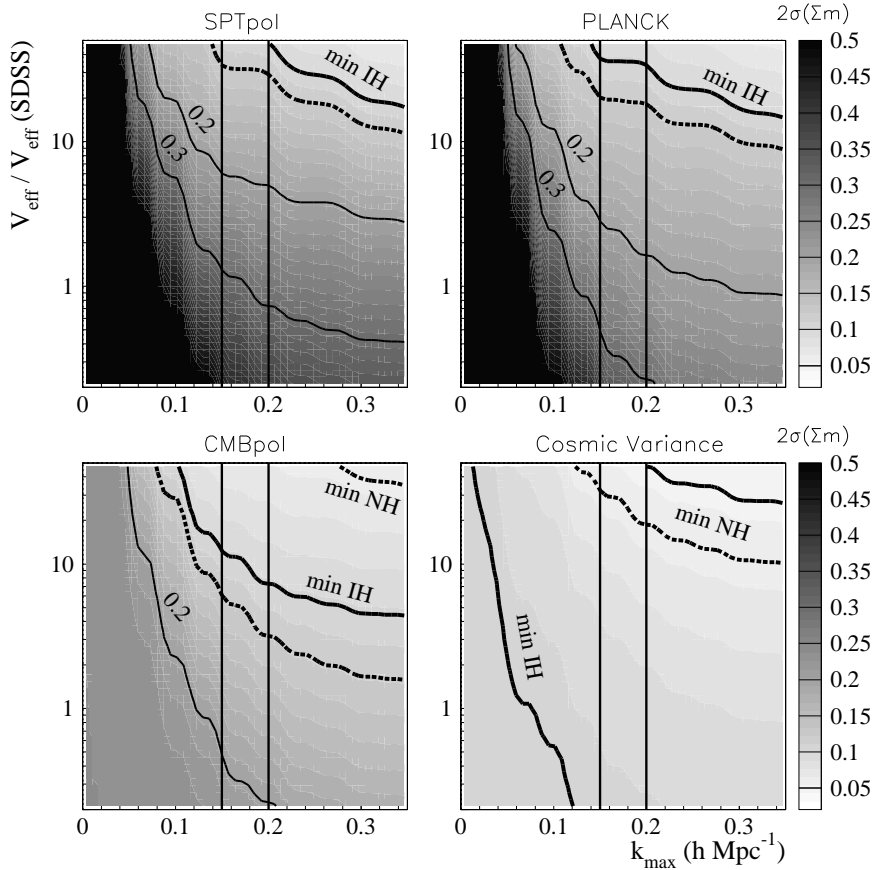}
\caption{\label{figall} Results from Ref.~\cite{Lesgourgues:2004ps} on
  the neutrino mass sensitivity of future redshift surveys, combined
  with future CMB experiment.  The level of gray gives the 2$\sigma$
  expected errors on the total mass $M_{\nu}$ (eV) for a fiducial
  value of 0.11 eV, as a function of the largest wavenumber $k_{\rm
  max}$ used in the analysis [in units of $h\,$Mpc$^{-1}$] and of the
  survey volume $V$ [in units of $V_{\rm SDSS} = 1 \, ({\rm
  Gpc}/h)^3$].  The vertical lines indicate the cut-off wavenumber
  $k_{\rm max}$ for the linear matter power spectrum at the
  conservative (optimistic) value $0.15(0.2)\, h$ Mpc$^{-1}$. The thin
  contours are (from bottom to top) for 0.3 and 0.2 eV, while the
  thick contours correspond to the minimum values of $M_\nu$ in the IH
  (lower lines) and NH (upper lines) schemes, assuming a squared mass
  difference $\Delta m^2_{\rm atm}= 2.6 \times 10^{-3}$ eV$^2$ (thick
  solid lines).  The thick dashed lines correspond to the same limits
  for a value $\Delta m^2_{\rm atm}= 3.7 \times 10^{-3}$ eV$^2$ which
  is now essentially ruled out by current data on neutrino
  oscillations.}
\end{figure}
For these four cases, Fig.\ \ref{figall} shows the predicted 2$\sigma$
error on the total neutrino mass $M_\nu$.  In these plots the
fiducial value of the total mass was fixed to $M_\nu=0.11$ eV,
and distributed according to the NH scheme.  For SDSS (or for any
survey with $z<1$) we expect the relevant value of $k_{\rm max}$ to be
around $0.15\, h$ Mpc$^{-1}$. However, depending on the overall
amplitude of the matter power spectrum (often parametrized by
$\sigma_8$, and still poorly constrained) and on future improvements
in our understanding of non-linear corrections, this value might
appear to be either too optimistic or too pessimistic: this is the
reason why it is interesting to leave it as a free parameter.  The
figure shows the importance of employing high-volume surveys, which
have the potential to improved the forecast errors even for the best
CMB experiments. Keeping $k_{\rm max} = 0.15\, h$ Mpc$^{-1}$, we see
that a two-sigma detection of the minimal mass in the IH scenario
would require $V=40~({\rm Gpc}/h)^3$ after {\sc Planck}, $V=12~({\rm
Gpc}/h)^3$ after CMBpol, while the ``ideal CMB experiment'' alone would
suffice. A two-sigma detection of the minimal mass in the NH
scenario seems to be unreachable for any plausible survey volume.

\begin{figure}
\includegraphics[width=.48\textwidth]{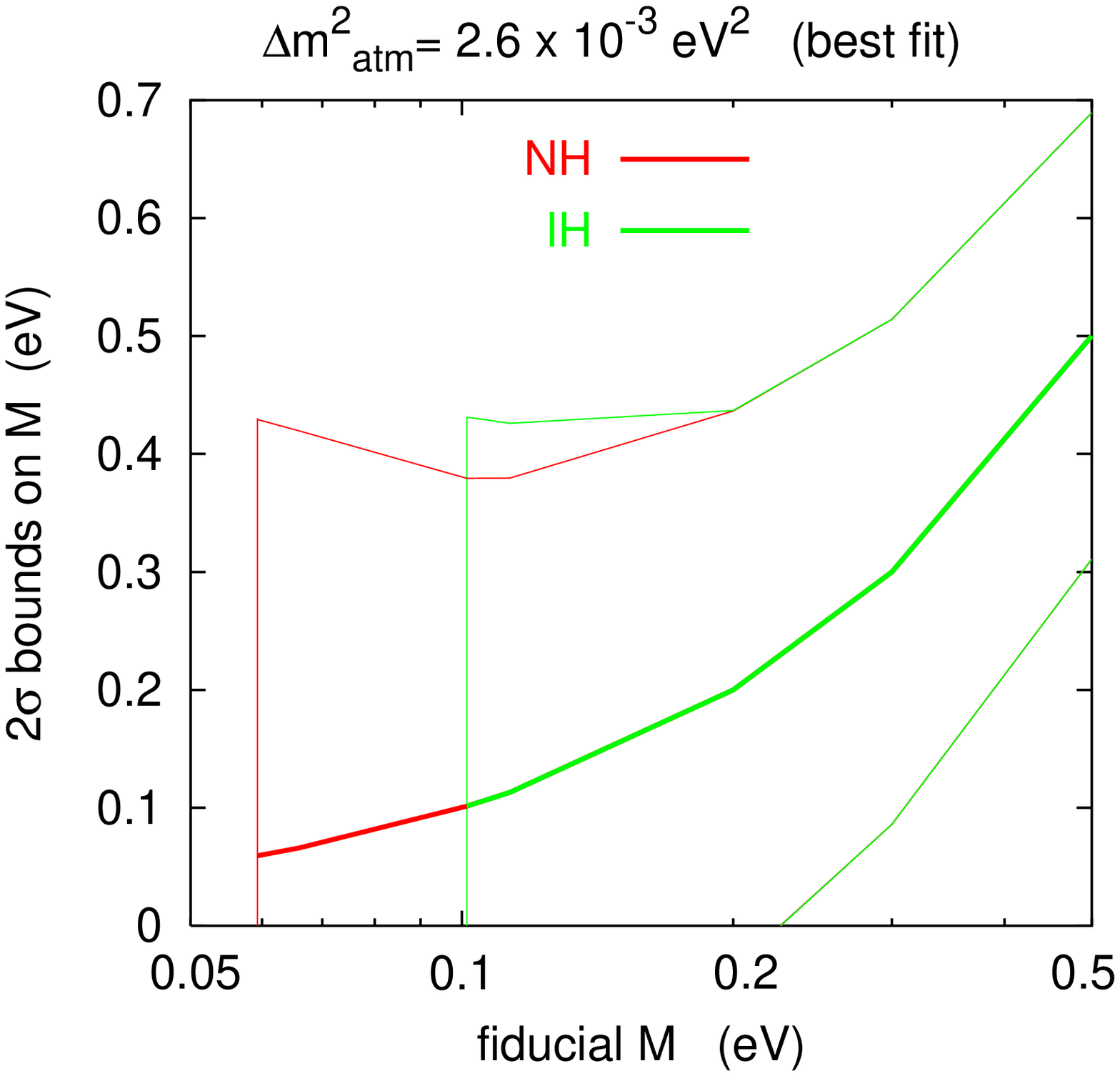}
\includegraphics[width=.48\textwidth]{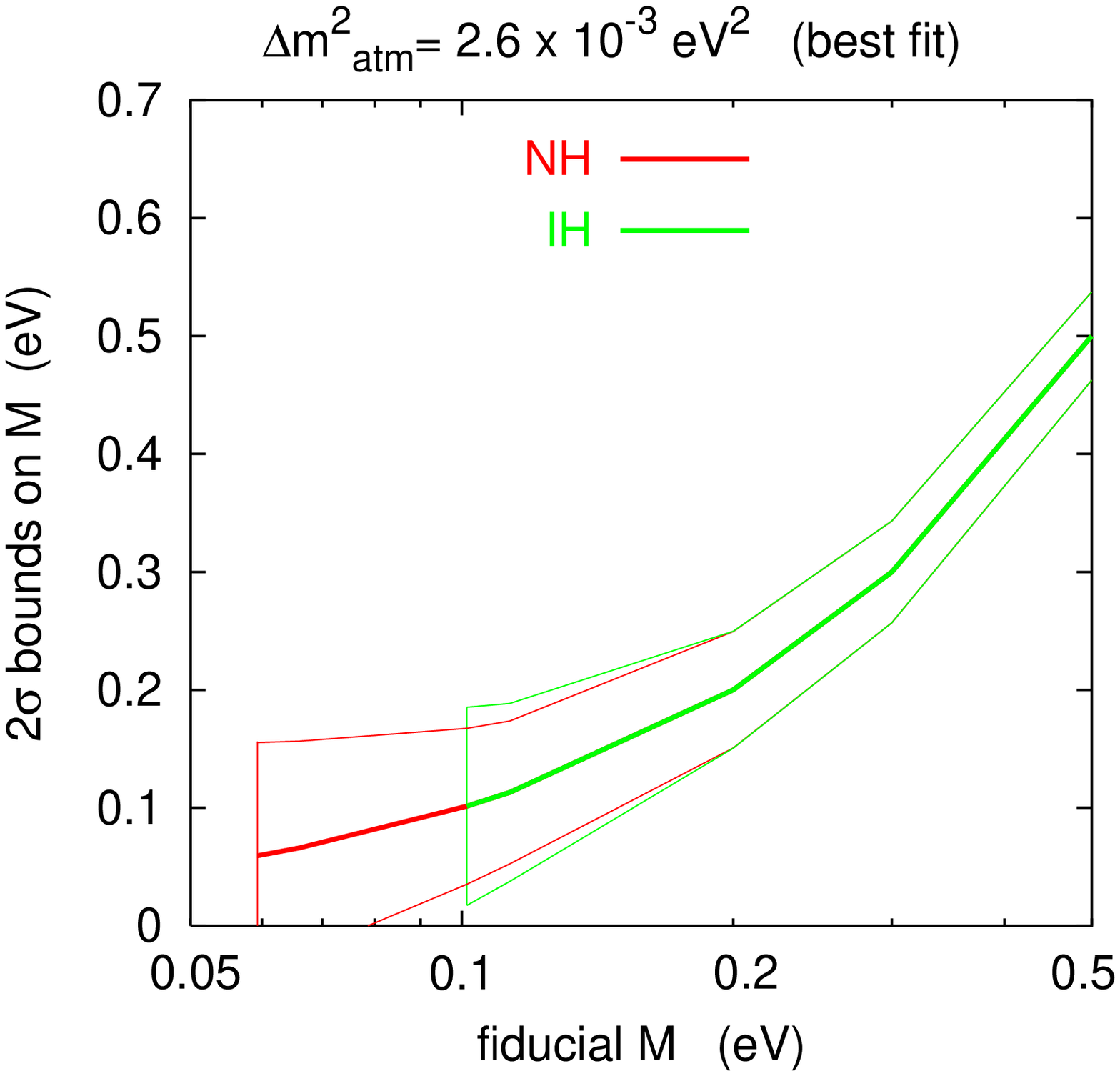}
\caption{\label{figIdeal} Predicted 2$\sigma$ error on the total
  neutrino mass as a function of the assumed fiducial value of
  $M_\nu$, using: (left) {\sc Planck}+SDSS, (right) an ideal CMB
  experiment (limited only by cosmic variance up to $l=2500$, both for
  temperature and polarization) and a redshift survey covering 75\% of
  the sky up to $z \simeq 0.8$ [$V_{\rm eff} = 40~({\rm
  Gpc}/h)^3$]. Both plots assume a limit $k_{\rm max}=0.15\, h$
  Mpc$^{-1}$ and a squared mass difference $\Delta m^2_{\rm
  atm}=2.6\times10^{-3}$~eV$^2$.  In each case, we show the results
  assuming either NH or IH. Figure from Ref.\
  \cite{Lesgourgues:2004ps}.}
\end{figure}

As in the previous subsection, we should mention that the results for
$\sigma(M_{\nu})$ depend significantly on the assumed fiducial value
$M_{\nu}$. This is illustrated in Fig.\ \ref{figIdeal}, taken from
Ref.\ \cite{Lesgourgues:2004ps}.  For the combination of either {\sc
Planck} with SDSS, or of the ``ideal CMB experiment'' with a larger
hypothetical survey, we can see that the 2$\sigma$ error is typically
twice larger for $M_{\nu}\sim 0.1$~eV than for $M_{\nu}\sim 0.5$~eV.

It is legitimate to wonder whether the most precise combination of
experiments would be sensitive to different neutrino mass splittings
between the three families, for a fixed value of the total
mass. Indeed, we have seen in Sec.\ \ref{subsec:sum_mass-effect}
that unlike CMB observables, the power spectrum of large scale
structure is in principle sensitive both to the total and the
individual values of the three masses. The analysis of Ref.\
\cite{Lesgourgues:2004ps} demonstrates however that the tiny
difference between the power spectrum obtained in the NH and IH cases
for a fixed value of the total mass (close to $0.1$ eV) is far too
small for being detected even with the combination of a huge redshift
survey of volume $V=40~({\rm Gpc}/h)^3$ with the ``ideal CMB
experiment''.

The issue of degeneracies between the neutrino mass and other
parameters, such as those describing the dark energy evolution,
probably needs further investigation. Assuming that dark energy has a
constant equation-of-state parameter $w$, Ref.\ \cite{Hannestad:2005gj}
showed that future analyses of combined CMB and LSS data could be
affected by a significant degeneracy between $M_{\nu}$ and $w$, a
result which was not found in Ref.\ \cite{Lesgourgues:2004ps}. The
authors of Ref.\ \cite{Ichikawa:2005hi} focused on this degeneracy,
including a possible dependence of $w$ on redshift. They point out
that in order to help resolving the degeneracy, it is useful to take
into account the fact that the CMB and LSS data are not statistically
independent, as assumed for simplicity in all the works cited in this
section. Instead, it is well-known that due to the late Integrated
Sachs-Wolfe (ISW) effect (see Sec.\ \ref{subsec:params}), the
power spectrum of temperature anisotropies encodes some information on
neighboring structures (like galaxy clusters). So, there is a non-zero
cross-correlation between temperature and galaxy maps, which has
actually been already measured \cite{Fosalba:2003ge,Scranton:2003in},
still with a rather low level of significance. In the future, precise
measurements of this cross-correlation could be very useful for
neutrino mass extraction.

\subsection{CMB weak lensing}
\label{cmb-lensing}

\subsubsection{Principle}

It was recently realized that the CMB anisotropies encode even more
cosmological information than expected, because it should be possible
in a near future to measure the deflection field caused by the weak
lensing of CMB photons by the large scale structure of the neighboring
universe up to a redshift $z\sim 3$ \cite{Ber97,LE,LE2} (see
\cite{Lewis:2006fu} for a recent review on CMB lensing). The power
spectrum of the deflection field encodes some information concerning
structure formation mainly in the linear or quasi-linear regime, and
is therefore extremely useful for measuring parameters like the total
neutrino mass (or the dark energy equation-of-state) which mildly
affect the primary anisotropy \cite{Kaplinghat:2003bh}.  So, the next
generation of CMB experiments could output for free a LSS power
spectrum, without suffering from the systematics induced by
mass-to-light bias and by strong non-linear corrections on small
scales at $z\leq0.2$, as is the case for galaxy redshift surveys.

More precisely, weak lensing produces a deflection of light rays 
parametrized by a two-dimensional deflection field ${\bf d}({\bf n})$
such that an object located in a direction ${\bf n}$ will appear in a 
direction ${\bf n}+{\bf d}({\bf n})$. 
At leading order \cite{IE,Cooray:2005hm},
the deflection field can be written as the gradient of a lensing potential,
${\bf d}={\bf \nabla} \phi$. The lensing potential contains information on
the gravitational potential along the photon trajectory, convolved
by a window function which depends on the distance to the source
\begin{equation}
\phi(\hat{n};r_s) = \int_0^{r_s} dr \,\, W(r;r_s) \,\, 
\Phi(r\hat{n}, \eta=-r)~,
\label{projected_potential_CMB}
\end{equation}
where $r$ is the comoving coordinate along the line of sight, $r_s$ is
the source coordinate, $\Phi$ the gravitational potential, and $\eta$
the conformal time defined by $dt= - a \, d\eta$ and $\eta=0$ today
(with this definition, the photons emitted by the source and reaching
us today travel according to $r(\eta)=-\eta$). Using the geodesic
equation, it is possible to show that the window function $W(r;r_s)$
is equal to
\begin{equation}
W(r;r_s) = 2 \, \, \frac{r_s-r}{r \, r_s}~.
\label{window_lensing_CMB}
\end{equation}
In the present case, the source is the last scattering surface and
$r_s$ stands for its comoving radius. For a given cosmological model,
the primordial spectrum $\langle |\Phi|^2 \rangle$ and its time
evolution can be calculated; the same numerical code which computes
the CMB anisotropy spectrum $C_l^{XX'}$ and the matter power spectrum
$P(k)$ can also easily compute the deflection field power spectrum
$C_l^{dd}$, which encodes some information on LSS at redshifts of
order $z \sim 3$ or so.  Hence, if we could map the CMB deflection
field ${\bf d}$, we could measure $C_l^{dd}$ and compare it with
theoretical predictions in order to constrain cosmological parameters.

There are several methods on the market for extracting the deflection
map \cite{IE,QE1,QE2,QE3}, all based on the non-Gaussianity induced by
lensing \cite{Ber97}. These methods start from the assumption that
both the primary anisotropies and the deflection field are Gaussian;
they also assume that the noise present in the temperature and
polarization maps is Gaussian and uncorrelated with the signal.  None
of these assumptions is exactly true. Ref.\ \cite{Amb04} estimated to which
extent the lensing extraction will be biased, first, by the
non-Gaussianity of the lensing potential caused by the non-linear
growth of matter perturbations on small scales, and second, by the
imperfect cleaning of the CMB maps from the kinetic Sunyaev-Zel'dovich
effect, which also has a blackbody spectrum, induces non-Gaussianity,
and features spatial correlations with many of the structures
responsible for the lensing. Both effects were found to be relevant
(i.e.\ to induce a significant bias in the estimators). However, they
are small enough to preserve the validity of the method.
Ref.\ \cite{Lesgourgues:2004ew} estimated the impact on lensing
extraction of a small level of non-Gaussianity produced in the Early
Universe by inflation, reheating or some other mechanisms (for a
review, see \cite{NG}), and concluded that given the current bounds on
primordial non-gaussianity inferred from CMB maps, this
non-gaussianity cannot be large enough to threaten lensing extraction.

\subsubsection{Quadratic estimator method}

This field is rather technical, but we can stress the main ideas
without entering into many details. For simplicity, we will write the
main formulas in the flat-sky approximation, which amounts in
replacing the expansion of CMB maps in harmonic space by an expansion
in two-dimensional Fourier space, and to use two-dimensional vectors
${\bf l}$ instead of pairs of integers $(l,m)$. In the flat-sky
approximation the Fourier modes are fully described by the power
spectra $\tilde{C}^{ab}_l$ where $a$ and $b$ belong to the basis $\{T,
E, B\}$ of temperature, $T$-mode polarization and $E$-mode
polarization. Weak lensing correlates the lensed multipoles
\cite{Sel96,Ber97} according to
\begin{equation}
\langle a({\bf l}) b({\bf l'}) \rangle_{\rm CMB}
= (2 \pi)^2 \delta({\bf l}+{\bf l'}) \tilde{C}^{ab}_{l}
+ f^{ab}({\bf l}, {\bf l'}) \phi({\bf l}+{\bf l'})
\label{ab}
\end{equation}
where the average holds over different realizations (or different
Hubble patches) of a given cosmological model with fixed primordial
spectrum and background evolution (i.e.\ fixed cosmological
parameters). In this average, the lensing potential $\phi$ is also
kept fixed by convention, which makes sense because the CMB
anisotropies and the LSS that we observe in our past light-cone are
statistically independent, at least as long as we neglect the
integrated Sachs-Wolfe effect. The above functions $f^{ab}$ are defined
in \cite{QE2}. For instance, for $ab=TT$ one has
\begin{equation}
f^{TT}({\bf l}, {\bf l'})C
= C^{TT}_{l}
({\bf l}+{\bf l'}) \cdot {\bf l}
+
C^{TT}_{l'}
({\bf l}+{\bf l'}) \cdot {\bf l}'~.
\end{equation}
Let us introduce the basic principle of the quadratic estimator method
of Hu \& Okamoto \cite{QE1,QE2,QE3} (which is equivalent in terms of
precision to the alternative iterative estimator method of Hirata \&
Seljak \cite{IE} as long as CMB experiments will make noise-dominated
measurements of the B-mode, i.e.\ at least for the next decade).  By
inverting Eq.\ (\ref{ab}), one builds a quadratic combination of the
temperature and polarization observed Fourier modes
\begin{equation}
{\bf d}^{ab}({\bf L}) =
\frac{i {\bf L} A^{ab}_L}{L^2}
\int \!\!\! \frac{d^2 {\bf l}_1}{(2 \pi)^2}
a({\bf l_1}) b({\bf l_2}) g^{ab}({\bf l_1},{\bf l_2})
\label{defd}
\end{equation}
where ${\bf l_2} = {\bf L} - {\bf l_1}$, and in which
the normalization condition
\begin{equation}
A^{ab}_L = L^2 \left[ \frac{d^2 {\bf l}_1}{(2 \pi)^2}
f_{TT}({\bf l}_1,{\bf l}_2) g^{ab}({\bf l}_1,{\bf l}_2)
\right]^{-1}
\label{aaanorm}
\end{equation}
ensures that ${\bf d}^{ab}$ is
an unbiased estimator of the lensing potential
\begin{equation}
\langle {\bf d}^{ab}({\bf L})\rangle_{\rm CMB}
= i {\bf L} \phi({\bf L}) = {\bf d}({\bf L})~.
\label{avd}
\end{equation}
Note that, so far, the coefficients $g^{ab}({\bf l}_1,{\bf l}_2)$ are
still arbitrary. {}From the observed temperature and polarization maps,
one could compute each mode of ${\bf d}^{ab}$ and obtain various
estimates of the deflection modes, precise up to cosmic
variance and experimental errors. In order to quantify the total
error, it is necessary to compute the power spectra of the quadratic
estimators
\begin{equation}
\langle{\bf d}^{ab*}({\bf L}){\bf d}^{ab}({\bf L})\rangle =
(2\pi)^2 \delta({\bf L}-{\bf L}') C_L^{dd(ab)}
\end{equation}
where the average is now taken over both CMB and LSS realizations,
since $\phi({\bf L})$ is also a stochastic quantity. In this
definition, the power spectra are written with a superscript {\small
$dd(ab)$} in order to be distinguished from the actual power spectrum
of the true deflection field.  These spectra feature the four-point
correlation function of the observed (lensed) Fourier modes $\langle
a({\bf l}_1) b({\bf l}_2) a({\bf l}_3) b({\bf l}_4)\rangle$, which
should be expanded at order two in $\phi({\bf L})$ in order to catch
the leading non-Gaussian contribution.

The four-point correlation functions are composed as usual of a
connected and an unconnected piece. The connected piece is by
definition a function of the power spectra $C_{l}^{ab}$ in which we
now include all sources of variance: cosmic variance, lensing
contribution and experimental noise. The unconnected piece is a
function of the same spectra plus the deflection spectrum
$C_{l}^{dd}$, and as usual it can be decomposed in three terms
corresponding to the different pairings of the four indexes
\cite{TRI1}: $(l_1,l_2)$, $(l_3,l_4)$ or $(l_1,l_3)$, $(l_2,l_4)$ or
$(l_1,l_4)$, $(l_3,l_2)$.  The first term leads to considerable
simplifications when it is plugged into the expression of the
quadratic estimator power spectrum, and the result is simply
$C_l^{dd}$, as one would expect naively from squaring Eq.~(\ref{avd}).
The other terms lead to more complicated expressions that we will
write as a noise term
\begin{equation}
C_L^{dd(ab)} =
C_L^{dd} + [N_{\rm c}]^{ab}_{L} + N^{ab}_{L}~,
\label{clddab}
\end{equation}
which represent respectively the contribution from the connected piece
and from the two non-trivial terms of the unconnected piece
\cite{Coo02,TRI1}.  In order to get an efficient estimator, one should
adopt the set of coefficients $g^{ab}({\bf l}_1,{\bf l}_2)$ which
minimize the noise terms.  It is actually much easier to minimize the
connected term only, which leads to the simple results
\begin{equation}
g^{aa}({\bf l}_1,{\bf l}_2)
=
\frac{f_{aa}({\bf l},{\bf l}')}{2 C_{l}^{aa} C_{l'}^{aa}}
\quad {\rm and} \quad
[N_{\rm c}]^{aa}_{L} = A^{aa}_L
\label{naaeqaaa}
\end{equation}
for $a\!=\!b$ (for $a \!\neq\! b$ see \cite{QE2}).
With such a choice, the unconnected piece contribution
$N^{ab}_{L}$ can be shown to be smaller than $A^{aa}_L$,
but not completely negligible \cite{Coo02}.

The various estimators ${\bf d}^{ab}$ can be constructed for each pair
of modes, except for the pair $BB$, because the spectrum $C_l^{BB}$ is
dominated by lensing at least on small scales, which invalidates the
present method. Therefore, the quadratic estimator technique would not
be optimal for long-term CMB experiments with
cosmic-variance-dominated measurement of the $B$ mode
\cite{IE,Smi04}. For an experiment of given sensitivity, the five
other estimators can be combined into a final minimum variance
estimator, which gives the best possible estimate of the deflection
field by weighing each estimator accordingly to its noise level. The
sensitivity of the {\sc Planck} satellite is slightly above the
threshold for successful lensing extraction, but only at intermediate
angular scales, and with essentially all the signal coming from the
${\bf d}^{TT}$ estimator. The following generation of experiments --
such as the CMBpol or Inflation probe project -- should
obtain the lowest noise level from the ${\bf d}^{EB}$ estimator
\cite{QE2}.

\subsubsection{Neutrino mass from CMB lensing extraction.}
\label{subsubsec:mnucmblensing}

The authors of Refs.\ \cite{Kaplinghat:2003bh} and \cite{LPPP} studied
the limits that could be obtained on neutrino mass by using future CMB
experiments only, assuming that the lensing power spectrum could be
extracted using the quadratic estimator technique. In this way, it is
possible to combine information on CMB acoustic oscillations and on
the surrounding LSS of the Universe at redshifts of order $z \leq 3$,
avoiding many of the complicated features of galaxy redshift surveys
related to the mass-to-light bias and to the strongly non-linear
evolution of small-scale perturbations at small redshift.

The principle of Refs.\ \cite{Kaplinghat:2003bh,LPPP} is to treat the
power spectrum of primordial anisotropies (for temperature and
polarization) and of the deflection field as two independent
information on the cosmological parameters.  The expected error-bars
on the temperature and polarization multipoles are estimated from the
characteristics of a given CMB experiment, while those on on the
lensing multipoles are taken to be the dominant term $[N_{\rm
c}]^{ab}_L$ in Eq.\ (\ref{clddab}), which can be computed from the
same experimental characteristics using essentially Eqs.\
(\ref{naaeqaaa}) and (\ref{aaanorm}).  Then, the authors assume a
fiducial model and perform a Fisher matrix analysis in the same
fashion as described in the previous subsection.

The fiducial model of Ref.\ \cite{Kaplinghat:2003bh} includes ten free
parameters: the usual six parameters of the minimal $\Lambda$CDM
model, the neutrino mass $M_\nu$ in the case of three degenerate
neutrinos, the equation of state parameter $w$, the running of the
tilt $\alpha$ and the primordial Helium abundance $Y_{\rm He}$.  The
authors conclude that the {\sc Planck} satellite will output a
1$\sigma$ error-bar $\sigma(M_\nu)=0.15$ eV, which shows that the use
of the lensing extraction technique should improve the global
sensitivity in a rather spectacular way at least for this
parameter. In the case of the CMBpol project an impressive error of
$\sigma(M_\nu)=0.044$ eV is predicted.

This study was repeated in Ref.\ \cite{Song:2003gg}, with one extra
parameter representing the variation of the dark energy
equation-of-state with respect to the scale factor. We saw before that
this parameter could potentially introduce extra degeneracies
involving the neutrino mass, but the forecast errors are found to be
approximately the same: $\sigma(M_\nu)=0.16$ eV and
$\sigma(M_\nu)=0.046$ eV for {\sc Planck} and CMBpol, respectively.

Finally, the authors of \cite{LPPP} assumed two possible fiducial
models, one with eighth free parameters (the same as in Ref.\
\cite{Kaplinghat:2003bh} with the exception of $\alpha$ and $w$) and
one with eleven free parameters (same as in Ref.\
\cite{Kaplinghat:2003bh} plus the effective number of extra
relativistic degrees of freedom $N_{\rm eff}$). They consider six
experiments or combinations of experiments as listed in Table
\ref{m_nu_lensing_CMB}, and for each case they quote two numbers,
corresponding either to optimistic or pessimistic assumptions
concerning the subtraction of astrophysical foregrounds from the
observed CMB maps.
\begin{table}
\caption{Forecast error $\sigma(M_{\nu})$ in eV using various CMB
experiments or combinations of them, and assuming that the lensing map
is extracted following the quadratic estimator method. These numbers
are all taken from Ref.\ \cite{LPPP}, which assumes either an eight
parameter model (the usual six of $\Lambda$CDM plus $M_{\nu}$ and
$Y_{\rm He}$), or an eleven parameter model (the same ones plus
$\alpha$, $w$, $N_{\rm eff}$).  The fiducial value of the total mass
is taken to be $M_{\nu}=0.1$ eV, and in each case the two numbers
correspond to optimistic or pessimistic assumptions concerning the
foregrounds contamination of the primary CMB signal.}
\label{m_nu_lensing_CMB}
\begin{center}
\begin{tabular}{ccccc}
Experiment & 8 parameters & 11 parameters\\
\hline
BICEP+QUaD &0.31 - 0.36&0.36 - 0.40\\
BRAIN+ClOVER &0.34 - 0.43&0.42 - 0.51\\
{\sc Planck} &0.13 - 0.14&0.15 - 0.15\\
SAMPAN       &0.10 - 0.17&0.12 - 0.18\\
{\sc Planck}+SAMPAN&0.08 - 0.10&0.10 - 0.12\\
{\sc Inflation Probe}    &0.032 - 0.036&0.035 - 0.039\\
\hline
\end{tabular}
\end{center}
\vspace{0.5cm}
\end{table}

It can be seen in Table \ref{m_nu_lensing_CMB} that the combination
QUaD+BICEP benefits a lot from lensing extraction, since the error
$\sigma(M_{\nu})$ decreases from approximately 1.5 eV to at least 0.4
eV. Thus, with QUaD+BICEP it should be possible to reach in a near
future --using CMB only-- the same precision that we have today
combining many observations of different types (galaxy-galaxy
correlation function, Lyman-$\alpha$ forest) which are affected by
various systematics.  The situation is almost the same for
BRAIN+ClOVER, which should also achieve $\sigma(M_{\nu}) \sim 0.4$~eV
using lensing extraction.  {\sc Planck} should make a decisive
improvement, lowering the error to $\sigma(M_{\nu}) \sim 0.15$~eV, in
excellent agreement with the results of Ref.\
\cite{Kaplinghat:2003bh}. Note that without lensing extraction the
error would be multiplied by three (by four in the case with extra
free parameters). No significant difference between the forecast
errors in the eight and eleven parameter models was found in Ref.\
\cite{LPPP}.  SAMPAN alone is slightly more efficient than {\sc
Planck}, and the combination {\sc Planck}+SAMPAN is the first one to
reach $\sigma(M_{\nu}) \sim 0.1$ eV, even in the pessimistic case of
large foreground residuals and extra free parameters.  Finally, the
version of the {\sc Inflation Probe} satellite considered in
\cite{LPPP}, which is slightly more ambitious than the CMBpol
assumption of Ref.\ \cite{Kaplinghat:2003bh}, is able to reach
$\sigma(M_{\nu}) \sim 0.035$ eV both in the eight and eleven parameter
cases. In any case, the results seem to be relatively robust against
pessimistic assumptions concerning foreground contamination.

\subsection{Galaxy weak lensing (cosmic shear)}

\subsubsection{Principle}

Weak lensing changes the apparent shape of galaxies in various ways.
Among others, it changes slightly their apparent ellipticity: i.e.\
galaxies which would be apparently spherical without weak lensing can
look in fact elliptical, stret\-ched in one direction and squeezed in the
orthogonal direction. This effect, which is coherent over the angular
size of the lensing gravitational field, i.e. potentially over many
sources, is called cosmic shear.  It can be detected provided one has
a very dense sample of galaxies with enough resolution in order to
measure each individual shape. In order to be able to reconstruct the
lensing gravitational field, it is also necessary to know either the
redshift of each source galaxy, or at least their coarse-grained
number density $g(z,\hat{n})$ in redshift space and in direction
$\hat{n}$.  Let us assume for simplicity that we have an isotropic
sample with a spherical number density $g(z)$ (or $g(r)$, where $r$ is
the comoving radius).  We can divide the sample in small solid angles
or pixels. Since intrinsic ellipticities are randomly distributed, the
average ellipticity in each pixel will give a measurement of the
average cosmic shear, up to a shot noise term proportional to $N_{\rm
pix}^{-1/2}$, where $N_{\rm pix}$ is the number of galaxies per
pixel. Therefore, the size of the pixels must be chosen optimally in a
trade-off between high angular resolution and high signal-to-noise.

The cosmic shear field reconstructed by this method derives from the
same lensing potential $\phi$ as the deflection field discussed in the
previous subsection. The relation between the lensing potential $\phi$
and the gravitational potential $\Phi$ along the line of sight is the
same as in Eqs.\ (\ref{projected_potential_CMB}) and
(\ref{window_lensing_CMB}), except that one must now average over all
individual sources of comoving distance $r_s \in [0,r_s^{\rm max}]$.
In the continuous limit and for an ensemble of sources described by
the density function $g(r)$, this gives
\begin{equation}
\phi(\hat{n};g) =
\frac{\displaystyle \int_0^{r_s^{\rm max}} dr_s  
\,\, g(r_s) \,\, \phi(\hat{n};r_s)}
{\displaystyle \int_0^{r_s^{\rm max}} dr_s  \,\, g(r_s)}
=
\int_0^{r_s^{\rm max}} dr \,\, W(r;g) \,\, \Phi(r\hat{n}, \eta=-r)~,
\end{equation}
with
\begin{equation}
W(r;g) \equiv \frac{\displaystyle 
\int_r^{r_s^{\rm max}} dr_s  \,\, g(r_s) \,\, W(r;r_s)}
{\displaystyle \int_0^{r_s^{\rm max}} dr_s  \,\, g(r_s)}~.
\end{equation}
Like for CMB anisotropies, once we have the lensing potential map
$\phi(\hat{n};g)$ we can compute the power spectrum $C_l^{\phi \phi, g}$,
which gives us an information on the surrounding large scale structure
projected along the line of sight. Therefore, cosmic shear surveys
offer another opportunity to measure the two-point correlation
function of total matter fluctuations, complementary to the
three-dimensional Fourier power spectrum $P(k)$ obtained from galaxy
redshift surveys (probing lower redshifts) and to the CMB lensing
power spectrum (probing higher redshifts).

If a shear survey contains a large enough number of galaxies with
known redshifts, it is even possible to refine this technique by
splitting the sources into redshift (or distance) bins, from
$0<r_s<r_1$ to $r_{n-1}<r_s<r_n$. It is then possible to measure $n$
lensing potential maps $\phi(\hat{n},g_i)$, such that the bin
distribution $g_i(r_s)$ vanishes outside of the range
$r_{i-1}<r_s<r_i$. {}From these maps one can obtain $n$ power spectra
$C_l^{\phi \phi,g_i}$ and $n(n-1)/2$ cross-correlation spectra
$C_l^{\phi \phi,g_i g_j}$ (cross-correlations do not vanish because the
same structure can contribute to the lensing of sources in different
bins).  Since the highest redshift bin probes structures at larger
redshift, this method allows to some extent for a three-dimensional
reconstruction of the gravitational fluctuations. This technique is
called tomography, and brings some hope of tracking the evolution of
$P(k,z)$ as a function of redshift caused by dark energy \cite{Hu:1999ek}
and by neutrino masses (see Sec.\ \ref{subsubsec:Pkz}).

\subsubsection{Neutrino mass from cosmic shear surveys}
\label{subsec:cs}

There are various ongoing and planned cosmic shear surveys, like for
instance the Canada-France-Hawaii Telescope Legacy Survey\footnote{\tt
http://www.cfht.hawaii.edu/Science/CFHLS/} (CFHTLS)
\cite{Semboloni:2005ct}; the Dark Energy Survey\footnote{\tt
http://www.darkenergysurvey.org/} (DES) \cite{DES,DES1,DES2};
the SuperNova Acceleration Probe\footnote{\tt
http://snap.lbl.gov} (SNAP); the Panoramic Survey Telescope And Rapid
Response System\footnote{\tt http://pan-starrs.ifa.hawaii.edu/}
(Pan-STARRS); or the Large Synoptic Survey Telescope\footnote{\tt
http://www.lsst.org/lsst\_home.shtml} (LSST) \cite{LSST,LSST2}. The
sensitivity of this type of observation to neutrino masses has been
addressed first in \cite{Cooray:1999rv}, and later in
\cite{Abazajian:2002ck,Song:2003gg,Hannestad:2006as}. 
Here we will review the results of
\cite{Song:2003gg}, where four hypothetical experiments are
considered. The first two, called S300 and S1000, stand for future
satellite surveys with an average source density of 100 galaxies per
squared arc-minute and a sky coverage of 300 square degrees ($f_{\rm
sky}=0.0073$) and 1000 squared degrees ($f_{\rm sky}=0.024$),
respectively. The SNAP lensing survey could be close to one of
these two assumptions.
The last two, called G2$\pi$ and G4$\pi$, stand for
future ground-based surveys with an average source density of 65
galaxies per square arc-minute and a sky coverage of respectively
$f_{\rm sky}=0.5$ (as expected for LSST)
and $f_{\rm sky}=1$ (as expected for Pan-STARRS). 
The authors make some realistic
assumption concerning the redshift distribution of the sources (given
experimental limitations) and perform a Fisher matrix analysis for the
combination of the cosmic shear surveys with either {\sc Planck} or
CMBpol (see Table \ref{tableexp}), including in each case information
on the CMB lensing, extracted with the quadratic estimator method
described in Sec.\ \ref{cmb-lensing}. They assume a tomographic
analysis of the galaxy-ellipticity data with eight redshifts
bins. Thanks to the quadratic estimator method, the CMB experiments
also provide a measurement of the cosmic shear power spectrum,
sensitive to higher redshifts: here the information from CMB lensing
plays the role of a ninth redshift bin.

There are eleven free cosmological parameters in the analysis of Ref.\
\cite{Song:2003gg}: the usual six $\Lambda$CDM parameters, plus the
total neutrino mass (attributed to a single family), the primordial
helium fraction, the running of the scalar tilt, and two parameters
describing the dark energy equation of state: $w(a)=w_0 + (a_0-a)
w_a$. The results are summarized in Table \ref{cosmic-shear},
\begin{table}
\caption{Forecast error $\sigma(M_{\nu})$ in eV from reference
\cite{Song:2003gg}, using some hypothetical cosmic shear surveys
(S300, S1000, G2$\pi$ and G4$\pi$) described in the text, combined
with either {\sc Planck} or CMBpol (including, in each case, information
on the CMB lensing, extracted with the quadratic estimator method).
\label{cosmic-shear}}
\begin{center}
\begin{tabular}{ccccc}
\hline
& S300 & S1000 & G2$\pi$ & G4$\pi$ \\
\hline
Planck & 0.12 & 0.090 & 0.052 & 0.045 \\
CMBpol & 0.047 & 0.045 & 0.031 & 0.027 \\
\hline
\end{tabular}
\end{center}
\end{table}
from which one concludes that cosmic shear surveys constitute an
extremely powerful tool for constraining the total neutrino mass. The
sensitivities obtained with this method are the absolute best
forecasts derived so far. This is not surprising, given the ideal
properties of weak lensing observations: precise, high redshift
measurement of the matter power spectrum over a wide range of scales,
and tomography. 

Following Ref.\ \cite{Song:2003gg} and Table \ref{cosmic-shear}, the
most ambitious hypothetical experiment G4$\pi$ combined with the {\sc
Planck} data would reach $\sigma(M_{\nu}) \simeq 0.045$ eV, i.e.\ the
same sensitivity as CMBpol alone (see Sec.\
\ref{subsubsec:mnucmblensing}); while G4$\pi$ combined with CMBpol
would reach $\sigma(M_{\nu}) \simeq 0.027$ eV, which means that a
2$\sigma$ detection of the neutrino mass would occur even for the
smallest $M_{\nu}$ in the NH scheme, of order $(\Delta m_{\rm
atm}^2)^{1/2}\simeq 0.05$ eV.  The results of Table \ref{cosmic-shear}
are relatively conservative in the sense that in the Fisher matrix
computation, the cosmic shear power spectrum was limited to the
angular scales $\theta > 0.18$ deg (in a multipole expansion, $l <
1000$) below which non-linearities render the power spectrum
reconstruction more difficult. However, the authors of Ref.\
\cite{Song:2003gg} refer to a method for extracting information on
smaller scales proposed by Refs.\ \cite{Jain:2003tb,Bernstein:2003es}.
Ref.\ \cite{Song:2003gg} speculates that using this method, one could
gain a factor three (two) with respect to the results of Table
\ref{cosmic-shear} for the combination of {\sc Planck} (CMBpol) with
G2$\pi$ or G4$\pi$.  This would imply a real precision measurement of
the total neutrino mass with cosmological observables.

Finally, the recent analysis in Ref.\ \cite{Hannestad:2006as}
considers the combination of {\sc Planck} (no lensing extraction) with
two lensing surveys inspired by the SNAP and LSST projects, using up
to five tomography bins. The results are consistent with those of
Ref.\ \cite{Song:2003gg}.

\subsection{Galaxy cluster surveys}
\label{subsec:gcs}

The last method that we will discuss is expected to bring some extra
independent information, and could further improve 
the bounds discussed in the previous subsection.

In the past years, the study of the cluster abundance evolution (the
function $N(z)$, where $N$ is the number density of galaxy clusters
and $z$ the redshift) has lead to interesting constraints on the
normalization of the matter power spectrum at small scales $\sigma_8$
(defined in Sec.\ \ref{subsec:scb}), and on the matter density
fraction $\Omega_{\rm m}$. For a given cosmological model, the
function $N(z)$ can be predicted from non-linear simulations, and
compared with experimental data from galaxy cluster surveys. The
non-linear simulations generally introduce systematic errors, but
since cluster scales are not too deep inside the non-linear regime
even today, the method can be kept under control.

The function $N(z)$ --or better, its derivative $dN/dz$-- is affected
by the evolution of the linear growth factor on cluster scales.
Therefore, as emphasized recently in \cite{Wang:2005vr}, future precise
measurements of cluster abundances will be useful probes of the
neutrino free-streaming effect.

On the observational side, the prospects for galaxy cluster surveys
are very promising, since catalogs will be derived from various
techniques, by mapping the X--ray emission of the hot cluster gas, or
the scattering of CMB photons in this gas (Sunyaev--Zel'dovich effect,
SZE), or the cosmic shear measured from galaxy weak lensing (as
discussed in the previous subsection).  We have already seen that
ambitious projects are being planned for galaxy weak lensing; this is
also the case for X--ray and SZE observations.  The full data will
then provide: first, a measurement of the matter power spectrum $P(k,z)$
(modulo some bias and redshift-space distortion effects), which can be
measured in various redshift bins, like for weak lensing tomography;
and second, an estimate of the abundance evolution $dN/dz$, also at
various redshifts.

An expression of the Fisher matrix describing future $P(k,z)$ and
$dN/dz(z)$ observations has been proposed in \cite{Wang:2004pk}; this
forecast method was applied to neutrino mass extraction in \cite{Wang:2005vr},
using the expected instrumental sensitivity of the following projects:
for CMB temperature and polarization anisotropies, {\sc Planck} (no
lensing extraction); for X-ray observations, the Dark Universe
Observatory (DUO) survey (see \cite{duo} for the description of a
similar, previously proposed survey), which should see $\sim 11,500$
clusters in 6,150 deg$^2$; for SZE observations, the South Pole
Telescope (SPT) survey \cite{Ruhl}, yielding $\sim 20,000$ clusters in
4,000 deg$^2$; and for galaxy weak lensing, the LSST project (already
introduced in Sec.\ \ref{subsec:cs}), assuming $\sim 200,000$ clusters
in 18,000 deg$^2$. Among the three data sets, Ref.\ \cite{Wang:2005vr}
finds that the most powerful should be the LSST lensing survey: the
forecast error $\sigma(M_\nu)$ is typically 50\% larger for SPT, and
60 to 100\% larger for DUO. For the combination {\sc Planck}+LSST, it
is interesting to note that the inclusion of $dN/dz$ improves the
neutrino mass determination by a factor two: using only information
from the CMB and from $P(k,z)$, the authors find $\sigma(M_\nu)\simeq
0.09$ eV, while adding $dN/dz$ they get $\sigma(M_\nu)\simeq
0.04$ eV. Including the DUO and SPT data marginally improves this
bound.  

\begin{figure}[t]
\begin{center}
\vspace{-2cm}
\hspace{-1.5cm}
\includegraphics[width=.95\textwidth]{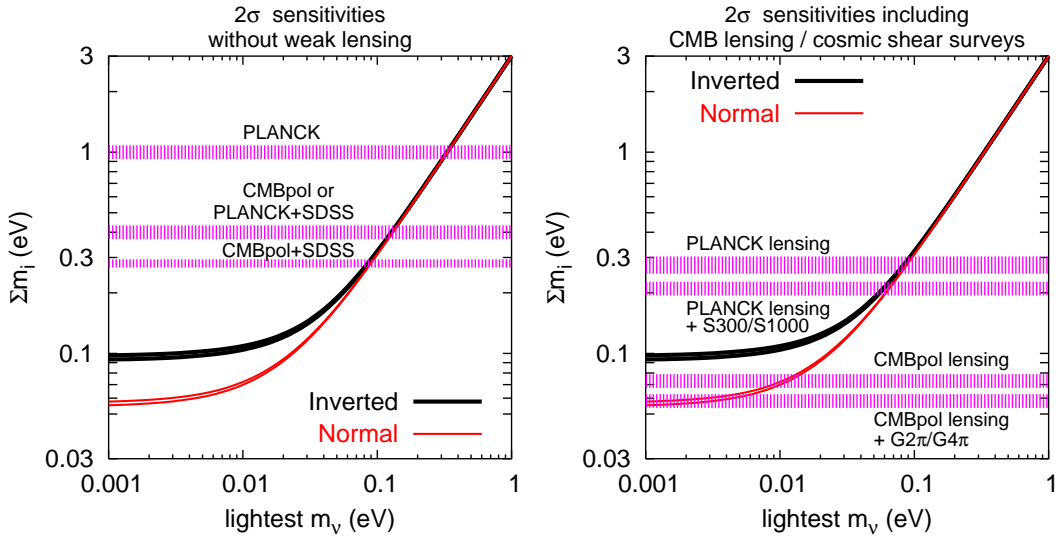}
\caption{\label{fig_future} Forecast 2$\sigma$ sensitivities to the
total neutrino mass from future cosmological experiments, as described
in Sec.\ \ref{sec:future}, compared to the values in agreement with
present neutrino oscillation data in Eq.\ (\ref{oscpardef}) (assuming
a future determination at the $5\%$ level). Left: sensitivities
expected for future CMB experiments (without lensing extraction),
alone and combined with the completed SDSS galaxy redshift
survey. Right: sensitivities expected for future CMB experiments
including lensing information, alone and combined with future cosmic
shear surveys.  Here CMBpol refers to a hypothetical CMB experiment
roughly corresponding to the {\sc Inflation Probe} mission.}
\end{center}
\end{figure}

\section{Conclusions}
\label{sec:concl}

Cosmology has played an important role in constraining neutrino
properties, providing information on these elusive particles that
complements the efforts of laboratory experiments, and neutrino
cosmology will remain an active research field in the next years. 

In this review, we were interested in the connection between neutrino
masses and cosmology. 
We have described in detail how the evolution of cosmological
perturbations is modified by massive neutrinos, and how their effects
leave an imprint in the different cosmological observables. 
We focused on the standard case of three active neutrino
species, known to be massive thanks to the experimental results
on flavour neutrino oscillations which, however, do not fix the
absolute neutrino mass scale. 

We saw how the analysis of cosmological observables can provide either
a positive indication or an upper bound on a total neutrino mass in
the range from 0.05 to a few eV indicated by oscillation data and
other laboratory results. We discussed why a unique cosmological bound
does not exist: the values depend on the assumed cosmological model
and the considered cosmological data. Within a minimal $\Lambda$MDM
model, the present cosmological bounds at $95\%$ CL are given in
Tables \ref{table:mass_CMB}-\ref{table:mass_LSS_2}, ranging from
$2$-$3$ eV with CMB data only to more stringent values when LSS data
are added: $0.9$-$1.7$ eV considering only the shape of the matter power
spectrum from 2dF or SDSS data, or $0.3$-$0.9$ eV fixing the
light-to-mass bias and/or adding Lyman-$\alpha$ forest data and/or the
SDSS measurement of the baryon oscillation peak. These values limit
the neutrino masses in the so-called degenerate region, as presented
in Fig.\ \ref{fig_current}. We also discussed how these cosmological
bounds could be relaxed or even disappear when adding some extra
cosmological parameters or in scenarios with non-standard relic
neutrinos. It is interesting to note that these mass bounds also apply
for other thermal relic particles, such as axions with masses in the
eV range as shown in \cite{Hannestad:2005df}.

Finally, we have discussed how the sensitivity to neutrino masses will
improve with new cosmological data, such as new CMB experiments or
larger galaxy redshift surveys. In particular we have reviewed the
excellent prospects of future weak lensing measurements, either
from CMB data or with cosmic shear surveys. We saw how these future
data will be sensitive to a total neutrino mass well below 1 eV, even
when considering extra cosmological parameters. 

We give a graphical summary of the forecast sensitivities to neutrino
masses of different cosmological data in Fig.\ \ref{fig_future},
compared to the allowed values of neutrino masses in the two possible
3-neutrino schemes. One can see from this figure that there are very
good prospects for testing neutrino masses in the degenerate and
quasi-degenerate mass regions above $0.2$ eV or so. A detection at a
significant level of the minimal value of the total neutrino mass in
the inverted hierarchy scheme will demand the combination of future
data from CMB lensing and cosmic shear surveys, whose more ambitious
projects will provide a 2$\sigma$ sensitivity to the minimal value in
the case of normal hierarchy (of order $0.05$ eV). The combination of
CMB observations with future galaxy cluster surveys (derived from the
same weak lensing observations, as well as X-ray and
Sunyaev--Zel'dovich surveys) should yield a similar sensitivity.

The information on neutrino masses from analyses of cosmological data
is complementary and can not replace the efforts in terrestrial
projects such as tritium beta decay and neutrinoless double beta decay
experiments (and vice versa). In particular, a positive result from
one of these three possibilities, which as reviewed in 
Sec.\ \ref{sec:numasses} do not measure the same combination of
neutrino masses and mixing parameters, should be cross-checked by the
others.  Of course any information on the absolute neutrino mass scale
will be a very important input for theoretical models of particle
physics beyond the Standard Model.

Note that light massive neutrinos could also play a role in the
generation of the baryon asymmetry of the Universe from a previously
created lepton asymmetry. In these leptogenesis scenarios, one can
also obtain quite restrictive bounds on light neutrino masses, which
are however model-dependent (see \cite{Buchmuller:2004nz} for a
recent review).

This review was entirely devoted to the case of light neutrinos with
masses of order eV or smaller. However, we would like to mention the
case of a sterile neutrino with a mass of the order of a few keV's and
a small mixing with the flavour neutrinos. It has been shown that
these keV neutrinos could also have interesting cosmological
consequences: they would play the role of dark matter and replace the
usual CDM component (see
\cite{Dodelson:1993je,Dolgov:2000ew,Abazajian:2005gj,Boyarsky:2006fg}
for an incomplete list of references) . Because of their large thermal
velocity (slightly smaller than that of active neutrinos), they would
behave as Warm Dark Matter and erase small-scale cosmological
structures.  Their mass can be bounded from below using Lyman-$\alpha$
forest data from quasar spectra, and from above using X-ray
observations
\cite{Abazajian:2001vt,Hansen:2001zv,Viel:2005qj,Abazajian:2005xn,Seljak:2006qw}. This
scenario, already disfavoured according to the results in
\cite{Seljak:2006qw}, is currently under careful examination.

\section*{Note added}

A few new analyses of cosmological data including non-zero neutrino
masses appeared after this review was accepted for publication. In
Ref.\ \cite{Seljak:2006bg} the authors used data on the Lyman-$\alpha$
forest from SDSS (improved thanks to the inclusion of the mean flux
constraints derived from a principal component analysis of quasar
spectra), SDSS and 2dF galaxy clustering, SNIa and CMB power spectra
(WMAP3 and others) to get an upper bound of $M_\nu<0.17$ eV (at $95\%$
CL) on the total mass of the three active neutrinos ($M_\nu<0.26$ eV
for four thermalized species). Instead, in Ref.\
\cite{Fukugita:2006rm} the limit $M_\nu<2.0$ eV for three neutrinos
was found using WMAP3 data only, in agreement with the result of the
WMAP team \cite{WMAP3:Spergel}.

\section*{Acknowledgments}
We would like to thank Elena Pierpaoli for many useful comments, as
well as Alessandro Cuoco, Alexander Dolgov, Kazuhide Ichikawa,
Gianpiero Mangano, Laurence Perotto, Georg Raffelt, Rafael Rebolo,
Jos\'e A.\ Rubi\~no-Mart\'{\i}n, Subir Sarkar and Pasquale Serpico for
interesting feedback on this manuscript. This work was supported by a
MEC-IN2P3 agreement and by the European Network of Theoretical
Astroparticle Physics ILIAS/N6 under contract number
RII3-CT-2004-506222. SP was supported by the Spanish grants
FPA2005-01269, BFM2002-00345 and GV/05/017 of Generalitat Valenciana,
as well as by a Ram\'{o}n y Cajal contract of MEC.

\end{document}